\documentclass[11pt]{article}
\pdfoutput=1
\usepackage[sumlimits,intlimits,namelimits]{amsmath}
\usepackage{amssymb}
\usepackage[english]{babel}
\usepackage{bbm}
\usepackage{parskip}
\usepackage[para]{footmisc}
\usepackage[margin=1cm]{caption}

\usepackage{graphicx,subfig}      
\DeclareGraphicsExtensions{.pdf,.png,.jpg}

\usepackage{cite}

\usepackage[makeroom]{cancel}
\usepackage[usenames,dvipsnames]{color}
\usepackage[dvipsnames]{xcolor}

\newcommand{\K}{\textrm{K}}
\newcommand{\beq}{\begin{equation}}
\newcommand{\eeq}{\end{equation}}
\newcommand{\bea}{\begin{eqnarray}}
\newcommand{\eea}{\end{eqnarray}}
\newcommand{\godo}{G_{\mathcal{O}^{\dagger}\mathcal{O}}}
\newcommand{\good}{G_{\mathcal{O}\mathcal{O}^{\dagger}}}
\newcommand{\goo}{G_{\mathcal{O}\mathcal{O}}}
\newcommand{\godod}{G_{\mathcal{O}^{\dagger}\mathcal{O}^{\dagger}}}

\newcommand{\rodo}{\rho_{\mathcal{O}^{\dagger}\mathcal{O}}}
\newcommand{\rood}{\rho_{\mathcal{O}\mathcal{O}^{\dagger}}}

\newcommand{\rhofano}{\rho_{\textrm{Fano}}}

\def\pa{\partial}

\def\le{\left}
\def\ri{\right}

\def\dif{\mathrm{d}}
\def\a{\alpha}
\def\b{\beta}
\def\g{\gamma}
\def\d{\delta}

\def\l{\lambda}

\def\F{\Phi}
\def\f{\phi}
\def\e{\epsilon}
\def\m{\mu}
\def\n{\nu}
\def\o{\omega}

\def\z{\zeta}

\def\k{\chi}

\global\parskip 6pt

\newcommand{\p}{\partial}

\newcommand{\qed}{\nobreak \ifvmode \relax \else\ifdim\lastskip<1.5em 
\hskip-\lastskip\hskip1.5em plus0em minus0.5em \fi \nobreak\vrule height0.75em 
width0.5em depth0.25em\fi}

\newcommand{\be}{\begin{eqnarray}}
\newcommand{\ee}{\end{eqnarray}}

\def\>{\rangle}
\def\<{\langle}
\def\d{\hbox{d}}

\newcommand{\executeiffilenewer}[3]{%
\ifnum\pdfstrcmp{\pdffilemoddate{#1}}%
{\pdffilemoddate{#2}}>0%
{\immediate\write18{#3}}\fi%
}

\graphicspath{{pics/}}

\usepackage{hyperref}

\def\){\right)}
\def\({\left( }
\def\]{\right] }
\def\[{\left[ }

\def\NO{\nonumber}

\def\bea{\begin{eqnarray}}
\def\eea{\end{eqnarray}}

\def\bal#1\eal{\begin{align}#1\end{align}}

\def\bald{\begin{aligned}}
\def\eald{\end{aligned}}

\def\bsub{\begin{subequations}}
\def\esub{\end{subequations}}

\def\beqx{\begin{displaymath}}
\def\eeqx{\end{displaymath}}

\newcommand{\bmat}{\left(\begin{array}}
\newcommand{\emat}{\end{array}\right)}

\def\a{\alpha}
\def\b{\beta}

\def\d{\delta}
\def\e{\epsilon}
\def\f{\phi}
\def\g{\gamma}

\def\j{\psi}
\def\k{\kappa}
\def\l{\lambda}
\def\m{\mu}
\def\n{\nu}
\def\o{\omega}
    \def\om{\omega}
\def\p{\pi}

\def\z{\zeta}

\def\F{\Phi}
\def\G{\Gamma}

    \def\Th{\Theta}

    \def\vth{\vartheta}

\def\ca{{\cal A}}

\def\cc{{\cal C}}

\def\cg{{\cal G}}

\def\cj{{\cal J}}

\def\co{{\cal O}}

\def\car{{\cal R}}
\def\cs{{\cal S}}

\def\cw{{\cal W}}

\def\bb#1{\ensuremath{\mathbb{#1}}} 

\def\bo{{\raise-.3ex\hbox{\large$\Box$}}}               
\def\pa{\partial}                                       
\def\face{{\raise.2ex\hbox{$\displaystyle \bigodot$}\mskip-2.2mu \llap {$\ddot
        \smile$}}}                                   
\def\>{\rangle}                                      
\def\<{\langle}                                      

\def\tx#1{\text{#1}}
\def\sbtx#1{{}_{\rm #1}}                           
\newcommand{\sub}[1]{\phantom{}_{(#1)}\phantom{}}    
\def\wt#1{\widetilde{#1}}                            
\def\Hat#1{\widehat{#1}}                             
\def\leftrightarrowfill{$\mathsurround=0pt \mathord\leftarrow \mkern-6mu
        \cleaders\hbox{$\mkern-2mu \mathord- \mkern-2mu$}\hfill
        \mkern-6mu \mathord\rightarrow$}        
\def\dvec#1{\vbox{\ialign{##\crcr
        \leftrightarrowfill\crcr\noalign{\kern-1pt\nointerlineskip}
        $\hfil\displaystyle{#1}\hfil$\crcr}}}           

\def\-{\hphantom{-}}

\begin{document}

\begin{titlepage}
\text{\ }\hfill{OUTP-16-27P}
\\
\text{\ }\hfill{FPAUO-16/16}
\\
\text{\ }\hfill{SISSA 61/2016/FISI}
\vspace*{.8cm}
\begin{center}
{\Large{\bf Two-point Functions in a Holographic Kondo Model}} \\[.5ex]
\vspace{.8cm}
Johanna Erdmenger$^{a,b,}$\footnote{jke@mpp.mpg.de},
Carlos Hoyos$^{c,}$\footnote{hoyoscarlos@uniovi.es},
Andy O'Bannon$^{d,}$\footnote{a.obannon@soton.ac.uk},\\
Ioannis Papadimitriou$^{e,}$\footnote{ioannis.papadimitriou@sissa.it},
Jonas Probst$^{f,}$\footnote{Jonas.Probst@physics.ox.ac.uk},
Jackson~M.~S.~Wu$^{g}$\footnote{jknw350@yahoo.com}
\vspace{0.6cm}\\
\em{$^a$Institut f\"ur Theoretische Physik und Astrophysik, Julius-Maximilians-Universit\"at W\"urzburg,\\ Am Hubland, D-97074 W\"urzburg, Germany.\\
\vskip.2cm
$^b$Max-Planck-Institut f\"ur Physik (Werner-Heisenberg-Institut),\\
F\"ohringer Ring 6, D-80805 Munich, 
Germany.
\\ \vskip.2cm
$^c$Department of Physics, Universidad de Oviedo, Avda.~Calvo Sotelo 18, 33007, Oviedo, Spain.
\\ \vskip.2cm
$^d$STAG Research Centre, Physics and Astronomy, University of Southampton,
Highfield, Southampton SO17 1BJ, United Kingdom.
\\ \vskip.2cm
$^e$ SISSA and INFN - Sezione di Trieste, Via Bonomea 265, I 34136
Trieste, Italy. 
\\ \vskip.2cm
$^f$ Rudolf Peierls Centre for Theoretical Physics, University of Oxford,
1 Keble Road, Oxford~OX1~3NP, United Kingdom.
\\ \vskip.2cm
$^g$Department of Physics and Astronomy, University of Alabama, \\ Tuscaloosa, AL 35487, USA.
}
\end{center}
\vspace*{0.5cm}
\begin{abstract}
We develop the formalism of holographic renormalization to compute two-point functions in a holographic Kondo model. The model describes a $(0+1)$-dimensional impurity spin of a gauged $SU(N)$ interacting with a $(1+1)$-dimensional, large-$N$, strongly-coupled Conformal Field Theory (CFT). We describe the impurity using Abrikosov pseudo-fermions, and define an $SU(N)$-invariant scalar operator $\mathcal{O}$ built from a pseudo-fermion and a CFT fermion. At large $N$ the Kondo interaction is of the form $\mathcal{O}^{\dagger} \mathcal{O}$, which is marginally relevant, and generates a Renormalization Group (RG) flow at the impurity. A second-order mean-field phase transition occurs in which $\mathcal{O}$ condenses below a critical temperature, leading to the Kondo effect, including screening of the impurity. Via holography, the phase transition is dual to holographic superconductivity in $(1+1)$-dimensional Anti-de Sitter space. At all temperatures, spectral functions of $\mathcal{O}$ exhibit a Fano resonance, characteristic of a continuum of states interacting with an isolated resonance. In contrast to Fano resonances observed for example in quantum dots, our continuum and resonance arise from a $(0+1)$-dimensional UV fixed point and RG flow, respectively. In the low-temperature phase, the resonance comes from a pole in the Green's function of the form $-i \langle {\cal O} \rangle^2$, which is characteristic of a Kondo resonance.
\end{abstract}
\vspace*{.25cm}
\end{titlepage}

\tableofcontents
\addtocontents{toc}{\protect\setcounter{tocdepth}{1}}
\renewcommand{\theequation}{\arabic{section}.\arabic{equation}}
\setcounter{footnote}{0}

\section{Introduction and Summary}
\label{sec:intro}
\setcounter{equation}{0}

The Kondo model of a magnetic impurity interacting with a Fermi liquid of electrons, proposed by Jun Kondo in 1964~\cite{PTP.32.37}, has been seminal for both experimental and theoretical physics. In experimental physics, the Kondo model explains the thermodynamic and transport properties of many systems, including certain types of quantum dots~\cite{Goldhaber1998,Cronenwett24071998} and certain metals doped with magnetic impurities~\cite{PTP.32.37,0034-4885-37-2-001,Gruner1978591}. Most famously, for doped metals the Kondo model successfully describes the logarithmic rise of the electrical resistivity $\rho$ with decreasing temperature $T$. In theoretical physics, the Kondo model provides perhaps the simplest example of a renormalization group (RG) flow exhibiting asymptotic freedom, the dynamical generation of a scale, namely the Kondo temperature, $T_K$, and a non-trivial infra-red (IR) fixed point describing the screening of the impurity by the electrons. As a result, the Kondo model has played a central role in the development of many techniques in theoretical physics: Wilson's numerical RG~\cite{Wilson:1974mb,PhysRevB.21.1003,PhysRevB.21.1044}, integrability~\cite{PhysRevLett.45.379,Wiegmann:1980,RevModPhys.55.331,doi:10.1080/00018738300101581,0022-3719-19-17-017,1994cond.mat..8101A,ZinnJustin1998, PhysRevB.58.3814}, large-$N$ limits~\cite{PhysRevB.35.5072,RevModPhys.59.845,1997PhRvL..79.4665P,1998PhRvB..58.3794P,2006cond.mat.12006C,2015arXiv150905769C}, Conformal Field Theory (CFT)~\cite{Affleck:1990zd,Affleck:1990by,Affleck:1990iv,Affleck:1991tk,PhysRevB.48.7297,Affleck:1995ge}, and more. For reviews of many of these, see for example refs.~\cite{Hewson:1993,doi:10.1080/000187398243500}.

Indeed, given the successes of these techniques, the single-impurity Kondo model is often called a ``solved problem.'' However, in reality many fundamental questions about the Kondo model remain unanswered, such as how to measure (or even \textit{define}) the size of the Kondo screening cloud, how entanglement entropy (EE) depends on the size of a spatial subsystem, or how observables evolve after a (quantum) quench, \textit{i.e.}\ after the Kondo model is ``kicked'' far from equilibrium.

Moreover, many generalizations of the original Kondo model remain impervious to the existing techniques. For example, what if we replace the electron Fermi liquid with (strongly) interacting degrees of freedom, such as a Luttinger liquid? What if multiple impurities interact not only with the electrons, but also with each other? Answers to these questions are urgently needed to understand important experimental systems. For example, a heavy fermion compound can be described as a dense lattice of impurities in which the competition between the Kondo and inter-impurity interactions leads to a quantum critical phase very similar to the  ``strange metal'' phase of the cuprate superconductors. Understanding the strange metal phase may be the key to understanding the mechanism of high-temperature superconductivity. The Kondo lattice therefore remains a major unsolved problem.

Motivated by these questions, in a series of papers we have developed an alternative Kondo model, based on holographic duality~\cite{Erdmenger:2013dpa,O'Bannon:2015gwa,Erdmenger:2015spo,Erdmenger:2015xpq}. Holography equates certain strongly-interacting quantum field theories (QFTs) with weakly-coupled theories of gravity in one higher dimension. Holography is therefore a natural tool for studying impurities coupled to strongly-interacting degrees of freedom, and is particularly well-suited for studying EE and far-from-equilibrium evolution.

Our holographic model is based on the large-$N$~\cite{PhysRevB.35.5072,RevModPhys.59.845,1997PhRvL..79.4665P,1998PhRvB..58.3794P,2003PhRvL..90u6403S,2004PhRvB..69c5111S,2006cond.mat.12006C,2015arXiv150905769C} and CFT~\cite{1998PhRvB..58.3794P,Affleck:1990zd,Affleck:1990by,Affleck:1990iv,Affleck:1991tk,PhysRevB.48.7297,Affleck:1995ge} approaches to Kondo physics. The large-$N$ approach involves replacing the $SU(2)$ spin symmetry with $SU(N)$ and then sending $N\to\infty$, keeping $T_K$ fixed. Following many previous large-$N$ Kondo models~\cite{0022-3719-19-17-017,PhysRevB.35.5072,1998PhRvB..58.3794P,2003PhRvL..90u6403S,2004PhRvB..69c5111S}, we restrict to an impurity spin in a totally anti-symmetric representation of $SU(N)$, whose Young tableau is a single column with $\mathcal{Q} < N$ boxes, and describe the impurity spin using Abrikosov pseudo-fermions, $\chi$, constrained to obey $\chi^{\dagger} \chi = \mathcal{Q}$. The Kondo coupling between the impurity spin and the electrons is then of the form $\lambda{\cal O}^{\dagger}{\cal O}$, where $\lambda$ is the Kondo coupling constant and ${\cal O}=\psi^{\dagger} \chi $, with $\psi$ an electron. At large $N$, the screening of the impurity appears as the formation of the condensate $\langle {\cal O} \rangle \neq 0$ below a critical temperature $T_c \simeq T_K$~\cite{0022-3719-19-17-017,PhysRevB.35.5072,2003PhRvL..90u6403S,2004PhRvB..69c5111S}. We thus refer to the phases with $\langle \mathcal{O}\rangle=0$ and $\langle \mathcal{O}\rangle\neq 0$ as ``unscreened'' and ``screened,'' respectively. Crucially, the logarithmic rise of $\rho$ with $T$, which normally occurs when $T \gg T_K$, is absent at large $N$. However, the large-$N$ limit is useful at low temperatures, $T \leq T_K$, where $\lambda$ is large and hence conventional perturbation theory in $\lambda$ breaks down. When $T \ll T_K$, $\rho$ exhibits power-law scaling in $T$, with a power determined by the dimension of the leading irrelevant operator about the IR fixed point~\cite{1998PhRvB..58.3794P,PhysRevB.48.7297,Affleck:1995ge}.

The CFT approach to Kondo physics begins with the observation that the impurity couples only to the electron $s$-wave spherical harmonic, so non-trivial physics only occurs in the radial direction about the impurity~\cite{Affleck:1990zd,Affleck:1990iv,Affleck:1995ge}. The low-energy physics is therefore effectively one-dimensional. Linearizing about the Fermi momentum then produces a relativistic electron dispersion relation, with the Fermi velocity playing the role of the speed of light. The low-energy effective theory thus consists of free, relativistic fermions in one dimension, interacting with the impurity at the origin. That theory is a boundary CFT, which has an infinite number of symmetry generators, namely those of a single Virasoro algebra, plus Kac-Moody algebras for charge, spin, and channel (or flavor)~\cite{Affleck:1990zd,Affleck:1995ge}. These infinite accidental symmetries make the CFT approach very powerful. For example, together with the boundary conditions these symmetries determine the IR spectrum completely~\cite{Affleck:1990zd,Affleck:1990by,Affleck:1990iv,Affleck:1995ge}. The CFT approach also provides novel results for low-$T$ scaling exponents~\cite{Affleck:1990zd,Affleck:1990by,Affleck:1990iv,PhysRevB.48.7297,Affleck:1995ge}.

Our holographic model combines the large-$N$ and CFT approaches, and adds two more ingredients. First, we \textit{gauge}\ the $SU(N)$ spin symmetry, so that the impurity spin becomes an $SU(N)$ Wilson line. Second, we make the $SU(N)$ 't Hooft coupling large, so that the gauge degrees of freedom (adjoint fields) are strongly-interacting. These two ingredients are necessary to produce a tractable gravitational dual, with a small number of light fields in a classical limit. Indeed, \textit{all} holographic quantum impurity models to date use these two ingredients, as reviewed in refs.~\cite{Erdmenger:2013dpa,O'Bannon:2015gwa}.

To be specific, our holographic model includes four fields. First is an asymptotically $AdS_3$ metric, with Einstein-Hilbert action with negative cosmological constant, which is dual to the stress-energy tensor. Second is a Chern-Simons gauge field, $A$, dual to Kac-Moody currents, $J$, representing our electrons $\psi$. Third is a Maxwell gauge field, $a$, restricted to a co-dimension one, asymptotically $AdS_2$ brane, localized in the field theory direction, and dual to the Abrikosov pseudo-fermion charge $j \equiv \chi^{\dagger} \chi$. Fourth is a complex scalar field, $\Phi$, also restricted to the brane, charged under both $A$ and $a$, and dual to ${\cal O}=\psi^\dagger\chi$. In refs.~\cite{Erdmenger:2013dpa,O'Bannon:2015gwa} we treated the matter fields as probes of a BTZ black brane.

Our model is a novel impurity RG flow in both holography and condensed matter physics. In holography, our model is novel as a holographic superconductor~\cite{Hartnoll:2008vx,Hartnoll:2008kx} in an $AdS_2$ subspace of a higher-dimensional AdS space. Indeed, a general lesson of our model is that holographic superconductors in $AdS_2$ describe impurity screening. In condensed matter physics, our model describes a novel impurity RG flow between two strongly-interacting fixed points, unlike the original Kondo model, where the UV fixed point is trivial and the IR fixed point may or may not be trivial, depending on the number of channels~\cite{Nozieres:1980,Affleck:1990by,Affleck:1990iv,Affleck:1995ge}. More specifically, our $\lambda$ runs in the same way as the original Kondo model, but our model has a second coupling, the 't Hooft coupling, which does not run, and is large. Our holographic model not only reproduces expected large-$N$ Kondo physics, such as condensation of ${\cal O}$, screening of the charge $Q$, power-law scaling of $\rho$ with $T$ at low $T$~\cite{Erdmenger:2013dpa}, etc., but also exhibits novel phenomena due to the large 't Hooft coupling, as described below.

Indeed, using our holographic model, we have begun to address some of the open questions about Kondo physics. For example, in ref.~\cite{O'Bannon:2015gwa}, we introduced a second impurity in our holographic model, as a first step towards building a holographic Kondo lattice. We found evidence that the competition between Kondo and inter-impurity (RKKY) interactions may lead to a quantum phase transition. In ref.~\cite{Erdmenger:2015spo} we calculated the impurity entropy in our holographic model, by calculating the change in EE due to the impurity, for an interval of length $\ell$ centered on the impurity. Calculating the EE holographically required calculating the back-reaction of the $AdS_2$ matter fields on the metric~\cite{Erdmenger:2015spo,Erdmenger:2014xya}. The impurity screening reduced the impurity entropy, \textit{i.e.}\ reduced the number of impurity degrees of freedom, consistent with the $g$-theorem~\cite{Affleck:1991tk,Friedan:2003yc}. On the gravity side, the reduction in degrees of freedom appeared as a reduction in the volume of the bulk spacetime around the $AdS_2$ brane, similar to the deficit angle around a cosmic string. Furthermore, at low $T$ the EE decayed exponentially in $\ell$ as $\ell$ increased. The decay rate provides one definition of the Kondo screening length, which made a particularly intuitive appearance in the gravity theory, as a distance the $AdS_2$ brane ``bends.'' 

In this paper, we take a first step toward addressing another major open problem in Kondo physics: out-of-equilibrium evolution. In particular, we work in the probe limit, and compute response functions, namely the retarded Green's functions involving ${\cal O}$, $j$, and $J$, in the regime of linear response to small, time-dependent perturbations. We then compute the spectral functions, \textit{i.e.}\ the anti-Hermitian parts of the retarded Green's functions. We also separately calculate the poles in the Green's functions, dual to the quasi-normal modes (QNMs) of the fields in the BTZ black brane background. Generically, these poles give rise to peaks in the spectral functions.

We presented some of our results in a companion paper~\cite{Erdmenger:2016vud}. In this paper we will present full details of calculations and further results. In particular, we have three main results.

Our first main result is technical: we perform the holographic renormalization (holo-ren)~\cite{Henningson:1998gx,Balasubramanian:1999re,deBoer:1999tgo,deHaro:2000vlm,Bianchi:2001de,Bianchi:2001kw,Martelli:2002sp,Skenderis:2002wp,Papadimitriou:2004ap,Papadimitriou:2010as,Papadimitriou:2016yit} of our model. The main challenge here is the well-known fact that a YM field diverges near the asymptotically $AdS_2$ boundary, unlike YM fields in higher-dimensional AdS spaces. That divergence can alter the asymptotics of fields coupled to the YM field, and indeed alters the asymptotics of our field $\Phi$. The asymptotic region is dual to the UV of the field theory~\cite{Aharony:1999ti}, so we learn that in our holographic model $j$ acts like an irrelevant operator, and in particular, changing the value of $\langle j \rangle$, which controls the impurity's spin, changes the dimension of ${\cal O}$ at the UV fixed point. Such behavior does not occur in non-holographic Kondo models, and so, by process of elimination, must be due to the strongly-interacting degrees of freedom we added. Strong-coupling effects can also appear in the IR, for example the leading irrelevant operator about the IR fixed point likely has non-integer dimension~\cite{Erdmenger:2013dpa}.

Our holo-ren draws from, and extends, several previous examples of holo-ren: for fields dual to irrelevant operators~\cite{vanRees:2011fr,vanRees:2011ir}, for our holographic two-impurity Kondo model~\cite{O'Bannon:2015gwa}, and for asymptotically conical (rather than asymptotically AdS) black holes~\cite{An:2016fzu}. The holo-ren provides covariant boundary counterterms, enabling us to compute renormalized correlators, including the thermodynamic free energy and two-point functions. The holo-ren also allows us to identify the Kondo coupling $\lambda$ from a boundary condition on $\Phi$~\cite{Erdmenger:2013dpa,O'Bannon:2015gwa,Erdmenger:2015spo,Erdmenger:2015xpq}.

As in many large-$N$ Kondo models, our holographic model exhibits a large-$N$, second-order, mean-field phase transition~\cite{Erdmenger:2013dpa,O'Bannon:2015gwa,Erdmenger:2015spo,Erdmenger:2015xpq}. For all $T$, one class of static solutions obeying the boundary conditions includes $\Phi =0$, dual to the unscreened phase, with $\langle \mathcal{O} \rangle = 0$. When $T \leq T_c$, another class of solutions appears, with $\Phi \neq 0$, dual to the screened phase, with $\langle \mathcal{O} \rangle \neq 0$. For all $T\leq T_c$, the $\Phi \neq 0$ solution has lower free energy than the $\Phi=0$ solution, so a phase transition occurs at $T_c$, with mean-field exponent: for $T$ just below $T_c$, $\langle \mathcal{O} \rangle \propto (T_c -T)^{1/2}$~\cite{Erdmenger:2013dpa}.

In the unscreened phase, the holo-ren reveals that the only non-trivial retarded Green's function in our model is $\langle \mathcal{O}^{\dagger} \mathcal{O} \rangle$, with all other one- and two-point functions completely determined by $\langle \mathcal{O} \rangle$, the Ward identities for the currents $j$ and $J$, and the particle-hole transformation $\mathcal{Q} \to N-\mathcal{Q}$. For example, $\langle \mathcal{O} \mathcal{O}^{\dagger}\rangle$ can be obtained from $\langle \mathcal{O}^{\dagger} \mathcal{O} \rangle$ by taking $\mathcal{Q} \to N-\mathcal{Q}$. We denote $\langle \mathcal{O}^{\dagger} \mathcal{O} \rangle$'s Fourier transform as $\godo$, which we compute as a function of complex frequency $\omega$, and the associated spectral function as $\rodo \equiv - 2 \,\textrm{Im}\,\godo$, which we compute for real $\omega$. We are able to compute $\godo$ analytically, by obtaining an exact solution to $\Phi$'s Klein-Gordon equation (with gauge covariant derivatives) in $AdS_2$, with boundary condition involving the Kondo coupling $\lambda$.

The defect's asymptotic $AdS_2$ isometry is dual to a $(0+1)$-dimensional conformal symmetry. When $T>T_c$, the only breaking of that conformal symmetry is through $T$ and the running of $\lambda$. For static solutions, we can approach the UV fixed point by sending $T\to\infty$, which also sends $\lambda \to 0$ due to asymptotic freedom. When $\lambda \to 0$, $\Phi$'s boundary condition reduces to Dirichlet~\cite{Erdmenger:2013dpa,O'Bannon:2015gwa}, guaranteeing that $\godo$ indeed takes the form required by $(0+1)$-dimensional conformal symmetry~\cite{Faulkner:2009wj,Sachdev:2015efa}.

More generally, our model falls into one of the three known classes of models whose large-$N$ fixed points exhibit $(0+1)$-dimensional conformal symmetry. The first are holographic $AdS_2$ models, such as our model. The second are large-$N$ quantum impurity models, including large-$N$ Kondo models (without holography)~\cite{1998PhRvB..58.3794P}. The third are so-called Sachdev-Ye-Kitaev (SYK) models, namely fermions on a lattice without kinetic terms and with long-range many-body interactions, in a large-$N$ limit~\cite{Sachdev:1992fk,kitaev,Sachdev:2015efa,Polchinski:2016xgd,You:2016ldz,Fu:2016yrv,Jevicki:2016bwu,Maldacena:2016hyu,Jensen:2016pah,Jevicki:2016ito,Gu:2016oyy,Gross:2016kjj,Berkooz:2016cvq,Fu:2016vas,Witten:2016iux}. For all three classes, $(0+1)$-dimensional conformal symmetry completely determines any Green's function, such as $\godo$, in terms of scaling dimension and global symmetry charges~\cite{Sachdev:2015efa,Jensen:2016pah,Davison:2016ngz}.

However, in our model, as $T$ decreases and $\lambda$ grows, $(0+1)$-dimensional conformal symmetry is broken. As $T \to T_c$ from above, in the complex $\omega$-plane the lowest pole in $\godo$, meaning the pole closest to the origin, which we denote $\omega^*$, moves towards the origin. When $T=T_c$, $\omega^*$ reaches the origin, and when $T<T_c$, $\omega^*$ moves into the upper half of the complex $\omega$ plane, signaling the instability of the unscreened phase when $T<T_c$, as expected~\cite{Erdmenger:2013dpa}. In contrast, in the standard (non-holographic) Kondo model, at large $N$ and at leading order in perturbation theory in $\lambda$, the lowest pole sits exactly at the origin of the complex $\omega$ plane for all $T \geq T_c$~\cite{Coleman2015}. By process of elimination, our results for the movement of $\omega^*$ must arise from the additional degrees of freedom of our holographic model, and in particular must be a strong coupling effect, since we do not rely on perturbation theory in either $\lambda$ or the 't Hooft coupling.

A pole in a retarded Green's function (for complex $\omega$) leads to a peak in the associated spectral function (for real $\omega$). Our second main result is for $\rodo$ in the unscreened phase: $\omega^*$ produces the only significant feature in $\rodo$, namely a peak, and specifically a \textit{Fano resonance}. Fano resonances occur when one or more resonance appears within a continuum of states (in energy). In such cases, scattering states have two options: they can either scatter off the isolated resonance(s) (\textit{resonant scattering}), or they can bypass these resonances (\textit{non-resonant scattering}). The classic example is light scattering off the excited states of an atom. In spectral functions, the interference between the two options leads to a Fano resonance, which generically is asymmetric, with a minimum and a maximum (see fig.~\ref{fig:fanopics} (a)), and is determined by three parameters: the position, the width, and the \textit{Fano} or \textit{asymmetry} parameter, $q$, which controls the distance between the minimum and maximum. In physical terms, $q^2$ is proportional to the probability of resonant scattering over the probability of non-resonant scattering. For an introduction to Fano resonances, see for example ref.~\cite{RevModPhys.82.2257}.

In our case, the continuum comes from the $(0+1)$-dimensional fixed point dual to the $AdS_2$ subspace, where the scale invariance implies any spectral function must be power law in $\omega$, \textit{i.e.}\ a continuum. Our resonance arises from our relevant deformation, \textit{i.e.}\ our Kondo coupling, which necessarily breaks scale invariance. Moreover, the asymmetry of our Fano resonances is possible because particle-hole symmetry is generically broken when $|\mathcal{Q}-N/2|\neq 0$.

We expect asymmetric Fano resonances in any system with the same three ingredients, namely an effectively $(0+1)$-dimensional UV fixed point, resonances that appear when scale invariance is broken, and particle-hole symmetry breaking. In fact, Fano resonances have appeared in such systems, though they are often not identified as such. For example, Fano resonances appear in spectral functions of charged bosonic operators in the non-holographic large-$N$ Kondo model~\cite{1998PhRvB..58.3794P} and in holographic duals of extremal AdS-Reissner-Nordstrom black branes, whose near-horizon geometry is $AdS_2$~\cite{Faulkner:2009wj,Faulkner:2011tm,Sachdev:2015efa}. Indeed, we expect Fano resonances in $AdS_2$ models generically, such as Sachdev-Ye-Kitaev models~\cite{Sachdev:1992fk,kitaev,Sachdev:2015efa,Polchinski:2016xgd,Maldacena:2016hyu,Witten:2016iux}, if some deformation breaks scale invariance and produces a resonance. Specifically $(0+1)$ dimensions is special because any resonance must necessarily be immersed in a continuum, unlike higher dimensions, where the two may be separated in momentum and/or real space.

Fano resonances have been produced experimentally in side-coupled QDs~\cite{RevModPhys.82.2257,2000PhRvB..62.2188G}, that is, by coupling the discrete states in a QD to a continuum of states in a quantum wire. Crucially, however, in these cases $(0+1)$-dimensional scale invariance apparently plays no role: before the coupling between QD and quantum wire, spectral functions on the QD would be a sum of Lorentzians, not a scale-invariant continuum. Our Fano resonances therefore have a different physical origin from those in QDs, and are more characterisitc of $(0+1)$-dimensional fixed points, as explained above.

In the screened phase, the symmetry breaking condensate $\langle \mathcal{O} \rangle \neq 0$ induces operator mixing, so that generically all two-point functions are non-trivial. However, the holo-ren shows that all four scalar correlators are equivalent: $\godo = \good = \goo = \godod$, so we will discuss only $\godo$, which we compute numerically. Our third main result is: in the screened phase, the lowest pole in $\godo$, $\omega^*$, is purely imaginary, and moves down the imaginary axis as $T$ decreases. In fact, $\omega^* \propto - i \langle \mathcal{O}\rangle^2$ for $T$ just below $T_c$. The spectral function $\rodo$ then exhibits a Fano resonance symmetric under $\omega \to -\omega$.

This result is consistent with expectations from the standard (non-holographic) Kondo model. At finite $N$, an essential feature of the Kondo effect is the \textit{Kondo resonance}, a peak in the spectral function of the conduction electrons, with five characteristic features. First, for all $T$ the peak is localized in energy exactly at the Fermi energy. Second, for all $T$ the peak is localized in real space at the impurity. Third, as $T$ approaches $T_K$ from above, the peak's height rises logarithmically in $T$. Fourth, when $T$ reaches $T_K$, the peak's height saturates and remains for all lower $T$ at a value fixed by the impurity's representation (the Friedel sum rule). Fifth, as $T$ drops below $T_K$ and then continues to decrease, the peak narrows, and at $T=0$ has width $\propto \, T_K$. The Kondo resonance is a many-body effect (\textit{i.e.}\ is not obvious from the Kondo Hamiltonian) signaling the emergence of the highly-entangled state in which the conduction electrons act collectively to screen the impurity. For more details about the Kondo resonance, see for example the textbooks refs.~\cite{Hewson:1993,Phillips2012,Coleman2015}.

The features of the Kondo resonance change in the large-$N$ limit, as explained in ref.~\cite{Coleman2015} and references therein. In particular, the Kondo resonance is \textit{absent} in the unscreened phase ($T>T_c$), and appears only in the screened phase ($T<T_c$). If we introduce Abrikosov pseudo-fermions $\chi$, then due to operator mixing induced by $\langle \mathcal{O} \rangle \neq 0$, the Kondo resonance can be transmitted from the electron spectral function to other spectral functions. In particular, in $\godo$ the Kondo resonance appears as a pole of the form $\omega \propto - i \langle \mathcal{O}\rangle^2$. As mentioned above, for $T$ just below $T_c$, we indeed find a pole in $\godo$ of precisely that form, providing compelling evidence for a Kondo resonance in our model.

This paper is organized as follows. In section~\ref{sec:review} we review our holographic Kondo model. In section~\ref{sec:holorg} we perform the holo-ren of our model. In section~\ref{sec:fano} we review Fano resonances. We present our results for the unscreened phase in section~\ref{sec:highT}, and for the screened phase in section~\ref{sec:lowT}. We conclude in section~\ref{sec:discussion} with discussion of our results and suggestions for future research.

\section{Review: Holographic Kondo Model}
\label{sec:review}
\setcounter{equation}{0}

As mentioned above, our holographic model combines the CFT and large-$N$ approaches to the Kondo effect. In this section we will review these briefly and then introduce the action and equations of motion of our holographic model, and the transition between the unscreened to screened phases. For more details on the CFT, large-$N$, and holographic approaches to the Kondo effect, see refs.~\cite{Erdmenger:2013dpa,O'Bannon:2015gwa,Erdmenger:2015spo,Erdmenger:2015xpq}.

The CFT approach to the Kondo effect~\cite{Affleck:1990zd,Affleck:1990by,Affleck:1990iv,Affleck:1991tk,PhysRevB.48.7297,Affleck:1995ge} begins with a $(1+1)$-dimensional effective description: relativistic fermions that are free except for a Kondo interaction with the impurity at the boundary of space. In that description, left-moving fermions ``bounce off'' the boundary and become right-moving, interacting with the impurity in the process. By extending the half line to the entire real line, reflecting the right-movers to the ``new'' half of the real line, and re-labeling them as left-movers, we obtain a simpler description: left-movers alone, interacting with the impurity at the origin. The Hamiltonian (density) is then, in units where the Fermi velocity acting as speed of light is unity,
\beq
H=\frac{1}{2\pi} \psi^{\dagger}_{\alpha} i \partial_x \psi_{\alpha}+\lambda \, \delta(x) \, S^A \psi_{\alpha}^{\dagger} T^A_{\alpha\beta} \psi_{\beta}, \label{KondoCFT}
\eeq
where $\psi_{\a}^{\dagger}$ creates a left-moving electron with spin $\a$, $\lambda$ is the classically marginal Kondo coupling, $T^A_{\alpha\beta}$ are the generators of the $SU(2)$ spin symmetry ($A=1,2,3$) in the fundamental representation, and $S^A$ is the spin of the impurity, which is localized at $x=0$, hence the $\delta(x)$. The left-moving fermions form a chiral CFT, invariant under a single Virasoro algebra as well as $SU(2)_1$ and $U(1)$ Kac-Moody algebras, representing spin and charge, respectively (the $U(1)$ acts by shifting $\psi_{\alpha}$'s phase). With $k>1$ channels of fermions, the Kac-Moody algebra is enhanced to $SU(2)_k \times SU(k)_2 \times U(1)$.

The one-loop beta function for $\lambda$ is negative. As a result, a non-trivial RG flow occurs only for an anti-ferromagnetic Kondo coupling, $\lambda>0$. Due to asymptotic freedom, the UV fixed point is a trivial chiral CFT, namely free left-moving fermions and a decoupled impurity. The Virasoro and Kac-Moody symmetries and (trivial) boundary conditions then determine the spectrum of eigenstates completely~\cite{Affleck:1990zd,Affleck:1990by,Affleck:1990iv,Affleck:1995ge}. The IR fixed point will again be a chiral CFT, whose spectrum of eigenstates can be obtained from those in the UV by fusion with the impurity representation~\cite{Affleck:1990by}.

Our holographic Kondo model will also employ a large-$N$ limit~\cite{PhysRevB.35.5072,RevModPhys.59.845,1997PhRvL..79.4665P,1998PhRvB..58.3794P,2006cond.mat.12006C,2015arXiv150905769C}, which is based on replacing the $SU(2)$ spin symmetry with $SU(N)$ and then sending $N \to \infty$ with $N \lambda$ fixed. In particular, we will employ the large-$N$ description of the Kondo effect as symmetry breaking at the impurity's location~\cite{0022-3719-19-17-017,PhysRevB.35.5072,2003PhRvL..90u6403S,2004PhRvB..69c5111S}, which begins by writing $S^A$ in terms of Abrikosov pseudo-fermions,
\begin{align}
	S^A=\chi_{\alpha}^{\dagger} T^A_{\alpha\beta}\chi_{\beta},
	\label{chidef}
\end{align}
where $\chi_{\alpha}^\dagger$ creates an Abrikosov pseudo-fermion. We construct a state in the impurity's Hilbert space by acting on the vacuum with a number $\mathcal{Q}$ of the $\chi_{\alpha}^\dagger$. Because the $\chi^{\dagger}_\a$ anti-commute, such a state will be a totally anti-symmetric tensor product of the fundamental representation of $SU(N)$ with rank $\mathcal{Q}$. To obtain an irreducible representation, we must fix the rank $\mathcal{Q}$ by imposing a constraint,
\begin{align}
	\chi_{\alpha}^\dagger\chi_{\alpha}=\mathcal{Q}.
	\label{qconstraint}
\end{align}
Due to the anti-commutation, Abrikosov pseudo-fermions can only describe totally anti-symmetric representations of $SU(N)$, so that $\mathcal{Q} \in \{0,1,2,\ldots,N\}$. Following our earlier work~\cite{Erdmenger:2013dpa,O'Bannon:2015gwa,Erdmenger:2015spo,Erdmenger:2015xpq}, we will only consider totally anti-symmetric impurity representations.

Plugging eq.~\eqref{chidef} into the Kondo interaction term in eq.~\eqref{KondoCFT}, and using $\chi_{\alpha}$'s anti-commutation relations as well as the completeness relation satisfied by the fundamental-representation $SU(N)$ generators,
\begin{align}
	T^A_{\alpha\beta}T^A_{\gamma\delta}=\frac{1}{2}\le(\delta_{\alpha\delta}\delta_{\beta\gamma}-\frac{1}{N}\delta_{\alpha\beta}\delta_{\gamma\delta}\ri), \label{completeness}
\end{align}
we can re-write the Kondo interaction as
\begin{align}
	\lambda \, S^A\psi^{\dagger}_{\gamma} T^A_{\gamma\delta}\psi_{\delta}=\lambda \left(\chi^{\dagger}_{\alpha} T^A_{\alpha\beta}\chi_{\beta}\right)\left(\psi^{\dagger}_{\gamma} T^A_{\gamma\delta}\psi_{\delta}\right)=\frac{1}{2}\lambda \left(-\mathcal{O}^{\dagger}\mathcal{O}+\mathcal{Q}-\frac{\mathcal{Q}}{N}\left(\psi^{\dagger}_{\alpha}\psi_{\alpha}\right)\right), \label{KondoAbrikosov}
\end{align}
where the scalar operator $\mathcal{O} \equiv \psi^{\dagger}_\a \chi_{\alpha}$ is $(0+1)$-dimensional, \textit{i.e.} is a function of time $t$ only, because $\chi_{\alpha}$ cannot propagate away from the impurity's location, $x=0$. Clearly, $\mathcal{O}$ is a singlet of the spin $SU(N)_k$ symmetry, is in the same $SU(k)_N \times U(1)$ representation as $\psi^{\dagger}_\a$, and has the same auxiliary $U(1)$ charge as $\chi_{\alpha}$. Classically $\psi_\a$ has dimension $1/2$ and $\chi_{\alpha}$ has dimension zero, so $\mathcal{O}$ has dimension $1/2$. The Kondo interaction eq.~\eqref{KondoAbrikosov} is thus classically marginal, \textit{i.e.}\ $\lambda$ is classically dimensionless.

We can introduce Abrikosov pseudo-fermions for any $N$, but let us now take the large-$N$ limit. In eq.~\eqref{KondoAbrikosov} the $\mathcal{Q}$ and $(\mathcal{Q}/N) \psi^{\dagger}_\a \psi_\a$ terms are then sub-leading in $N$ relative to the $\mathcal{O}^\dagger\mathcal{O}$ term, so the Kondo interaction reduces to $- \lambda \mathcal{O}^\dagger\mathcal{O}/2$. The solution of the large-$N$ saddle point equations reveals a second-order mean-field phase transition: below a critical temperature $T_c$, on the order of but distinct from $T_{\K}$, $\langle \mathcal{O} \rangle \neq 0$~\cite{0022-3719-19-17-017,PhysRevB.35.5072,2003PhRvL..90u6403S,2004PhRvB..69c5111S}, spontaneously breaking the channel symmetry down to $SU(k-1)$ and the $U(1)$ charge and $U(1)$ auxiliary symmetry down to the diagonal $U(1)$. Of course, spontaneous symmetry breaking in $(0+1)$ dimensions is impossible for finite $N$: the phase transition is an artifact of the large-$N$ limit. Corrections in $1/N$ change the phase transition to a smooth cross-over~\cite{0022-3719-19-17-017}. The large-$N$ limit describes many characteristic phenomena of the Kondo effect only when $T \leq T_c$, where $\langle \mathcal{O}\rangle \neq 0$, including the screening of the impurity by the electrons, and a phase shift of the electrons.

As described in section~\ref{sec:intro}, to obtain a classical Einstein-Hilbert holographic Kondo model, we want to combine the CFT and large-$N$ approaches and gauge the $SU(N)_k$ spin symmetry, which introduces the 't Hooft coupling, which we want to be large. Of course, the $SU(N)_k$ symmetry is anomalous, and so should not be gauged. To suppress the anomaly, we work in the probe limit: when $N\to \infty$ we hold $k$ fixed, so that $k \ll N$, and then compute expectation values only to order $N$. In the probe limit the $SU(N)_k$ anomaly does not appear~\cite{Buchbinder:2007ar,Erdmenger:2013dpa}, so that in effect $SU(N)_k \to SU(N)$.

Each $SU(N)$-invariant, single-trace, low-dimension (\textit{i.e.}\ dimension of order $N^0$) operator is dual to a field in the gravity dual. The stress-energy tensor is dual to the metric. The $SU(N)$ currents are not $SU(N)$-invariant, and hence have no dual fields. The $SU(k)_N \times U(1)$ Kac-Moody currents are dual to an $SU(k)_N \times U(1)$ Chern-Simons gauge field~\cite{Kraus:2006wn}, which we call $A$. The $U(1)$ charge $j=\chi^{\dagger}_{\alpha} \chi_{\alpha}$ is dual to a $U(1)$ gauge field, which we call $a$, localized to $x=0$. The complex scalar $\mathcal{O}$ is bi-fundamental under $SU(k)_N \times U(1)$ and the $U(1)$ with charge $j$, and is dual to a complex scalar field, $\Phi$, also localized to $x=0$, and bi-fundamental under $A$ and $a$. For simplicity, following refs.~\cite{Erdmenger:2013dpa,O'Bannon:2015gwa,Erdmenger:2015spo,Erdmenger:2015xpq} we will take $k=1$, so that the $SU(k)_N \times U(1)$ Kac-Moody symmetry reduces to $U(1)$. The Chern-Simons gauge field $A$ is then Abelian, with field strength $F=dA$. Similarly, $a$ has field strength $f=da$.

To describe a $(1+1)$-dimensional CFT with non-zero $T$, we use the BTZ black brane metric (with asymptotic $AdS_3$ radius set to unity),
\begin{align}
\label{ads3metric}
	ds^2_{\textrm{BTZ}} = \frac{1}{z^2}\le(\frac{1}{h(z)}\dif z^2 -h(z)\dif t^2+\dif x^2\ri), \qquad h(z) = 1-\frac{z^2}{z_H^2},
\end{align}
where $z$ is the radial coordinate, with the boundary at $z=0$ and horizon at $z=z_H$, $t$ and $x$ are the CFT time and space directions, and $\mu,\nu=z,t,x$. The CFT's temperature is dual to the black brane's Hawking temperature, $T=1/(2\pi z_H)$. The fields $a$ and $\Phi$ are localized to $x=0$, \textit{i.e.}\ to the submanifold spanned by $t$ and $z$, whose induced metric is asymptotically $AdS_2$,
\beq
\label{ads2metric}
ds^2_{AdS_2}=g_{mn} dx^m dx^n =  \frac{1}{z^2}\le(\frac{1}{h(z)}\dif z^2 -h(z)\dif t^2\right),
\eeq
where $m,n=t,z$. The determinant of the metric in eq.~\eqref{ads2metric} is $g= - 1/z^4$.

The classical action of the holographic Kondo model of refs.~\cite{Erdmenger:2013dpa,O'Bannon:2015gwa,Erdmenger:2015spo,Erdmenger:2015xpq} is the simplest action quadratic in the fields. We will split the bulk action into two terms, namely the Chern-Simons action for $A$, $S_{\textrm{CS}}$, and the bulk terms for the fields in the asymptotically $AdS_2$ submanifold, $S_{AdS_2}$,
\begin{subequations}
\label{action}
\beq
S=S_{\textrm{CS}}+S_{AdS_2},
\eeq
\beq
S_{\textrm{CS}} = -\frac{N}{4\pi}\int\limits_{AdS_3} A\wedge\dif A,
\eeq
\beq	
S_{AdS_2} = -N\int\limits_{AdS_2}\dif^2x\: \sqrt{-g}\le[\frac{1}{4} f^{mn}f_{mn} + \ \le(D^m\F\ri)^\dagger \le(D_m\F\ri) +M^2 \Phi^{\dagger} \Phi \ri], \label{actionAdS2}
\eeq
\end{subequations}
where $D_m$ is a gauge-covariant derivative,
\beq
D_m\F =\le(\pa_m+iA_m-ia_m \ri) \F, \label{covariantderiv}
\eeq
and $M^2$ is $\Phi$'s mass-squared. We will discuss the value of $M^2$, and the boundary terms that must be added to $S$ for holo-ren, in section~\ref{sec:holorg}. We will also discuss the equations of motion following from eq.~\eqref{action}, and their solutions, in detail in section~\ref{sec:holorg}. In the remainder of this section we will focus on features of the equations of motion and their solutions relevant for our model's phase structure.

We split $\Phi$ into a modulus and phase, $\Phi = e^{i\psi} \phi$. Furthermore, throughout this paper we work in a gauge with $A_z=0$ and $a_z=0$. As shown in refs.~\cite{Erdmenger:2013dpa,O'Bannon:2015gwa,Erdmenger:2015spo,Erdmenger:2015xpq}, a self-consistent gauge choice and ansatz that can describe a static state with $\mathcal{Q} \neq 0$ and possibly $\langle {\cal O} \rangle \neq 0$ includes $A_x(z)$, $a_t(z)$, and $\phi(z)$, with all other fields set to zero. The equations of motions for these fields are
\begin{subequations}
\label{eoms}
\beq
 \label{Aeom2}
\partial_z A_x = -4 \pi \delta(x) \sqrt{-g} \, g^{tt} \, a_t \, \phi^2,
\eeq
\beq
\label{aeom2}
\partial_z \left( \sqrt{-g} \, g^{zz} g^{tt} \, \partial_z a_t \right) = 2 \sqrt{-g} \, g^{tt} \, a_t \, \phi^2,
\eeq
\beq
\label{phieom2}
\partial_z \left( \sqrt{-g} \, g^{zz} \, \partial_z \phi \right) = \sqrt{-g} \, g^{tt} \, a_t^2 \, \phi + \sqrt{-g} \, M^2 \, \phi.
\eeq
\end{subequations}
Crucially, $A_x(z)$ does not appear in $a_t(z)$ or $\phi(z)$'s equation of motion, eqs.~\eqref{aeom2} and~\eqref{phieom2}. As a result, the only way that $a_t(z)$ and $\phi(z)$ ``know'' they live on a defect in a higher-dimensional spacetime is through the blackening factor, $h(z)$. In particular, if $T=0$ then the defect's metric is precisely that of $AdS_2$. Moreover, $A_x(z)$ has trivial dynamics (as expected for a Chern-Simons gauge field): we only need to solve for $a_t(z)$ and $\phi(z)$, and then insert those solutions into eq.~\eqref{Aeom2} to obtain $A_x(z)$.

\begin{figure}[h!]
\begin{center}
\includegraphics[width=0.9\textwidth]{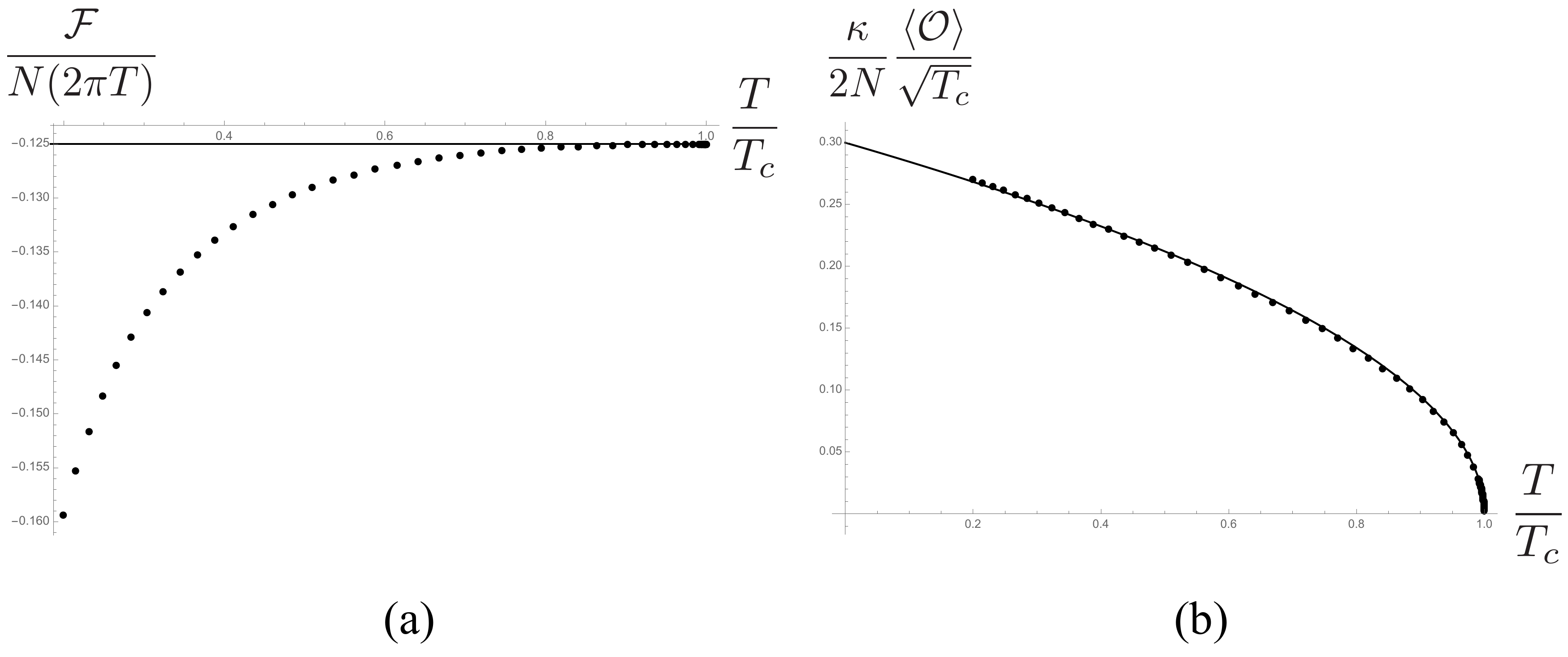}
\caption{\label{fig:phasetrans}(a) The free energy $\mathcal{F}$, normalized by $1/\left(N(2 \pi T)\right)$, as a function of $T/T_c$ for $Q=0.5$. The solid line is for the unscreened phase, where $\langle \mathcal{O} \rangle = 0$, and which has $\mathcal{F}/\left(N (2 \pi T)\right)=-Q^2/2=-0.125$. The dots represent our numerical results for the screened phase, where $\langle \mathcal{O} \rangle \neq 0$. Clearly the screened phase always has lower $\mathcal{F}$, and hence is thermodynamically preferred, for all $T \leq T_c$. (b) The dots are our numerical results for $\frac{\kappa}{2N}\langle \mathcal{O} \rangle/\sqrt{T_c}$ as a function of $T/T_c$ in the screened phase with $Q=0.5$. The solid line is a numerical fit to $0.30 \left(T_c-T\right)^{1/2}$. The agreement between the numerical results and the fit indicates second-order mean field behavior, $\langle \mathcal{O}\rangle \propto \left(T_c-T\right)^{1/2}$.}
\end{center}
\end{figure}

As mentioned in section~\ref{sec:intro}, our holographic Kondo model exhibits a phase transition as $T$ decreases through a critical temperature $T_c$, just like the standard (non-holographic) Kondo model at large $N$. For any $T$, eqs.~\eqref{aeom2} and~\eqref{phieom2} admit the solution $a_t(z) = \mu - Q/z$ and $\phi(z)=0$. These solutions are dual to states with $\langle \mathcal{O} \rangle=0$. However, when $T \leq T_c$ a second branch of solutions exists that have $\phi(z) \neq 0$. Given that $\phi(z)$ is dual to $\mathcal{O}^{\dagger}+\mathcal{O}$, these $\phi(z)\neq0$ solutions are dual to states with $\langle \mathcal{O}^{\dagger}+\mathcal{O} \rangle\neq0$, which implies $\langle \mathcal{O}^{\dagger}\rangle = \langle \mathcal{O} \rangle = \langle \mathcal{O}^{\dagger}+\mathcal{O}\rangle/2 \neq 0$. We will therefore just refer to $\langle \mathcal{O} \rangle \neq 0$ henceforth. To determine which state is thermodynamically preferred, we must determine which state has lower free energy $\mathcal{F}$, which we compute holographically from the on-shell Euclidean action: for details, see refs.~\cite{Erdmenger:2013dpa,O'Bannon:2015gwa,Erdmenger:2015spo,Erdmenger:2015xpq}. Fig.~\ref{fig:phasetrans} (a) shows $\mathcal{F}/\left(N (2 \pi T)\right)$ as a function of $T/T_c$ for $Q=0.5$, for the two branches of solutions. Clearly the solution with $\phi(z) \neq 0$ has lower $\mathcal{F}$, and hence is thermodynamically preferred, for all $T \leq T_c$. Fig.~\ref{fig:phasetrans} (b) shows our numerical results for $\kappa/(2N)\langle \mathcal{O} \rangle/\sqrt{T_c}$ as a function of $T/T_c$ for $Q=0.5$, where $\kappa$ is our holographic Kondo coupling constant, defined in the boundary term eq.~\eqref{kondo}. Fig.~\ref{fig:phasetrans} (b) also shows a numerical fit revealing second-order mean-field behavior: $\langle \mathcal{O}\rangle \propto \left(T_c-T\right)^{1/2}$ when $T \lesssim T_c$. Clearly our model exhibits a second-order mean-field transition when $T$ drops through $T_c$. In section~\ref{sec:highT} we will show $T_c\propto \, T_K$, where the proportionality constant depends only on $Q$: see in particular fig.~\ref{fig:criticalT}.

As mentioned above, at large $N$ the screening of the impurity, and other characteristic Kondo phenomena, such as a phase shift of the electrons, occurs only when $T \leq T_c$, where $\langle \mathcal{O}\rangle \neq 0$. We will thus refer to states with $\langle\mathcal{O} \rangle = 0$ as the \textit{unscreened} phase and states with $\langle \mathcal{O} \rangle \neq 0$ as the \textit{screened} phase.

What does the screening look like on the gravity side of the correspondence? The flux of $a_t(z)$ controls the ``size'' of the impurity's representation, by controlling the number of boxes in the associated Young tableau. To see how, consider the $a_t(z)$'s general asymptotic form, $a_t(z) = \mu - Q/z+\ldots$, where $\ldots$ represents terms that vanish as $z \to 0$. The parameter $\mu$ acts as a chemical potential for $j = \chi^{\dagger}\chi$, and in particular a non-zero $\mu$ breaks particle-hole symmetry. The particle-hole symmetric value of the charge is $\mathcal{Q}=N/2$, which thus corresponds to $\mu=0$. In general the parameter $Q$ depends monotonically on $\mu$. For example, for the solution $a_t(z) = \mu - Q/z$ mentioned above, regularity of $a_t(z)$ at the horizon, $a_t(z_H)=0$, requires $Q = \mu z_H$. As a result, $Q=0$ corresponds to $\mathcal{Q} = N/2$, while $Q>0$ corresponds to $\mathcal{Q}>N/2$, and $Q<0$ corresponds to $\mathcal{Q}<N/2$. A totally anti-symmetric representation must have $0\leq \mathcal{Q}\leq N$, which should translate to limits on $Q$. Our model is too crude to determine the exact relation between $Q$ and $\mathcal{Q}$, and includes nothing to impose limits on $Q$, although these features could potentially be incorporated, following similar models~\cite{Camino:2001at,Gomis:2006sb,Kachru:2009xf,Harrison:2011fs}. They only feature we will need, however, is that $Q$ is monotonically related to $\mathcal{Q}-N/2$.

For any solution, the flux of $a_t(z)$ at the boundary is $Q$. When $\phi(z)=0$, the flux of $a_t(z)$ is constant from the boundary to the horizon. However, when $\phi(z)\neq 0$, the flux of $a_t(z)$ is transferred from $a_t(z)$ to $A_x(z)$, because $\Phi$ is bi-fundamental. Recalling that the holographic coordinate $z$ corresponds to energy scale, where the boundary corresponds to the UV and increasing $z$ corresponds to moving towards the IR~\cite{Susskind:1998dq,Peet:1998wn}, solutions with $\phi(z)\neq0$ thus describe an impurity whose size shrinks as we move towards the IR~\cite{Erdmenger:2013dpa}. In other words, the impurity is screened, as advertised.

What does the phase shift look like on the gravity side of the correspondence? The phase shift is encoded in $A_x(z)$~\cite{Erdmenger:2013dpa}. In particular, eq.~\eqref{Aeom2} shows that $\partial_z A_x(z)\neq0$ if and only if both $a_t \neq 0$ and $\phi(z)\neq 0$. If we imagine compactifying $x$ into a circle, then $A_x(z)\neq0$ implies a non-zero Wilson loop around the $x$ direction, $\oint A \neq 0$, which is dual to a phase shift for our strongly-coupled ``electrons,'' or more generally for any object charged under our $U(1)$ channel symmetry. Non-zero $\partial_z A_x(z)$ means the phase shift grows as we move towards larger $z$, \textit{i.e.} as we move towards the IR, as expected.

In short, our holographic model captures some of the essential phenomena of the \textit{large-$N$} Kondo effect, namely impurity screening and a phase shift at $T\leq T_c$, when $\langle \mathcal{O}\rangle \neq 0$. In the following we will show that our holographic model also captures another essential phenomenon: the Kondo resonance.

\section{Holographic Renormalization and Two-point Functions}
\label{sec:holorg}
\setcounter{equation}{0}

In this section we derive general expressions for the renormalized holographic two-point functions of the Kondo model described by the action in eq.~\eqref{action}, in both the unscreened and screened phases. Before we embark on the technical aspects of this calculation, it is instructive to outline the main steps involved, and to highlight several subtleties that this specific model presents.

A particularly economical way of computing holographic two-point functions is to read them off directly from the linearized fluctuation equations, bypassing the usual step of evaluating the on-shell action to quadratic order in the fluctuations. This is possible due to the holographic identification of the radial canonical momenta, which on-shell become functions of the induced fields, with the one-point functions of the dual operators in the theory with Dirichlet boundary conditions~\cite{Papadimitriou:2004ap}. To obtain the two-point functions it suffices to expand the canonical momenta to linear order in the induced fields. As in standard linear response theory, the coefficients of the linear terms in this expansion are identified with the corresponding response functions, i.e. the \textit{unrenormalized} two-point functions~\cite{Papadimitriou:2004rz}. Inserting the covariant expansions of the canonical momenta to linear order in the fluctuations in the second order fluctuation equations results in a system of first order non-linear Riccati equations for the response functions~\cite{Papadimitriou:2004rz, Papadimitriou:2013jca}. Like the system of second order linear equations for the fluctuations, the system of Riccati equations for the response functions is generically coupled, and can only be solved numerically. However, in contrast to the second order linear equations, the general solution of the Riccati equations contains only one integration constant per response function, since the arbitrary sources have already been eliminated, which is determined by imposing regularity in the bulk of the spacetime. Generically, the fact that the arbitrary sources have been eliminated from the Riccati equations renders them better suited for a numerical evaluation of the two-point functions.

Both the on-shell action and the response functions obtained from the Riccati equations are generically divergent and need to be evaluated with a radial cutoff near the AdS boundary. Moreover, local covariant boundary counterterms need to be determined in order to renormalize these quantities. However, two important subtleties arise in obtaining the correct boundary counterterms in our holographic Kondo model, both directly related to the special asymptotic behavior of the $AdS_2$ gauge field. In contrast to gauge fields in $AdS_4$ and above, in $AdS_3$ and $AdS_2$ the asymptotically leading mode of an abelian gauge field is the conserved charge $Q$, instead of the chemical potential, $\m$~\cite{Cvetic:2016eiv}. The same phenomenon is observed with higher rank antisymmetric $p$-forms in higher dimensions~\cite{thermo2}. In such cases, consistency of the boundary counterterms requires that they be a function of the canonical momentum conjugate to the gauge field, rather than the gauge field itself~\cite{An:2016fzu,Cvetic:2016eiv}.

Moreover, the requirement that the charge $Q$ be kept fixed leads to an asymptotic second class constraint in phase space, which further complicates the computation of the boundary counterterms~\cite{Cvetic:2016eiv}. Relaxing the constraint,~\textit{i.e.} changing the value of $Q$ in this case, changes the form of the asymptotic solutions for the scalar field. In order to have a well-defined space of asymptotic solutions, therefore, we must restrict the phase space asymptotically to the subspace defined by constant $Q$. However, if we want to compute correlation functions of the operator dual to $Q$, which as we will discuss later is not a local operator, then we must allow for infinitesimal deformations away from the asymptotic constraint surface. The boundary counterterms then take the form of a Taylor expansion in the infinitesimal deformation away from the constraint surface, with the coefficient of the $n$-th power renormalizing the $n$-point function of the operator dual to $Q$.

In our holographic Kondo model, a further complication arises due to the double-trace boundary conditions we need to impose on the scalar field in order to introduce the Kondo coupling. The response functions obtained directly from the Riccati equations correspond to the two-point functions in the theory defined by Dirichlet boundary conditions on the scalar field and Neumann boundary conditions on the $AdS_2$ gauge field,~\textit{i.e.} keeping $Q$ fixed. In the large-$N$ limit, however, the renormalized two-point functions in the theory with double-trace boundary conditions on the scalar field are algebraically related to those in the theory with Dirichlet boundary conditions on the scalar field. The precise relation is obtained by identifying additional \textit{finite} boundary terms required to impose the double-trace boundary condition on the scalar field, and then carefully examining the variational problem.

In this section we will address all the above subtleties as we go along. We start by reformulating the Kondo model in eq.~\eqref{action} in radial Hamiltonian formalism, which allows us to introduce the radial canonical momenta, the linear response functions, and the Hamilton-Jacobi equation we must solve in order to determine the boundary counterterms. We then proceed to derive the Riccati equations for the linear response functions, determine their general asymptotic solutions in the UV (\textit{i.e.} near the asymptotically AdS boundary), and determine the most general regular asymptotic solution in the IR (\textit{i.e.} deep in the bulk). The arbitrary integration constants appearing in the UV expansions parameterize the renormalized two-point functions, and their value is determined by matching the solution, numerically, to the regular asymptotic solution in the IR. Subsequently we determine the boundary counterterms necessary to renormalize the free energy, as well as the one- and two-point functions in the theory with Dirichlet boundary conditions on the scalar field. Finally, the renormalized two-point functions with a non-zero Kondo coupling are obtained by adding further boundary terms that implement the double-trace boundary condition on the scalar field.

\subsection{Radial Hamiltonian Formulation of the Kondo Model}

To describe our holographic Kondo model in radial Hamiltonian language, we re-write the induced metric in eq.~\eqref{ads2metric} in the form
\be\label{gf-metric}
ds_{AdS_2}^2=\tx dr^2+\g \, \tx dt^2, 
\ee 
where the radial coordinate $z$ of eq.~\eqref{ads2metric} is related to the canonical radial coordinate $r$ of eq.~\eqref{gf-metric} as 
\be\label{r-coord}
r=\log\Big(1+\sqrt{1-z^2/z_{H}^2}\Big)-\log(2z),
\ee
with $r\in [r_H,+\infty)$, and $r_H=-\log (2z_H)$, and asymptotically, $\g=-e^{2r}+\co(1)$ as $r\to+\infty$.

In these coordinates the action in eq.~\eqref{action} may be written as
\bal
\label{gf-lagrangian}
S=&-\frac{N}{4\p}\int \tx d^2x\;\bar{\e}^{ij}(-A_{i}\dot{A}_{j}+2A_r \pa_{[i}A_{j]})\NO\\
&-N\int \tx dt \sqrt{-\g}\left(\rule{0cm}{0.4cm}\frac12\g^{-1}f_{rt}^2+|D_r\F|^2+\g^{-1}|D_t\F|^2+M^2\F^\dag\F\right),
\eal
where $(i,j)=(t,x)$, a dot denotes differentiation with respect to $r$, $\dot{A}_j \equiv \partial_r A_j$, and $\bar{\e}^{ij}\equiv \e^{zij}$. From eq.~\eqref{gf-lagrangian} we obtain the radial canonical momenta:
\bal\label{momenta}
&\p^i_A=\frac{\d S}{\d \dot A_i}=-\frac{N}{4\p}\bar{\e}^{ij}A_j,&& \p^t_a=\frac{\d S}{\d \dot a_t}=-N\sqrt{-\g}\;\g^{-1}(\dot a_t-\pa_t a_r),\NO\\
&\p_{\F}=\frac{\d S}{\d \dot\F}=-N\sqrt{-\g}\;(D_r\F)^\dag,&& 
\p_{\F^\dag}=\frac{\d S}{\d\dot\F^\dag}=-N\sqrt{-\g}\;D_r\F.
\eal
In terms of the modulus and phase, $\Phi = e^{i\psi} \phi$, the scalar field's canonical momenta become
\be\label{momenta-phase}
\p_{\f}=\frac{\d S}{\d \dot\f}=-2N\sqrt{-\g}\;\dot\f,\qquad
\p_{\j}=\frac{\d S}{\d\dot\j}=-2N\sqrt{-\g}\;\f^2(A_r-a_r+\dot\j).
\ee
No radial derivatives of the components $a_r$ and $A_r$ appear in eq.~\eqref{gf-lagrangian}, so they correspond to non-dynamical Lagrange multipliers. Moreover, the canonical momentum of the Chern-Simons field in eq.~\eqref{momenta} amounts to a primary constraint, which implies that the canonical momentum and the gauge connection $A_i$ are not independent variables on phase space.

The Legendre transform of the action in eq.~\eqref{gf-lagrangian} gives the radial Hamiltonian, 
\bea
H&=&\int \tx d^2x\dot A_i\p^i_A+\int \tx dt(\dot a_t\p^t_a+\dot\f\p_\f+\dot\j\p_\j)-S\NO\\
&=&\int \tx d^2x\; A_r \left(-\p_\j\d(x)+\frac{N}{4\p} \bar{\e}^{ij}2\pa_{[i}A_{j]}\right)+\int \tx dt\; a_r\left(-\pa_t\p^t_a+\p_\j\right)\NO\\
&&-\frac1N\int \tx dt\;\frac{1}{\sqrt{-\g}}\left(\frac12\g(\p^t_a)^2+\frac14\p_{\f}^2+\frac14\f^{-2}\p_\j^2\right)
\NO\\
&&+N\int \tx d t\; \sqrt{-\g}\left(\rule{0cm}{0.4cm}\g^{-1}(\pa_t\f)^2+\g^{-1}\f^2(A_t-a_t+\pa_t\j)^2+M^2\f^2\right).
\label{hamiltonian}
\eea

Hamilton's equations for the non dynamical fields $a_r$ and $A_r$ result in the first class constraints 
\be\label{constraints}
\p_\j=i(\F\p_\F-\F^\dag\p_{\F^\dag})=\pa_t\p^t_a,\qquad \frac{N}{4\p} \bar{\e}^{ij}2\pa_{[i}A_{j]}=-2\,\pa_i\p_A^i=\p_\j\d(x),
\ee
which reflect the $U(1)$ gauge invariances associated with the $AdS_2$ and Chern-Simons gauge fields, respectively. We will see below that these constraints lead to Ward identities in the dual field theory.  

Hamilton-Jacobi theory connects the canonical momenta with the regularized on-shell action $\cs$ through the relations\footnote{The expression in eq.~\eqref{momenta} for the Chern-Simons momentum implies that $\cs$ cannot be a local covariant functional of $A_i$. This is consistent with the fact that $A_i$ parameterizes the full phase space, and only a particular component of $A_i$, depending on the boundary conditions, will be identified with the source of the dual current operator.}
\be\label{HJ-momenta}
\p_A^i=\frac{\d\cs}{\d A_i},\qquad \p_a^t=\frac{\d\cs}{\d a_t},\qquad \p_\F=\frac{\d\cs}{\d \F},\qquad \p_{\F^\dag}=\frac{\d\cs}{\d \F^\dag},
\ee
or for the modulus and phase of the scalar field, $\p_\f=\frac{\d\cs}{\d \f}$ and $\p_\j=\frac{\d\cs}{\d \j}$. The regularized on-shell action $\cs$, also known as Hamilton's principal function in this context, is identified via the holographic dictionary with the regularized generating function of connected correlation functions in the theory defined by Dirichlet boundary conditions on the scalar and Chern-Simons fields, and Neumann boundary conditions on the $AdS_2$ gauge field. The canonical momenta, therefore, correspond to the regularized one-point functions with arbitrary sources. The regularized two-point functions are thus obtained by differentiation of the canonical momenta with respect to the induced fields. As we will see in the next subsection, this property allows us to rewrite the fluctuation equations in terms of Riccati equations, which are first order, and whose solution gives directly the regularized two-point functions.  

Since $\cs$ is identified with the regularized on-shell action as a function of the induced fields on a radial cutoff, its divergent asymptotic form determines the boundary counterterms that are required to renormalize the theory. The asymptotic form of $\cs$ can be determined in covariant form by solving the radial Hamilton-Jacobi equation
\be
H+\frac{\pa \cs}{\pa r}=0\quad  \Leftrightarrow\quad  H+\dot\g\,\frac{\d \cs}{\d \g}=0,
\ee
or more explicitly
\bal\label{HJ}
&-\frac1N\int \tx dt\;\frac{1}{\sqrt{-\g}}\left(\frac12\g\left(\frac{\d\cs}{\d a_t}\right)^2+\frac14\left(\frac{\d\cs}{\d \f}\right)^2+\frac14\f^{-2}\left(\frac{\d\cs}{\d\j}\right)^2\right)
\NO\\
&\hskip0.5in+N\int \tx dt\; \sqrt{-\g}\Big(\g^{-1}(\pa_t\f)^2+\g^{-1}\f^2(A_t-a_t+\pa_t\j)^2+M^2\f^2\Big)+\dot\g\frac{\d \cs}{\d \g}=0,
\eal
together with the two constraints
\be\label{HJ-constraints}
\frac{\d\cs}{\d\j}=\pa_t\left(\frac{\d\cs}{\d a_t}\right),\qquad \d(x)\frac{\d\cs}{\d\j}=\frac{N}{4\p} \bar{\e}^{ij}\,2\,\pa_{[i}A_{j]}=-2\,\pa_i\left(\frac{\d\cs}{\d A_i}\right),
\ee
which reflect the $U(1)$ gauge invariances associated with the $AdS_2$ and Chern-Simons gauge fields, respectively.

\subsection{Linear Response Functions from Riccati Equations}

In this subsection we use the relation between the radial canonical momenta and the one-point functions in order to rewrite the second order fluctuation equations in the form of Riccati equations, which are first order. For convenience, we will work with the complex scalar field $\F$ and its complex conjugate $\F^{\dagger}$, rather than its modulus and phase. In the coordinates of eq.~\eqref{gf-metric}, and in the radial gauge $A_r=a_r=0$, the equations of motion associated with the action in eq.~\eqref{gf-lagrangian} are
\begin{subequations}
	\label{eoms-fg}
	\bal
	&\frac{1}{2\p}\bar\e^{ij}\pa_i A_j+\d(x)\sqrt{-\g}\,J_r=0,\label{eq:Ar}\\
	&\frac{1}{2\p}\bar\e^{ij}\dot A_j-\d(x)\d^{it}\sqrt{-\g}\,\g^{-1}J_t=0,\label{eq:Am}\\
	&\pa_r(\sqrt{-\g}\,\g^{-1}\dot a_t)+\sqrt{-\g}\,\g^{-1}J_t=0,\label{eq:at}\\
	&\g^{-1}\pa_t\dot a_t-J_r=0,\label{eq:ar}\\
	&\pa_r(\sqrt{-\g}\,\dot\F)+\sqrt{-\g}\g^{-1}\(\pa_t+i(A_t-a_t)\)^2\F-\sqrt{-\g}\,M^2\F=0,\label{eq:Phi}
	\eal
\end{subequations}
where we have defined a current associated with $\F$,
\be
J_m\equiv-i\Big(\F^\dag D_m\F-(D_m\F)^\dag\F\Big). \label{u1current}
\ee

We solve first for the Chern-Simons gauge field. Eliminating $J_r$ from eqs.~\eqref{eq:Ar} and~\eqref{eq:ar} and $J_t$ from eqs.~\eqref{eq:Am} and~\eqref{eq:at}, results respectively in the two conditions 
\begin{subequations}
	\bal
	&\pa_r\(\bar\e^{ij}A_j+2\p\d(x)\d^{it}\sqrt{-\g}\,\g^{-1}\dot a_t\)=0,\\
	&\pa_i\(\bar\e^{ij} A_j+2\p\d(x)\d^{it}\sqrt{-\g}\,\g^{-1}\dot a_t\)=0.
	\eal
\end{subequations}
The general solution for the Chern-Simons gauge field thus takes the form
\be\label{CS-sol-cov}
\bar\e^{ij}A_j=\frac{2\p}{N}\p_a^t\d(x)\d^{it}+\bar\e^{ij}A_{(0)j}(t,x),
\ee
where $A_{(0)i}(t,x)$ is a flat connection on the $AdS_3$ boundary, \textit{i.e.}~$\bar\e^{ij}\pa_iA_{(0)j}=0$. This implies that the two components $A_{(0)i}(t,x)$ are not both arbitrary sources, in contrast to what happens for a Maxwell gauge field. As we shall see below, in order to obtain a well-defined variational problem for the Chern-Simons gauge field, we must add the appropriate boundary term~\cite{Kraus:2006wn,Kiritsis:2010xc,Jensen:2010em,Andrade:2011sx,Chang:2014jna}. 

In our model, a key observation that will play a role in the choice of boundary conditions for the Chern-Simons gauge field is that the $AdS_2$ fields source only $A_x$, while $A_t$ is independent of the radial coordinate. This implies that we can use a residual $U(1)$ gauge transformation, \textit{i.e.}~preserving the radial gauge $A_r=0$, to set $A_t$ to zero, so that the Chern-Simons gauge field decouples from the equations of motion for the $AdS_2$ fields. In that choice of gauge, the Chern-Simons gauge field takes the simple form 
\be\label{CS-sol}
A_x=-2\p\d(x)\sqrt{-\g}\,\g^{-1}\dot a_t+A_{(0)x},\qquad A_t=A_{(0)t}=0,
\ee
where $A_{(0)x}$ is a function of $x$ only, but is otherwise arbitrary. However, when we discuss the variational problem for the Chern-Simons gauge field, we will reinstate $A_{(0)t}$.

We now solve for the $AdS_2$ fields. We want to find a real and static background solution, and then consider time-dependent fluctuations about that solution. The most generic real and static background solution includes $a_t^0(r)$ and $\f_0(r)$, whose equations of motion are
\begin{subequations}
	\label{eoms-bg}
	\bal
	&\ddot a_t^0-\frac12\g^{-1}\dot\g \dot a_t^0-2a_t^0\f_0^2=0,\label{eq:at-bg}\\
	&\ddot\f_0+\frac12\g^{-1}\dot\g\dot\f_0-(\g^{-1}(a_t^0)^2+M^2)\f_0=0\label{eq:Phi-bg}.
	\eal
\end{subequations} 
We have been able to solve these equations analytically (\textit{i.e.}\ without numerics) only for $\f_0(r)=0$. Solutions with $\f_0(r)\neq 0$ were obtained numerically in refs.~\cite{Erdmenger:2013dpa,O'Bannon:2015gwa}.  

We now introduce fluctuations $\delta a_t$, $\delta \F$, and $\delta \F^{\dagger}$ about the static background solution, linearize their equations of motion, and Fourier transform from time $t$ to frequency $\o$ via $\pa_t\to -i\o$, to obtain
\begin{subequations}
	\label{eoms-fl-FT}
	\bal
	&\o\g^{-1}\d\dot a_t=\f_0(\d\dot\F-\d\dot\F^\dag)-\dot\f_0(\d\F-\d\F^\dag)\label{eq:ar-fl-FT}\\
	&\d\ddot\F+\frac12\g^{-1}\dot\g\d\dot\F-\g^{-1}(\o+a_t^0)^2\d\F-M^2\d\F=\g^{-1}\f_0(\o+2a_t^0)\d a_t,\label{eq:Phi-fl-FT}\\
	&\d\ddot\F^\dag+\frac12\g^{-1}\dot\g\d\dot\F^\dag-\g^{-1}(-\o+a_t^0)^2\d\F^\dag-M^2\d\F^\dag=\g^{-1}\f_0^\dag(-\o+2a_t^0)\d a_t.\label{eq:Phi-dag-fl-FT}
	\eal
\end{subequations} 
We will consider these equations in the unscreened and screened phases separately. 

\subsubsection{Response Functions in the Unscreened Phase}
\label{subsec:unscreenedresponse}

In the unscreened phase, where $\phi_0=0$, eq.~\eqref{eq:ar-fl-FT} becomes trivial, and eqs.~\eqref{eq:Phi-fl-FT} and~\eqref{eq:Phi-dag-fl-FT} decouple. These second-order equations for the fluctuations $\d\F$ and $\d\F^\dag$ can be turned into first-order equations for the two-point functions as follows. The canonical momenta in eqs.~\eqref{momenta} and~\eqref{HJ-momenta} imply that on-shell the radial velocities become functions of the induced fields. To linear order in the fluctuations we thus have  
\be\label{response-unscreened}
\d\dot\F=\car_{\F^\dag\F}\d\F,\qquad \d\dot\F^\dag=\car_{\F\F^\dag}\d\F^\dag,
\ee
where the response functions $\car_{\F^\dag\F}$ and $\car_{\F\F^\dag}$ depend only on the background $a_t^0$ and $\f_0$, as well as $\o$. Hermitian conjugation implies that $\car_{\F\F^\dag}^\dag(\o)=\car_{\F^\dag\F}(-\o)$. Inserting these expressions into the two decoupled fluctuation equations, eqs.~\eqref{eq:Phi-fl-FT} and~\eqref{eq:Phi-dag-fl-FT}, leads to the two Riccati equations \cite{Papadimitriou:2004rz,Papadimitriou:2013jca}
\begin{subequations}
	\label{riccati-unscreened-r}
	\bal
	&\dot\car_{\F^\dag\F}+\frac12\g^{-1}\,\dot\g\,\car_{\F^\dag\F}+\car_{\F^\dag\F}^2-\g^{-1}\(\o+a_t^0\)^2-M^2=0,\\
	&\dot\car_{\F\F^\dag}+\frac12\g^{-1}\,\dot\g\,\car_{\F\F^\dag}+\car_{\F\F^\dag}^2-\g^{-1}\(\o-a_t^0\)^2-M^2=0.
	\eal
\end{subequations}
Using eq.~\eqref{r-coord} to change the radial coordinate from $r$ back to $z$, and using the solution for the background gauge field $a_t^0=Q(1/z-1/z_H)$ these Riccati equations become 
\bsub
\label{riccati-unscreened-z}
\bal
&-zh^{1/2}\car_{\F^\dag\F}'+\frac12h^{-1/2}(2h-zh')\car_{\F^\dag\F}+\car_{\F^\dag\F}^2+z^2h^{-1}\(\o+Q(1/z-1/z_H)\)^2-M^2=0,\\
&-zh^{1/2}\car_{\F\F^\dag}'+\frac12h^{-1/2}(2h-zh')\car_{\F\F^\dag}+\car_{\F\F^\dag}^2+z^2h^{-1}\(\o-Q(1/z-1/z_H)\)^2-M^2=0,
\eal
\esub
where primes denote $\partial_z$, for example $\car_{\F^\dag\F}'\equiv \partial_z \car_{\F^\dag\F}$.

We want to solve eqs.~\eqref{riccati-unscreened-z} with in-going boundary conditions at the horizon. Eqs.~\eqref{riccati-unscreened-z} can be solved analytically, either directly, or by first transforming them into second-order linear homogeneous equations through the change of variables
\be
\car_{\F^\dag\F}=-z\,h^{1/2}\,y'_+/y_+,\qquad \car_{\F\F^\dag}=-z\,h^{1/2}\,y'_-/y_-,
\ee
where the functions $y_\pm$ satisfy the second order equations\footnote{Eq.~\eqref{eq-symm} is identical to the equation of motion in ref.~\cite{Faulkner:2011tm} (after their eq.~(5.20)), with the identifications $\z_0=z_H$, $qe_d=\pm Q$, $m R_2=M$.}
\be\label{eq-symm}
y''_\pm+\frac{2z}{z^2-z_H^2}y'_\pm+\Bigg(\frac{\(\o\pm Q(1/z-1/z_H)\)^2}{(1-z^2/z_H^2)^2}-\frac{\n^2+Q^2-1/4}{z^2(1-z^2/z_H^2)}\Bigg)y_\pm=0,
\ee
where $\n\equiv\sqrt{M^2-Q^2+1/4}$. The two linearly independent solutions of eq.~\eqref{eq-symm} are $y_\pm(z,\o;\n)$ and $y_\pm(z,\o;-\n)$ where
\be
\hskip-0.4cm y_\pm(z,\o;\n)=\frac{(z/z_H)^{\frac12+\n}(1-z/z_H)^{\frac{i\o z_H}{2}}}{(1+z/z_H)^{\frac12+\n+\frac{i\o z_H}{2}}}{}_2F_1\(\frac12+\n\mp iQ +i\o z_H,\frac12+\n\pm iQ,1+2\n;\frac{2z}{z+z_H}\)\hskip-0.12cm.
\ee
The linear combination of $y_\pm(z,\o;\n)$ and $y_\pm(z,\o;-\n)$ that satisfies in-going boundary condition at the horizon is
\bsub
\label{ingoing-symm}
\bal
& y^{in}_\pm(z,\o;\n)=\frac1\n\Big(y_\pm(z,\o;\n)+c_\pm(\o;\n)y_\pm(z,\o;-\n)\Big),\\
& c_\pm(\o;\n)\equiv-\frac{\G(1+2\n)\G(\frac12-\n\pm iQ-i\o z_H)\G(\frac12-\n\mp iQ)}{2^{2\n}\G(1-2\n)\G(\frac12+\n\pm iQ-i\o z_H)\G(\frac12+\n\mp iQ)}.
\eal
\esub
The general in-going solutions of eqs.~\eqref{riccati-unscreened-z} are therefore
\be
\label{eq:unscreenedresponseresults}
\car_{\F^\dag\F}=-z\,h^{1/2}\,\frac{y'^{in}_+(z,\o;\nu)}{y^{in}_+(z,\o;\nu)},\qquad \car_{\F\F^\dag}=-z\,h^{1/2}\,\frac{y'^{in}_-(z,\o;\nu)}{y^{in}_-(z,\o;\nu)}.
\ee

As explained in detail in refs.~\cite{Erdmenger:2013dpa,O'Bannon:2015gwa,Erdmenger:2015spo,Erdmenger:2015xpq}, to guarantee that $\mathcal{O}$ is dimension $1/2$, and hence our Kondo coupling $\mathcal{O}^{\dagger} \mathcal{O}$ is classically marginal, we must choose $M^2 = -1/4+Q^2$, so that $\nu =0$. In the limit $\n\to 0$, the solution in eq.~\eqref{ingoing-symm} has the asymptotic behavior
\be
y^{in}_\pm(z,\o; 0)= 2z^{1/2}\(\log (z/z_H)+\Th_\pm(\o)\) + \ldots,
\ee
where $\ldots$ represents terms that vanish faster than those shown as $z \to 0$, and
\be
\label{eq:thetapmdef}
\Th_\pm(\o)\equiv H\left(-\frac12\pm iQ-i\o z_H\right)+H\left(-\frac12\mp iQ\right)+\log 2,
\ee
and $H(n)$ denotes the $n$th harmonic number. The response functions' asymptotic expansions are then
\be
\label{respfuncasymp}
\car_{\F^\dag\F}=-\frac12-\frac{1}{\log(z/z_H)+\Th_+(\o)}+\co(z),\qquad
\car_{\F\F^\dag}=-\frac12-\frac{1}{\log(z/z_H)+\Th_-(\o)}+\co(z).
\ee
One of our main tasks in the remainder of this section is to determine how the coefficients in the asymptotic expansion in eq.~\eqref{respfuncasymp} can be translated into the two-point functions of $\co$ and $\co^\dag$.

\subsubsection{Response Functions in the Screened Phase}
\label{subsec:screenedresponse}

In the screened phase, where $\phi_0\neq0$, eqs.~\eqref{eoms-fl-FT} are three coupled equations for the three fluctuations. They can be turned into a system of coupled Riccati equations by introducing response functions as
\begin{subequations}
	\label{response-screened}
	\bal
	\d\dot\F&=\car_{\F^\dag\F}\d\F+\car_{\F^\dag\F^\dag}\d\F^\dag+\g^{-1}\car_{\F^\dag a}\d a_t,\\
	\d\dot\F^\dag&=\car_{\F\F}\d\F+\car_{\F\F^\dag}\d\F^\dag+\g^{-1}\car_{\F a}\d a_t.
	\eal
\end{subequations}
We could similarly introduce response functions for $\d\dot a_t$, however eq.~\eqref{eq:ar-fl-FT} implies that they are completely determined by the response functions in eq.~\eqref{response-screened}. Inserting eq.~\eqref{response-screened} into the fluctuation equations eqs.~\eqref{eq:Phi-fl-FT} and~\eqref{eq:Phi-dag-fl-FT} leads to a system of six coupled Riccati equations.

Although the six response functions defined in eq.~\eqref{response-screened} will be useful for extracting the two-point functions in the following, we will now show that in fact they can be mapped to only four independent response functions. The in-going boundary condition then forces one of these four to vanish identically, leaving only three non-trivial, independent response functions.

We first re-express eq.~\eqref{eoms-fl-FT} in terms of the fluctuations of the modulus and phase, $\d\f$ and $\d\j$, respectively, which leads to two coupled second-order equations for the gauge invariant fluctuations, $(\d a_t+i\o\d\j)$ and $\d\f$:   
\bsub
\label{decoup-fluct-screened}
\bal
&\pa_r\Big(\frac{\d \dot a_t+i\o\d\dot\j}{1+\frac12\g^{-1}\f_0^{-2}\om^2}\Big)-\frac{\frac12\g^{-1}\dot\g \left(\d \dot a_t+i\o\d\dot\j\right)}{1+\frac12\g^{-1}\f_0^{-2}\om^2}-2\f_0^2\left(\d a_t+i\o\d\j\right)=4\f_0a_t^0\d\f,\\
&\d\ddot\f+\frac12\g^{-1}\dot\g\d\dot\f-\g^{-1}\o^2\d\f-\left(M^2+\g^{-1}\left(a_t^0\right)^2\right)\d\f=2\f_0\g^{-1}a_t^0\left(\d  a_t+i\o\d\j\right).
\eal
\esub
Given a solution for $(\d \dot a_t+i\o\d\dot\j)$, we can extract $\d a_t$, and hence also $\d\j$, by re-writing eq.~\eqref{eq:ar-fl-FT} as
\be\label{b-id}
\d\dot a_t=\frac{1}{1+\frac12\g^{-1}\f_0^{-2}\om^2}\left(\d \dot a_t+i\o\d\dot\j\right).
\ee
We can turn eq.~\eqref{decoup-fluct-screened} into a system of Riccati equations by introducing four response functions,
\be\label{min-response-screened}
\d\dot a_t=\car_{11}\left(\d  a_t+i\o\d\j\right)+\g\car_{12}\d\f,\qquad
\d\dot\f=\frac12(\car+\g\car_{12})\g^{-1}\left(\d  a_t+i\o\d\j\right)+\frac12\car_{22}\d\f,
\ee
where, with the benefit of hindsight, we have parameterized $\d\dot\f$ so that $\car$ will satisfy a homogeneous equation. Using the identities $\d\F=\d\f+i\f_0\d\j$ and $\d\F^\dag=\d\f-i\f_0\d\j$, we can express the six response functions introduced in eq.~\eqref{response-screened} in terms of only four independent response functions, namely those in eq.~\eqref{min-response-screened}, as advertised:
\begin{subequations}
	\label{building-blocks}
	\bal
	\car_{\F a}&=\frac12\(\car+\g\car_{12}-\o\f_0^{-1}\car_{11}\),\qquad \car_{\F^\dag a}=\frac12\(\car+\g\car_{12}+\o\f_0^{-1}\car_{11}\)\\
	\car_{\F\F}&=\frac14\(\car_{22}-\o^2\g^{-1}\f_0^{-2}\car_{11}-2\f_0^{-1}\dot\f_0+\o\g^{-1}\f_0^{-1}\car\),\\
	\car_{\F^\dag\F^\dag}&=\frac14\(\car_{22}-\o^2\g^{-1}\f_0^{-2}\car_{11}-2\f_0^{-1}\dot\f_0-\o\g^{-1}\f_0^{-1}\car\),\\
	\car_{\F^\dag\F}&=\frac14\(\car_{22}+2\o\f_0^{-1}\car_{12}+\o^2\g^{-1}\f_0^{-2}\car_{11}+2\f_0^{-1}\dot\f_0+\o\g^{-1}\f_0^{-1}\car\),\\
	\car_{\F\F^\dag}&=\frac14\(\car_{22}-2\o\f_0^{-1}\car_{12}+\o^2\g^{-1}\f_0^{-2}\car_{11}+2\f_0^{-1}\dot\f_0-\o\g^{-1}\f_0^{-1}\car\).
	\eal
\end{subequations}

Inserting eq.~\eqref{min-response-screened} into eqs.~\eqref{decoup-fluct-screened} then leads to Riccati equations 
\bsub
\label{riccati-screened-r}
\bal
&\dot\car_{11}-\frac12\g^{-1}\dot\g\car_{11}+\Big(1+\frac{\om^2}{2\g\f_{0}^{2}}\Big)\car_{11}^2+\frac12\car_{12}(\car+\g\car_{12})-2\f_0^2=0,\\
&\dot\car_{12}+\frac12\g^{-1}\dot\g\car_{12}+\Big(1+\frac{\om^2}{2\g\f_{0}^{2}}\Big)\car_{11}\car_{12}+\frac12\car_{12}\car_{22}-4\f_0\g^{-1}a_t^0=0,\\
&\dot\car_{22}+\frac12\g^{-1}\dot\g\car_{22}+\Big(1+\frac{\om^2}{2\g\f_{0}^{2}}\Big)\car_{12}(\car+\g\car_{12})+\frac12\car_{22}^2-2\left(M^2+\g^{-1}(a_t^0)^2+\g^{-1}\om^2\right)=0,\\
&\label{eq:4ricc4}\dot\car-\(\frac12\g^{-1}\dot\g-\Big(1+\frac{\om^2}{2\g\f_{0}^{2}}\Big)\car_{11}-\frac12\car_{22}\)\car=0.
\eal
\esub	
We can solve eq.~\eqref{eq:4ricc4} by direct integration,
\be\label{R}
\car=C(\o)\sqrt{-\g}\exp\(-\int \tx dr'\[\Big(1+\frac{\om^2}{2\g\f_{0}^{2}}\Big)\car_{11}+\frac12\car_{22}\]\),
\ee
where $C(\o)$ is an integration constant. In appendix~\ref{appendix:IRexp} we show that the in-going boundary conditions for the fluctuations on the horizon require $C(\o)=0$, and hence $\car=0$. We have thus shown that only three non-trivial, independent response functions remain, as advertised. Setting $\car=0$, and using eq.~\eqref{r-coord} to change the radial coordinate from $r$ to $z$, eqs.~\eqref{riccati-screened-r} become
\begin{subequations}
	\label{riccati-screened-z}
	\bal
	&\hskip-0.2cm-zh^{1/2}\car_{11}'-\frac{(2h-zh')}{2h^{1/2}}\car_{11}+\(1-\frac{z^2\o^2}{2h\f_{0}^2}\)\car_{11}^2-\frac12hz^{-2}\car_{12}^2-2\f_0^2=0,\\
	&\hskip-0.2cm-zh^{1/2}\car_{12}'+\frac{(2h-zh')}{2h^{1/2}}\car_{12}+\(1-\frac{z^2\o^2}{2h\f_{0}^2}\)\car_{11}\car_{12}+\frac12\car_{12}\car_{22}+4\f_0z^2h^{-1}a_t^0=0,\\
	&\hskip-0.2cm-zh^{1/2}\car_{22}'+\frac{(2h-zh')}{2h^{1/2}}\car_{22}-\(1-\frac{z^2\o^2}{2h\f_{0}^2}\)\frac{h}{z^2}\car_{12}^2+\frac12\car_{22}^2+\frac{2z^2}{h}\((a_t^0)^2+\om^2\)-2M^2=0.
	\eal
\end{subequations}
Using eqs.~\eqref{riccati-screened-z}, we derive the near-horizon asymptotic expansions of $\car_{11}$, $\car_{12}$, and $\car_{22}$ in appendix~\ref{appendix:IRexp}, and the near-boundary asymptotic expansions in appendix~\ref{appendix:UVexp}. Eqs.~\eqref{riccati-screened-z} are first-order, hence the solution for each response function has one integration constant, which we fix using the in-going boundary condition at the horizon (more specifically, by demanding that the near-horizon expansion coincides with that in eq.~\eqref{Rs_IR_NLO}).

In the screened phase we have been able to obtain the background solutions $a_t^0$ and $\phi_0$ only numerically. We have thus solved eqs.~\eqref{riccati-screened-z} only numerically, by integrating them from the horizon to the boundary, subject to the near-horizon behavior in eqs.~\eqref{Rs_IR_NLO}.  We then extract the two-point functions from the near-boundary asymptotic expansions of the solutions, as we discuss in the next subsection.

\subsection{Holographic Renormalization}
\label{sec:holorgsub}

To extract the physical one- and two-point functions from the solutions for the background and the response functions, we must perform holographic renormalization (holo-ren)~\cite{Henningson:1998gx,Balasubramanian:1999re,deBoer:1999tgo,deHaro:2000vlm,Bianchi:2001de,Bianchi:2001kw,Martelli:2002sp,Skenderis:2002wp,Papadimitriou:2004ap,Papadimitriou:2010as}. For a recent review of holo-ren, see ref.~\cite{Papadimitriou:2016yit}. Holo-ren consists of deriving the appropriate boundary counterterms that render the variational problem well posed for the desired boundary conditions, as well as determining the resulting holographic dictionary, relating physical observables to the solutions in the bulk.

As we mentioned in section~\ref{sec:intro} and at the beginning of this section, the holo-ren of our holographic Kondo model involves a number of subtleties, stemming from the unusual form of the Fefferman-Graham (FG) expansion of gauge fields in $AdS_2$ and the related second class constraint eq.~\eqref{Q-condition}, as well as the mixed boundary conditions we impose on the complex scalar $\F$ to introduce the Kondo coupling. In the remainder of this section we will address these issues systematically.

We saw above that the functional $\cs$ defined through eq.~\eqref{HJ-momenta} coincides with the regularized on-shell action, which we will denote as $S\sbtx{reg}$, and satisfies the Hamilton-Jacobi equation, eq~\eqref{HJ}. In particular, the divergent parts of $\cs$ and $S\sbtx{reg}$ coincide, allowing us to determine the counterterms by solving the Hamilton-Jacobi equation. Since we are only interested in the divergent part of $\cs$, we can simplify the Hamilton-Jacobi equation eq.~\eqref{HJ} by dropping terms that affect only the finite parts of $\cs$. Using the leading form of the asymptotic expansions~\eqref{UV-asymptotics-nl} in appendix~\ref{appendix:UVexp}, and the general solution for the  Chern-Simons field in eq.~\eqref{CS-sol-cov}, a simple power counting argument shows that we can ignore any terms that involve $A_t$, $\j$, or  the time derivatives of any fields, and moreover, we can take $\g\to-e^{2r}$. To determine the counterterms, we can thus use the ``reduced'' Hamiltonian
\be
\hskip-0.3cm H_{\rm reduced}\(\p_a^t,\p_\f,a_t,\f;\g\)=-\frac1N\int \tx dt\;\frac{1}{\sqrt{-\g}}\Big(\frac12\g (\p_a^t)^2+\frac14\p_\f^2\Big)+N\int \tx dt \sqrt{-\g}\(\g^{-1}a_t^2+M^2\)\f^2,
\ee
and solve the simplified Hamilton-Jacobi equation
\be
H_{\rm reduced}\Big(\p_a^t=\frac{\d\cs}{\d a_t},\p_\f=\frac{\d\cs}{\d\f},a_t,\f;\g\Big)+2\g\frac{\d\cs}{\d\g}=0,
\ee 
in order to determine the divergent part of $S\sbtx{reg}$ in the form $\cs[a_t,\f;\g]$.

At this point we encounter the first subtlety in the holo-ren of our model, namely, the leading term of the $AdS_2$ gauge field's FG expansion in eq.~\eqref{UV-asymptotics-nl} is the charge term, $Q e^r$, and not the chemical potential term, $\m(t)$. This is a generic feature of gauge fields in $AdS_2$ and $AdS_3$, as well as rank-$p$ antisymmetric tensor fields in $AdS_{d+1}$ with $p\geq d/2$~\cite{thermo2}. Following ref.~\cite{Cvetic:2016eiv}, we will argue that in this case, consistency with the symplectic structure of the theory, as well as locality, requires the counterterms to be a local function of the canonical momentum $\p_a^t$, and not of the gauge potential $a_t$. As a result, in practice we should determine not $\cs$, but its Legendre transform,
\be\label{leg1}
\wt\cs[\p_a^t,\f;\g]=\cs-\int \tx dt\; \p_a^t a_t, 
\ee
by solving the Legendre transform's Hamilton-Jacobi equation,
\be\label{HJ2}
H_{\rm reduced}\Big(\p_a^t,\p_\f=\frac{\d\wt\cs}{\d\f},a_t=-\frac{\d\wt\cs}{\d\p_a^t},\f;\g\Big)+2\g\frac{\d\wt\cs}{\d\g}=0.
\ee 

Our ansatz to solve eq.~\eqref{HJ2} is
\be\label{HJ-ansatz}
\wt\cs_\cg=N\int \tx dt\sqrt{-\g}\; \cg(u,v),
\ee
so that we now need to solve for $\cg(u,v)$, where
\be
u\equiv\frac12\(\p_a^t/N\)^2,\quad v\equiv\f^2.
\ee
By construction, $\wt\cs_\cg$ agrees with the Legendre transform of $S\sbtx{reg}$, up to finite terms. Inserting our ansatz eq.~\eqref{HJ-ansatz} into eq.~\eqref{HJ2} gives us an equation for $\cg(u,v)$,
\be\label{master}
\cg+u-v\(\cg_v^2+2u \cg_{u}^2-M^2\)=0,
\ee
where $\cg_u \equiv \partial_u \cg$ and $\cg_v \equiv \partial_v \cg$. Solving eq.~\eqref{master} asymptotically near the boundary, subject to the boundary conditions dictated by the FG expansions in appendix~\ref{appendix:UVexp}, and specifically eq.~\eqref{UV-asymptotics-nl}, unambiguously determines the divergent part of the on-shell action, and hence the counterterms required to renormalize the theory. 

Moreover, knowing $\cg(u,v)$ allows us to renormalize not only the on-shell action, but also the canonical variables, and hence the response functions through the identities 
\be
\label{canonical-relations}
a_t^\cg= -\frac{\d \wt\cs_\cg}{\d \p_a^t}=-\frac1N\sqrt{-\g}\;\cg_{u}\;\p_a^t,\quad
\p_\F^\cg=\frac{\d \wt\cs_\cg}{\d\F}=N\sqrt{-\g}\;\F^\dag \;\cg_v,\quad \p_{\F^\dag}^\cg=\frac{\d \wt\cs_\cg}{\d\F^\dag}=N\sqrt{-\g}\;\F \;\cg_v.
\ee
Linearizing these, and comparing to the definitions of the response functions in eq.~\eqref{response-screened}, gives 
\bal\label{responses-G}
&\car_{11}^\cg=-\frac{1}{\cg_{u}+2u\cg_{uu}}, &&\car_{\F\F^\dag}^\cg=\car_{\F^\dag\F}^\cg=\frac{2u v\cg_{u v}^2}{\cg_{u}+2u\cg_{uu}}-\(\cg_v+v\cg_{vv}\),\NO\\
&\car_{\F a}^\cg=\frac{\cg_{u v}}{\cg_{u}+2u\cg_{uu}}\(\frac{\g\p_a^t\F^\dag}{N\sqrt{-\g}}\), &&\car_{\F\F}^\cg=\(\frac{2u \cg_{u v}^2}{\cg_{u}+2u\cg_{uu}}-\cg_{vv}\)(\F^\dag)^2,\NO\\
&\car_{\F^\dag a}^\cg=\frac{\cg_{u v}}{\cg_{u}+2u\cg_{uu}}\(\frac{\g\p_a^t\F}{N\sqrt{-\g}}\), &&\car_{\F^\dag\F^\dag}^\cg=\(\frac{2u \cg_{u v}^2}{\cg_{u}+2u\cg_{uu}}-\cg_{vv}\)\F^2,
\eal
where the superscript $\cg$ on $\car_{11}^\cg$ and the other response functions is merely a reminder that, by construction, they coincide with the exact response functions only asymptotically near the boundary.

A second subtlety in the holo-ren concerns the form of the solution $\cg(u,v)$ of eq.~\eqref{master} and is related once more to the asymptotic form of the $AdS_2$ gauge field. The near-boundary asymptotic expansions in appendix~\ref{appendix:UVexp} imply that as $r\to\infty$, $\p^t_a\sim N Q$ and hence $u\sim Q^2/2$. Although the equations of motion allow $Q(t)$ to be an arbitrary function of time, a well-defined space of asymptotic solutions exits only when the constraint eq.~\eqref{Q-condition} holds, which implies that $Q^2/2 = M^2/2+1/8\equiv u_o$ on the constraint surface. As a result, only Neumann boundary conditions are admissible for the $AdS_2$ gauge field $a_t$, \textit{i.e.}~keeping the charge $Q$ fixed.\footnote{The two-impurity holographic Kondo model of ref.~\cite{O'Bannon:2015gwa} involved a $U(2)$ gauge field and a charged scalar in $AdS_2$. Mixed boundary conditions were imposed on the $U(2)$ gauge field, which required the scalar mass $M$ to change dynamically in order to preserve the scalar field's asymptotic form, and hence obtain a well defined variational problem. In the present work we treat $M$ as a fixed parameter of the theory and so mixed boundary conditions on the $AdS_2$ gauge field are not allowed. We stress that these types of problems do not arise in the absence of charged matter. For example in the model of ref.~\cite{Cvetic:2016eiv}, with a $U(1)$ gauge field and dilatonic scalar in $AdS_2$, but no charged matter, $Q$ was a strictly conserved quantity, and both Neumann and Dirichlet boundary conditions were permitted for the gauge field.} The solution of eq.~\eqref{master} satisfying the boundary conditions dictated by the near-boundary asymptotics in eq.~\eqref{UV-asymptotics-nl} thus admits an expansion of the form 
\be\label{counterterms-exp}
\cg(u,v)=\sum_{n=0}^\infty g_n(v)(u-u_o)^n.
\ee
Crucially, the series in eq.~\eqref{counterterms-exp} need not be convergent, and should be understood as an asymptotic expansion only, truncated to a finite, but arbitrary, order. 

Eq.~\eqref{UV-asymptotics-nl} also implies that asymptotically near the boundary, $u-u_o\sim Q^2/2-u_o+\co(e^{-r}r^2\a^2)\sim Q\d Q+\co(e^{-r}r^2\a^2)$, so $u-u_0$ can receive two different potential contributions: $\d Q$, which dominates if non-zero, and the mode $\a$. When $\d Q\neq 0$, the order $m$ term in the expansion in eq.~\eqref{counterterms-exp} encodes the near-boundary divergences of the $m$-point function of the operator sourced by $\d Q$. These divergences enter two-point functions via the near-boundary asymptotic expansions of the response functions in eq.~\eqref{UV-exp-R}. If $\d Q=0$, however, then $u-u_o$ does not contribute to any such divergences, but also, no correlators of the operator sourced by $\d Q$ can be computed. In the latter case, therefore, the counterterms come entirely from $g_0(v)$. In that case, the near-boundary expansions of the response functions appears in eq.~\eqref{UV-exp-R-dQ=0}, which encode the two-point functions of only $\mathcal{O}$ and $\mathcal{O}^{\dagger}$.  

Inserting the expansion in eq.~\eqref{counterterms-exp} into the equation for $\cg$, eq.~\eqref{master}, leads to a tower of differential equations for the coefficients $g_n(v)$, the first three of which are
\bsub
\label{g-eqs}
\bal
&g_0+u_o-v(g_0'^2+2u_o g_1^2-M^2)=0,\\
&g_1+1-v(2g_0'g_1'+2g_1^2+8u_og_1g_2)=0,\\
&g_2-v(g_1'^2+2g_0'g_2'+8u_o g_2^2+12u_og_1 g_3+8g_1g_2)=0,
\eal
\esub
where primes denote $\partial_v$. We will only need to solve these equations asymptotically near the boundary, and only keeping terms up to a certain order, since higher orders will not contribute to the divergences of an $m$-point function with fixed $m$. In particular, the near-boundary asymptotic expansions in eq.~\eqref{UV-asymptotics-nl} allow us to parameterize $g_0(v)$ and $g_1(v)$ as
\be
\label{eq:g0g1param}
g_0=-u_o+h_0,\qquad g_1=-1+h_1,
\ee
where $h_0$ and $h_1$ behave as $v$ times non-negative integer powers of $\log v$ as $v \to 0$, as do $g_2$ and $g_3$. We present the explicit small-$v$ expansions of $h_0$, $h_1$ and $g_2$ in appendix~\ref{appendix:hol-ren}.

The near-boundary, or equivalently small-$v$, asymptotic solutions for $g_0(v)$, $g_1(v)$ and $g_2(v)$ in appendix~\ref{appendix:hol-ren} present yet another subtlety of the holo-ren of our model: our choice of the scalar field's mass, to guarantee that our Kondo coupling $\mathcal{O}^{\dagger}\mathcal{O}$ is classically marginal, leads to powers of $\log v$ in the small-$v$ expansions of $g_0(v)$, $g_1(v)$, and $g_2(v)$. However, such non-analytic in $v$ terms in the counterterms amount to subtracting a non-analytic function of the source of the dual scalar operator, and hence violate the locality of the counterterms. To restore locality, we are forced to sacrifice the radial covariance of the counterterms~\cite{Bianchi:2001de,Papadimitriou:2004rz}, \textit{i.e.} the counterterms will exhibit explicit dependence on the $r$ cutoff, which is the holographic manifestation of a conformal anomaly. This is manifest, for example, in the expressions for the counterterms in eq.~\eqref{final-counterterms} in appendix~\ref{appendix:hol-ren}.

Given a near-boundary asymptotic solution $\cg^{\rm ct}(u,v)$ of eq.~\eqref{master}, the counterterms are defined as
\be\label{counterterms}
\wt S_{\rm ct}=-N\int \tx dt\sqrt{-\g}\; \cg^{\rm ct}(u,v),
\ee
and hence the renormalized action evaluated at the radial cutoff is
\be\label{Sren}
\wt S\sbtx{ren}=\wt S\sbtx{reg}+\wt S\sbtx{ct}.
\ee
By construction, $\wt S\sbtx{ct}$ has the same divergences as $\wt S\sbtx{reg}$, hence $\wt S\sbtx{ren}$ remains finite as we remove the cutoff. Varying $\wt S\sbtx{ren}$ gives then the \textit{renormalized} canonical variables:
\bal\label{ren-action-1}
\d \wt S\sbtx{ren}&=\int \tx dt\; \Big(-a_t^{\rm ren}\d\p_a^t+\p_\F^{\rm ren}\d\F+\p_{\F^\dag}^{\rm ren}\d\F^\dag\Big)+\int \tx d^2x\;\p^{i\;{\rm ren}}_A\d A_i\NO\\
&=\int \tx dt \(-a_t^{\rm ren}\d\p_a^t+\p_\f^{\rm ren}\d\f+\p_{\j}^{\rm ren}\d\j\)+\int \tx d^2x\;\p^{i\;{\rm ren}}_A\d A_i,
\eal
\bal\label{ren-variables-cutoff}
&a_t^{\rm ren}=-\frac{\d \wt S\sbtx{ren}}{\d\p_a^t}=a_t+\frac1N\sqrt{-\g}\;\cg_{u}^{\rm ct}\;\p_a^t,&& \p^{i\;{\rm ren}}_A=\frac{\d \wt S\sbtx{ren}}{\d A_i}=\p_A^i=-\frac{N}{4\p}\bar{\e}^{ij}A_j,\NO\\
&\p_\F^{\rm ren}=\frac{\d \wt S\sbtx{ren}}{\d\F}=\p_{\F}-N\sqrt{-\g}\;\cg_{v}^{\rm ct}\F^\dag,&& 
\p_\f^{\rm ren}=\frac{\d \wt S\sbtx{ren}}{\d\f}=\p_\f-N\sqrt{-\g}\;\cg_{v}^{\rm ct}2\f,\NO\\
&\p_{\F^\dag}^{\rm ren}=\frac{\d \wt S\sbtx{ren}}{\d\F^\dag}=\p_{\F^\dag}-N\sqrt{-\g}\;\cg_{v}^{\rm ct}\F, && \p_\j^{\rm ren}=\frac{\d \wt S\sbtx{ren}}{\d\j}=\p_\j,
\eal
which are evaluated at the radial cutoff. As mentioned above, for the scalar field the canonical momentum is renormalized, while for the $AdS_2$ gauge field, $a_t$ itself is renormalized instead, due to the asymptotic behavior of gauge fields in $AdS_2$ and the fact that the counterterms are local functions of the canonical momentum $\p_a^t$~\cite{Cvetic:2016eiv}.    

We now want to plug the FG expansions of the fields into the renormalized canonical variables in eq.~\eqref{ren-variables-cutoff}. Crucially, however, we show in appendix~\ref{appendix:UVexp} that background solutions and fluctuations have \textit{different} FG expansions, so we must treat them separately.

The FG expansions for background solutions appear in eq.~\eqref{UV-asymptotics-nl}, reproduced here for convenience:
\bsub
\label{eq:FGexpan}
\bal
a_t &=e^rQ-2Q\Big(\frac13\a^2r^3+(\a^2-\a\b)r^2+(2\a^2-2\a\b+\b^2)r\Big)+\m+\cdots,\\
\f &=e^{-r/2}\left(-\a r+\b\right)+\cdots,\\
\j &=\j_-+\j_+ r^{-1}+\cdots,
\eal
\esub
where $\m$, $\a$, $\b$ and $\j_-$ are arbitrary functions of time, while $U(1)$ gauge invariance implies both that $Q$ is independent of time and $\psi_+=0$. The $\ldots$ represent terms that vanish as $r \to \infty$ faster than those shown, and which are completely determined by those shown, via the equations of motion. Inserting eq.~\eqref{eq:FGexpan} into eq.~\eqref{ren-variables-cutoff} and using the counterterms in eq.~\eqref{final-counterterms} allows us to remove the radial cutoff, and hence obtain the renormalized canonical variables in terms of the FG coefficients: 
\bal
\label{ren-variables}
&\Hat a_t^{\rm ren}\equiv \lim_{r\to\infty}a^{\rm ren}_t=\m+A_{(0)t}-\frac{2Q}{3\a}\(\b^3-3\a\b^2+6\a^2\b-6\a^3\), &&\Hat{\p}_a^t\equiv\lim_{r\to\infty}\p_a^t=NQ,\NO\\
&\Hat{\p}_\F^{\rm ren}\equiv\lim_{r\to\infty}(r e^{-r/2}\p_\F^{\rm ren})=N\b e^{-i\j_-},
&&\Hat\F \equiv \lim_{r\to\infty}(r^{-1} e^{r/2}\F)=-\a e^{i\j_-},\NO\\
&\Hat{\p}_{\F^\dag}^{\rm ren}\equiv\lim_{r\to\infty}(r e^{-r/2}\p_{\F^\dag}^{\rm ren})=N\b e^{i\j_-},  &&\Hat\F^\dag \equiv \lim_{r\to\infty}(r^{-1} e^{r/2}\F^\dag)=-\a e^{-i\j_-},\NO\\
&\Hat{\p}_\f^{\rm ren}\equiv\lim_{r\to\infty}(r e^{-r/2}\p_\f^{\rm ren})=2N\b, &&\Hat\f\equiv \lim_{r\to\infty}(r^{-1} e^{r/2}\f)=-\a.
\eal
We took $\m\to\m+A_{(0)t}$ in the expression for $\Hat a_t^{\rm ren}$, because the above asymptotic solutions for the $AdS_2$ fields were obtained in the gauge of eq.~\eqref{CS-sol}, where $A_{(0)t}=0$. However, in order to identify the correct one-point functions, the general dependence on all the sources must be reinstated.\footnote{$A_{(0)t}$ can be reinstated by letting $a_t\to a_t+A_{(0)t}$, recalling that $A_t=A_{(0)t}$ is constant and enters $a_t's$ equation of motion through the $U(1)$ current $J_t$, eq.~\eqref{u1current}, with gauge-covariant derivative in eq.~\eqref{covariantderiv}.} As we shall see, this contribution of $A_{(0)t}$ is crucial for obtaining the two-point functions.

For the fluctuations, we determine the response functions by linearizing eq.~\eqref{ren-variables-cutoff} in the fields induced at the radial cutoff. The complete analysis leading to the full set of renormalized response functions is carried out in appendix~\ref{appendix:hol-ren}. As an illustration, we quote here the results for the renormalized scalar response functions only, which take the form
\bal
\label{eq:renormresponsfunctemp}
&\car^{\rm ren}_{\F\F}=\car_{\F\F}+\cg^{\rm ct}_{vv}(\F^\dag)^2,\qquad\qquad\hskip0.1cm  
\car^{\rm ren}_{\F^\dag\F^\dag}=\car_{\F^\dag\F^\dag}+\cg^{\rm ct}_{vv}\F^2,\NO\\
&\car^{\rm ren}_{\F\F^\dag}=\car_{\F\F^\dag}+\(\cg^{\rm ct}_v+v\cg^{\rm ct}_{vv}\),\qquad
\car^{\rm ren}_{\F^\dag\F}=\car_{\F^\dag\F}+\(\cg^{\rm ct}_v+v\cg^{\rm ct}_{vv}\). 	
\eal
The FG expansions of the response functions appear in eq.~\eqref{Rff-UV}, reproduced here for convenience:
\be
\label{eq:responseFGexpantemp}
\car_{\F^\dag\F}=-\frac12+\frac1r+\frac{\Hat\car_{\F^\dag\F}}{r^2}+\cdots,\quad
\car_{\F\F^\dag}=-\frac12+\frac1r+\frac{\Hat\car_{\F\F^\dag}}{r^2}+\cdots,\quad
\car_{\F\F}=\frac{\Hat\car_{\F\F}}{r^2}+\cdots,
\ee
where $\Hat\car_{\F^\dag\F}$, $\Hat\car_{\F\F^\dag}$, and $\Hat\car_{\F\F}$ are functions of frequency $\omega$. The $\ldots$ represent terms that vanish as $r \to \infty$ faster than those shown, and which are completely determined by those shown, via the equations of motion. Inserting eq.~\eqref{eq:responseFGexpantemp} into eq.~\eqref{eq:renormresponsfunctemp} and using the counterterms in eq.~\eqref{final-counterterms} allows us to remove the radial cutoff, and hence obtain the renormalized response functions:
\bal\label{ren-Rs-cond-N}
&\Hat\car_{\F\F^\dag}^{\rm ren}=\lim_{r\to\infty} \left(r^2 \car_{\F\F^\dag}^{\rm ren}\right)=\Hat\car_{\F\F^\dag},\qquad
\Hat\car_{\F^\dag\F}^{\rm ren}=\lim_{r\to\infty} \left(r^2 \car_{\F^\dag\F}^{\rm ren}\right)=\Hat\car_{\F^\dag\F},\NO\\
&\Hat\car_{\F\F}^{\rm ren}=\lim_{r\to\infty} \left(r^2 \car_{\F\F}^{\rm ren}\right)=\Hat\car_{\F\F},\qquad\hskip0.3cm
\Hat\car_{\F^\dag\F^\dag}^{\rm ren}=\lim_{r\to\infty} \left(r^2 \car_{\F^\dag\F^\dag}^{\rm ren}\right)=\Hat\car_{\F^\dag\F^\dag}.
\eal
Eq.~\eqref{ren-Rs-cond-N} is valid in both the unscreened and screened phases, although the values for $\Hat\car_{\F^\dag\F}$, $\Hat\car_{\F^\dag\F^\dag}$, $\Hat\car_{\F\F}$ and $\Hat\car_{\F\F^\dag}$ are different in the two phases.

\subsection{Boundary Conditions and the Renormalized Generating Function}

The renormalized action $\wt S\sbtx{ren}$ cannot be identified with the generating function in the dual theory until we impose boundary conditions on the fields and add the corresponding {\em finite} boundary terms that impose these boundary conditions. The boundary conditions also dictate which combinations of the renormalized canonical variables in eq.~\eqref{ren-variables} are identified with the sources in the dual field theory. In this subsection we will introduce the finite boundary terms of our model, and then identify the sources in the dual field theory. We will then determine the Ward identities of the dual field theory, and finally, determine the renormalized two-point functions of our model, in terms of coefficients in the FG expansion of the response functions, eq.~\eqref{Rff-UV} or equivalently eq.~\eqref{eq:responseFGexpantemp}.

In our case, three finite boundary terms are required to have a well-posed variational problem that captures the desired physics. First, for the Chern-Simons gauge field alone, with no $AdS_2$ defect fields, a well-posed variational problem requires the boundary term\footnote{Changing the sign of the boundary term in eq.~\eqref{CS-BT} simply interchanges the role of $A_+$ and $A_-$ in the following.\\} ~\cite{Kraus:2006wn,Kiritsis:2010xc,Jensen:2010em,Andrade:2011sx,Chang:2014jna}
\be\label{CS-BT}
S_1=\frac{N}{8\p}\int\tx d^2x\; \sqrt{-\bar\g}\;\bar\g^{ij}A_i A_j=\frac{N}{8\p}\int\tx d^2x\; A_+A_-,
\ee
where $A_\pm\equiv A_x\pm A_t$, and $\bar\g_{ij}$ is the induced metric on a radial slice of $AdS_3$. Second, because the general solution for the Chern-Simons gauge field in eq.~\eqref{CS-sol-cov} receives a contribution from the $AdS_2$ fields, to guarantee a well-posed variational principle for the Chern-Simons gauge field in the presence of the $AdS_2$ defect we must add the finite boundary term   
\be\label{mixed-BT}
S_2=-\frac{1}{4}\int \tx d t\;\p_a^t A_-,
\ee
which couples the Chern-Simons and $AdS_2$ fields. Third, in order to introduce our Kondo coupling, we must add the finite boundary term\footnote{Instead of eq.~\eqref{kondo}, refs.~\cite{Erdmenger:2013dpa,O'Bannon:2015gwa} used the boundary term $(\k/N)\int \tx dt(\Hat\p_\f^{\rm ren})^2=(\k/4N)\int \tx dt(\Hat\p_\F^{\rm ren}+\Hat\p_{\F^\dag}^{\rm ren})^2$, which agrees with eq.~\eqref{kondo} for background solutions, but not for fluctuations. Unlike the boundary term used in refs.~\cite{Erdmenger:2013dpa,O'Bannon:2015gwa}, eq.~\eqref{kondo} preserves the $U(1)$ gauge invariance associated with the $AdS_2$ gauge field.\\} \cite{Erdmenger:2013dpa,O'Bannon:2015gwa}
\be\label{kondo}
S_\k=\frac{\k}{N}\int \tx dt\;\Hat{\p}_\F^{\rm ren}\Hat{\p}_{\F^\dag}^{\rm ren}.
\ee
Putting everything together, the generating functional of the dual theory is\footnote{The free energy obtained from $\cw_{\k}$, that is with the Legendre transform in eq.~\eqref{leg1} and the counterterms in eq.~\eqref{counterterms} with eq.~\eqref{final-counterterms}, agrees with the free energy computed in refs.~\cite{Erdmenger:2013dpa,O'Bannon:2015gwa}.}
\be\label{gen-fun}
\cw_{\k}\equiv\lim_{r\to\infty}(\wt S\sbtx{ren}+S_{\tt 1}+S_{\tt 2}+S_\k).
\ee

To identify the sources in the dual field theory, we consider the variational principle for $\cw_\k$,
\be\label{variational}
\d\cw_\k=\int \tx dt \(-\Hat{\mathfrak a}_t\d\Hat\p_a^t+\Hat\p_\F^{\rm ren}\d\Hat\F_\k+\Hat\p_{\F^\dag}^{\rm ren}\d\Hat\F_\k^\dag\)+\frac{N}{4\p}\int\tx d^2x \(A_{(0)+}+\frac{\p}{N}\Hat\p_a^t\d(x)\)\d A_{(0)-},
\ee
where we have defined
\be\label{sources}
\Hat{\frak a}_t\equiv \Hat a_t^{\rm ren}-A_{(0)t}-\frac14A_{(0)-},\qquad \Hat\F_\k\equiv\Hat\F+\frac{\k}{N}\Hat{\p}_{\F^\dag}^{\rm ren},\qquad
\Hat\F^\dag_\k\equiv\Hat\F^{\dag}+\frac{\k}{N}\Hat{ \p}_\F^{\rm ren}.
\ee
A well-posed variational problem for $\cw_\k$ requires that we keep fixed $\Hat\p_a^t$, $\Hat\F_\k$, $\Hat\F_\k^\dag$, and $A_{(0)-}$, hence we identify these as the sources of the dual operators. Keeping these fixed corresponds to a Neumann boundary condition for the $AdS_2$ gauge field, and a mixed (or Robin) boundary condition for the scalar field, in which $\alpha = \kappa \beta$~\cite{Klebanov:1999tb,Erdmenger:2013dpa}. Our holographic Kondo coupling is $\kappa$, related to the Kondo coupling $\lambda$ of the Kondo Hamiltonian in eq.~\eqref{KondoCFT} as $\kappa \propto N \,\lambda$. For more details about our holographic Kondo coupling and its RG running, see ref.~\cite{Erdmenger:2013dpa} and especially section 4 of ref.~\cite{O'Bannon:2015gwa}.

The one-point functions of the dual operators are then defined via
\bal\label{ren-ops}
&\<\ca_t\>\equiv-\frac{\d\cw}{\d \Hat\p_a^t}=\Hat{\frak a}_t, \qquad \<\cj_+\>\equiv\frac{\d\cw}{\d A_{(0)-}}=\frac{N}{4\p}\(A_{(0)+}+\frac{\p}{N}\Hat\p_a^t\d(x)\),\NO\\
&\<\co\>\equiv-\frac{\d\cw}{\d \Hat\F_\k}=-\Hat{\p}_{\F}^{\rm ren},\qquad
\<\co^\dag\>\equiv-\frac{\d\cw}{\d \Hat\F_\k^\dag}=-\Hat{\p}_{\F^\dag}^{\rm ren},\qquad
\<\bb O\>\equiv-\frac{\d\cw}{\d \Hat\f_\k}=-\Hat{\p}_\f^{\rm ren},
\eal
and are functions of the sources. The scalar operator $\bb O$ is defined as the conjugate to the real source $\Hat\f_\k=(\Hat\F_\k+\Hat\F_\k^\dag)/2$. Using eq.~\eqref{ren-variables}, we can express these in terms of the FG expansion coefficients in eq.~\eqref{UV-asymptotics-nl}, or equivalently eq.~\eqref{eq:FGexpan},
\bal\label{ren-vevs}
&\<\ca_t\>=\m-\frac{2Q}{3\a}\(\b^3-3\a\b^2+6\a^2\b-6\a^3\)-\frac14A_{(0)-}, \qquad \<\cj_+\>=\frac{N}{4\p}\(A_{(0)+}+\p Q\d(x)\),\NO\\
&\<\co\>=-N\b e^{-i\j_-},\qquad
\<\co^\dag\>=-N\b e^{i\j_-},\qquad \<\bb O\>=-2N\b.
\eal

In general, the Ward identities for the $U(1)$ currents dual to the Chern-Simons and $AdS_2$ gauge fields depend on the choice of boundary conditions, since different boundary conditions may preserve different symmetries. Since the Kondo deformation in eq.~\eqref{kondo} preserves the $U(1)$ symmetry on the impurity, the $U(1)$ constraints in eq.~\eqref{constraints} translate to the Ward identities
\be\label{WIDs}
\Hat\F_\k \<\co\>-\Hat\F_\k^\dag \<\co^\dag\>=\o\Hat\p_a^t,\qquad \pa_- \<\cj_+\>=\frac{N}{4\p}\pa_+ A_{(0)-}+\frac{NQ}{4}\pa_-\d(x),
\ee
where $\pa_\pm\equiv \pa_x\pm\pa_t$. The Ward identity for the Chern-Simons current $\cj_+$ is simply the condition $\pa_- A_{(0)+}=\pa_+ A_{(0)-}$, as in the absence of the $AdS_2$ defect. Eqs.~\eqref{WIDs} are operator identities, \textit{i.e.}\ they hold with arbitrary sources. Differentiating the Ward identities in eqs.~\eqref{WIDs} with respect to the sources leads to relations among higher-point functions.

We are finally ready to compute the main result of this section, namely the two-point functions of our model. To write the two-point functions involving $\cj_+$, we introduce chiral coordinates $x^{\pm}$ and their Fourier counterparts, the chiral momenta $p^{\pm}$. Varying our result for $\< \cj_+\>$ in eq.~\eqref{ren-ops}, and using the Ward identity $\pa_- A_{(0)+}=\pa_+ A_{(0)-}$, we find
\be\label{CS-2pt}
\<\cj_+(p_+,p_-)\cj_+(-p_+,-p_-)\>=\frac{N}{4\p}\frac{p_+}{p_-},
\ee
which is completely independent of the $AdS_2$ fields, \textit{i.e.}\ eq.~\eqref{CS-2pt} is identical to the previous results for chiral currents in $(1+1)$-dimensional CFTs in refs.~\cite{Kraus:2006wn,Kiritsis:2010xc,Jensen:2010em}. All two-point functions between $\cj_+$ and the impurity operators are zero, except for one:
\be\label{CS-A-2pt}
\<\cj_+(p_+,p_-)\ca_t(-p_+,-p_-)\>=-\frac{1}{4}.
\ee
Since the two-point functions in eqs.~\eqref{CS-2pt} and~\eqref{CS-A-2pt} are completely insensitive to the transition between the unscreened and screened phases, we will ignore them henceforth.

In the unscreened phase, besides eqs.~\eqref{CS-2pt} and~\eqref{CS-A-2pt}, the only non-trivial two-point function is the one between $\co$ and $\co^\dag$. To derive this two-point function we use the following identities, derived in appendix~\ref{appendix:hol-ren} (in the unscreened phase the response functions $\Hat\car_{\F\p_a^t}$ and $\Hat\car_{\F^\dag\p_a^t}$ vanish, and so the infinitesimal source $\d\Hat\p_a^t$ does not contribute to these expressions): 
\be\label{momentum-var}
\d\Hat\p_{\F^\dag}=-N(\Hat\car_{\F^\dag\F}\d\Hat\F+\Hat\car_{\F^\dag\F^\dag}\d\Hat\F^\dag),\qquad
\d\Hat\p_{\F}=-N(\Hat\car_{\F\F}\d\Hat\F+\Hat\car_{\F\F^\dag}\d\Hat\F^\dag),
\ee
where $\Hat\car_{\F^\dag\F}$, $\Hat\car_{\F^\dag\F^\dag}$, $\Hat\car_{\F\F}$ and $\Hat\car_{\F\F^\dag}$ are the renormalized scalar response functions, which appear as coefficients in the FG expansions in eq.~\eqref{eq:responseFGexpantemp}. The quantities in eq.~\eqref{momentum-var} represent the renormalized one-point functions, which in the regime of linear response are linearly proportional to the sources, where the proportionality factor is the renormalized two-point function. We thus need to write eq.~\eqref{momentum-var} in terms of the scalar sources. Using the scalar sources defined in eq.~\eqref{sources}, we find
\be\label{source-var}
\d\Hat\F_\k=(1-\k \Hat\car_{\F^\dag\F})\d\Hat\F-\k \Hat\car_{\F^\dag\F^\dag}\d\Hat\F^\dag,\qquad \d\Hat\F^\dag_\k=-\k \Hat\car_{\F\F}\d\Hat\F+(1-\k \Hat\car_{\F\F^\dag})\d\Hat\F^\dag.
\ee

In the unscreened phase, in appendix~\ref{appendix:UVexp} we find that $\Hat\car_{\F^\dag\F^\dag}=0$ and $\Hat\car_{\F\F}=0$, indicating that $\< \co(\o)^\dag\co^\dag(-\o)\>=0$ and $\< \co(\o)\co(-\o)\>=0$ respectively. Using that result, and by combining the variations in eqs.~\eqref{momentum-var} and~\eqref{source-var} in the unscreened phase, we then find
\be
\d\Hat\p_{\F^\dag}=-N\frac{\Hat\car_{\F^\dag\F}}{(1-\k \Hat\car_{\F^\dag\F})}\d\Hat\F_\k,\qquad
\d\Hat\p_{\F}=-N\frac{\Hat\car_{\F\F^\dag}}{(1-\k \Hat\car_{\F\F^\dag})}\d\Hat\F^\dag_\k,
\ee
from which we read off the two-point functions 
\be\label{eq:unscreened2pt}
\<\co^\dag(\o)\co(-\o)\>_\k=N\frac{\Hat\car_{\F^\dag\F}}{1-\k \Hat\car_{\F^\dag\F}},\qquad \<\co(\o)\co^\dag(-\o)\>_\k=N\frac{\Hat\car_{\F\F^\dag}}{1-\k \Hat\car_{\F\F^\dag}}.
\ee
We computed $\car_{\F^\dag\F}$ and $\car_{\F\F^\dag}$ in the unscreened phase in subsection~\ref{subsec:unscreenedresponse}, with the result in eq.~\eqref{eq:unscreenedresponseresults} and asymptotic expansions in eq.~\eqref{respfuncasymp}. Indeed, comparing the asymptotic expansions in eq.~\eqref{respfuncasymp} to the general FG expansions in eq.~\eqref{eq:responseFGexpantemp}, we find
\bal
\label{ren-responses-symm}
\begin{aligned}
	&\Hat\car_{\F^\dag\F}=H\left(-\frac12+ iQ-i\o z_H\right)+H\left(-\frac12-iQ\right)-\log (z_H \Lambda/2),\\
	&\Hat\car_{\F\F^\dag}=H\left(-\frac12- iQ-i\o z_H\right)+H\left(-\frac12+ iQ\right)-\log (z_H \Lambda/2),
\end{aligned}
\eal
where $1/\Lambda$ is the near-boundary cutoff in $z$. Plugging eq.~\eqref{ren-responses-symm} into eq.~\eqref{eq:unscreened2pt} then gives our main result for the unscreened phase, the renormalized two-point functions $\<\co^\dag(\o)\co(-\o)\>_\k$ and $\<\co(\o)\co^\dag(-\o)\>_\k$ as functions of the field theory parameters $Q$, $T$, and $\omega$. We explore the physics of these two-point functions in detail in section~\ref{sec:highT}.

In the screened phase the variations of the renormalized one-point functions are 
\bsub\label{momentum-var-screened}
\bal
\d\Hat\p_{\F^\dag}&=-N(\Hat\car_{\F^\dag\F}\d\Hat\F+\Hat\car_{\F^\dag\F^\dag}\d\Hat\F^\dag)+\Hat\car_{\F^\dag\p_a^t}\d\Hat\p_a^t,\\
\d\Hat\p_{\F}&=-N(\Hat\car_{\F\F}\d\Hat\F+\Hat\car_{\F\F^\dag}\d\Hat\F^\dag)+\Hat\car_{\F\p_a^t}\d\Hat\p_a^t,\\
\d \Hat{\frak a}_t&=-\(\Hat\car_{\p_a^t\F}\d\Hat\F+\Hat\car_{\Hat\p_a^t\F^\dag}\d\Hat\F^\dag+\Hat\car_{\p_a^t\p_a^t}\d\Hat \p_a^t\)-\frac14A_{(0)-}.
\eal
\esub
All response functions in these expressions are determined in appendix \ref{appendix:hol-ren}, and their explicit forms in terms of the coefficients of the FG expansions are shown in eq.~\eqref{ren-responses}.

To evaluate the two-point functions involving the scalar operators we again need to determine the infinitesimal sources $\d\F_\k$ and $\d\F^\dag_\k$ in terms of the sources of the undeformed theory. From the definitions in eq.~\eqref{sources}, and the expressions for the response functions in terms of the FG coefficients in the screened phase in eq.~\eqref{ren-responses}, we find
\be\label{eq:screenedsources}
\d\Hat\F_\k=\Big(1-\frac{\k}{2}\Hat\car_{22}\Big)\d\Hat\f-\frac{\k}{2N}\Big(\Hat\car_{12}^\infty+\frac{\o}{\a_0}\Big)\d\Hat\p_a^t,\quad
\d\Hat\F^\dag_\k=\Big(1-\frac{\k}{2}\Hat\car_{22}\Big)\d\Hat\f-\frac{\k}{2N}\Big(\Hat\car_{12}^\infty-\frac{\o}{\a_0}\Big)\d\Hat\p_a^t,
\ee
where $\Hat\car_{12}^\infty$ is defined in eq.~\eqref{R-infinity}, and recall $\d\Hat\f=(\d\Hat\F+\d\Hat\F^\dag)/2$. However, linearizing the first Ward identity in eq.~\eqref{WIDs} around a screened phase background solution gives
\be\label{linearWard}
\d\Hat\F-\d\Hat\F^\dag=-\frac{\k}{N}\frac{\o}{\a_0}\d\Hat\p_a^t,
\ee
so eq.~\eqref{eq:screenedsources} can be re-written as
\bsub
\label{eq:sourcesscreened}
\bal
\d\Hat\F_\k&=\Big(1-\frac{\k}{4}\Hat\car_{22}\Big)\d\Hat\F-\frac{\k}{4}\Hat\car_{22}\,\d\Hat\F^\dag-\frac{\k}{2N}\Hat\car_{12}^\infty\,\d\Hat\p_a^t,\\
\d\Hat\F^\dag_\k&=-\frac{\k}{4}\Hat\car_{22}\,\d\Hat\F+\Big(1-\frac{\k}{4}\Hat\car_{22}\Big)\d\Hat\F^\dag-\frac{\k}{2N}\Hat\car_{12}^\infty\,\d\Hat\p_a^t.
\eal
\esub
Eqs.~\eqref{eq:sourcesscreened} can be inverted to obtain $\d\Hat\F$ and $\d\Hat\F^\dag$ in terms of $\d\Hat\F_\k$, $\d\Hat\F^\dag_\k$ and $\d\Hat\p_a^t$:
\bsub
\label{inversion-screened}
\bal
\d\F&=\frac{1}{1-\frac{\k}{2}\Hat\car_{22}}\(\Big(1-\frac{\k}{4}\Hat\car_{22}\Big)\d\Hat\F_k+\frac{\k}{4}\Hat\car_{22}\,\d\Hat\F_\k^\dag+\frac{\k}{2N}\Hat\car_{12}^\infty\,\d\Hat\p_a^t\),\\
\d\F^\dag&=\frac{1}{1-\frac{\k}{2}\Hat\car_{22}}\(\frac{\k}{4}\Hat\car_{22}\d\Hat\F_\k+\Big(1-\frac{\k}{4}\Hat\car_{22}\Big)\d\Hat\F_k^\dag+\frac{\k}{2N}\Hat\car_{12}^\infty\,\d\Hat\p_a^t\).
\eal
\esub
Inserting eq.~\eqref{inversion-screened} into eq.~\eqref{momentum-var-screened} for the scalar one-point functions and making use of the linearized Ward identity eq.~\eqref{linearWard}, we obtain
\be
\d\Hat\p_\F=\d\Hat\p_{\F^\dag}=\frac{1}{1-\frac{\k}{2}\Hat\car_{22}}\(-\frac{N}{4}\Hat\car_{22}(\d\Hat\F_\k+\d\Hat\F_\k^\dag)-\frac12\Hat\car_{12}^\infty\,\d\Hat\p_a^t\),
\ee
from which we read off the two-point functions,
\bsub
\label{eq:screened2ptresult}
\bal
&\<\co(\o)\co(-\o)\>_\k=\<\co(\o)\co^\dag(-\o)\>_\k=\<\co^\dag(\o)\co(-\o)\>_\k=\<\co^\dag(\o)\co^\dag(-\o)\>_\k=\frac{N\Hat\car_{22}/4}{1-\k\Hat\car_{22}/2},\\
&\<\co(\o)\ca_t(-\o)\>_\k=\<\co^\dag(\o)\ca_t(-\o)\>_\k=\frac{\Hat\car_{12}^\infty/2}{1-\k\Hat\car_{22}/2}.\label{2pt-functions-mixed}
\eal
\esub

Moreover, inserting eq.~\eqref{inversion-screened} in the gauge field one-point function in eq.~\eqref{momentum-var-screened} and using the expressions in eqs.~\eqref{ren-responses} gives 
\bal
\d \Hat{\frak a}_t&=\frac12\(\Hat\car_{12}^\infty+\o/\a_0\)\d\Hat\F+\frac12\(\Hat\car_{12}^\infty-\o/\a_0\)\d\Hat\F^\dag-\frac1N\Hat\car_{11}^\infty\d\Hat\p_a^t-\frac14A_{(0)-}\\
&=\frac{\Hat\car_{12}^\infty/2}{1-\k\Hat\car_{22}/2}\(\d\Hat\F_\k+\d\Hat\F_\k^\dag\)-\frac1N\(\Hat\car_{11}^\infty+\frac{\k}{2}\(\frac{\o}{\a_0}\)^2-\frac{\frac\k2(\Hat\car_{12}^\infty)^2}{1-\k\Hat\car_{22}/2}\)\d\Hat\p_a^t-\frac14A_{(0)-},\NO
\eal
which reproduce the two-point functions in eqs,~\eqref{CS-A-2pt} and \eqref{2pt-functions-mixed}, and from which we read off the two-point function
\be\label{eq:screened2ptresult-A}
\<\ca_t(\o)\ca_t(-\o)\>_\k=\frac1N\(\Hat\car_{11}^\infty+\frac{\k}{2}\(\frac{\o}{\a_0}\)^2-\frac{\frac\k2(\Hat\car_{12}^\infty)^2}{1-\k\Hat\car_{22}/2}\).
\ee

As mentioned at the end of subsection~\ref{subsec:screenedresponse}, in the screened phase we have been able to obtain the background solutions $a_t^0$ and $\phi_0$ only numerically, and hence have only solved eq.~\eqref{riccati-screened-z} for $\car_{22}$, $\car_{12}$ and $\car_{11}$ numerically. From those numerical solutions we then extract the FG expansion coefficients $\Hat\car_{22}$, $\Hat\car_{12}$ and $\Hat\car_{11}$ using the near boundary expansions in eqs.~\eqref{UV-exp-R}, and thus obtain the two-point function via eqs.~\eqref{eq:screened2ptresult} and \eqref{eq:screened2ptresult-A}. We present our numerical results for the scalar two-point functions in the screened phase in section~\ref{sec:lowT}.

\section{Review: Fano Resonances}
\label{sec:fano}
\setcounter{equation}{0}

A spectral function $\rho$ is defined as the anti-Hermitian part of a retarded Green's function, $G$:
\beq
\rho \equiv i \left [ G - G^{\dagger} \right ] = -2 \, \textrm{Im} \,G.
\eeq
In our system, we are interested in
\beq
\godo \equiv \<\co^\dag(\o)\co(-\o)\>_\k,
\eeq
and the associated spectral function $\rodo = - 2 \,\textrm{Im}\,\godo$. Given the anti-Hermitian part of a Green's function, a Kramers-Kroning relation completely determines the Hermitian part. The latter therefore contains no additional information, so we will compute only the former, \textit{i.e.}\ spectral functions. In general, for real $\omega$, when $\omega>0$ the spectral function is proportional to the probability amplitude to excite a particle, whereas when $\omega<0$ the spectral function is proportional to (minus) the probability amplitude to excite an anti-particle (hole). Unitarity implies the positivity property $\omega \rodo \geq 0$ for real $\omega \in (-\infty,\infty)$, so that $\rodo \geq0$ when $\omega>0$ and $\rodo \leq0$ when $\omega<0$.

Spectral functions exhibit Fano resonances when a continuum (in energy) of states scatter off a resonant state, or discrete set of resonant states, with energy somewhere in the continuum. The resonant states are always localized in energy, and usually (but not always) localized in real space, \textit{i.e.}\ they are often associated with some ``impurity''. Numerous examples of Fano resonances appear throughout physics, but a classic example is the scattering of light (the continuum) off the excited states of an atom (the resonant states). As mentioned in section~\ref{sec:intro}, Fano resonances have also been observed in quantum impurity models in one spatial dimension, including side-coupled QDs~\cite{RevModPhys.82.2257,2000PhRvB..62.2188G}. For a brief review of Fano resonances, see for example ref.~\cite{RevModPhys.82.2257}.

The Fano spectral function is
\beq
\label{eq:fanospec}
\rhofano=\frac{\left(\omega-\omega_0+q \, \frac{\Gamma}{2}\right)^2}{\left(\omega-\omega_0\right)^2+\left(\frac{\Gamma}{2}\right)^2},
\eeq
where $\omega_0$ fixes the position of the Fano resonance, $\Gamma$ fixes the width, $q$ is called the ``Fano'' or ``asymmetry'' parameter, and we have fixed the normalization so that $\lim_{\omega\to \pm \infty} \rhofano=1$. The $\rhofano$ in eq.~\eqref{eq:fanospec} can be re-written in an illuminating way:
\beq
\label{eq:rhofano2}
\rhofano = 1 + \frac{\left(q^2-1\right)\left(\frac{\Gamma}{2}\right)^2}{\left(\omega-\omega_0\right)^2+\left(\frac{\Gamma}{2}\right)^2} + \frac{2q\left(\frac{\Gamma}{2}\right)\left(\omega-\omega_0\right)}{\left(\omega-\omega_0\right)^2+\left(\frac{\Gamma}{2}\right)^2},
\eeq
where on the right-hand-side, the first term in the sum (the $1$) represents the continuum, the second term is a Lorentzian representing the resonant state, and the third term is the ``mixing'' or ``interference'' term arising from the interaction between the two. Indeed, the essential physics of Fano resonances is that the incoming scattering states, from the continuum, have two paths through the system: they can either scatter off the resonant state (``resonant scattering'') or they can bypass the resonant state (``non-resonant scattering''). The interference between the two paths generically produces an asymmetric resonance, the Fano resonance. The Fano parameter $q$ characterizes the amount of mixing or interference. More precisely, $q^2$ is proportional to a ratio of probabilities: $q^2 \propto$ the probability of resonant scattering over the probability of non-resonant scattering.

\begin{figure}[h!]
\begin{center}
\includegraphics[width=0.7\textwidth]{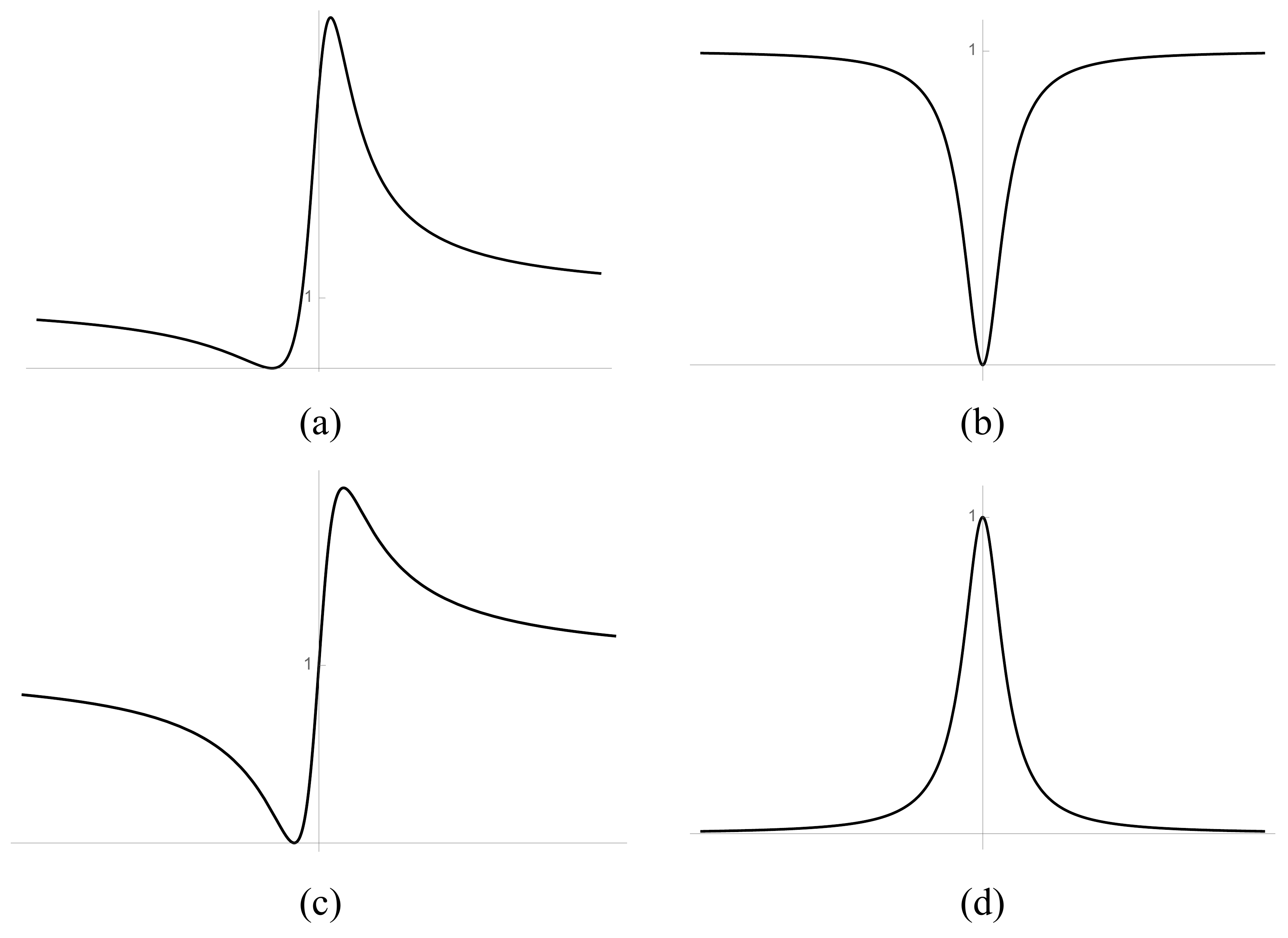}
\caption{\label{fig:fanopics}The Fano spectral function, $\rhofano$ in eq.~\eqref{eq:fanospec}, as a function of $\left(\omega-\omega_0\right)/\left(\Gamma/2\right)$ (so the origin is at $\omega=\omega_0$, where $\rhofano=q^2$), for (a) a generic value of $q$, where the Fano resonance is asymmetric, (b) $q=0$, where the Fano resonance becomes a symmetric dip or \textit{anti-resonance}, and (c) $q=1$, where the minimum and maximum becomes symmetric. (d) Shows $\rhofano/q^2$ as a function of $\left(\omega-\omega_0\right)/\left(\Gamma/2\right)$ in the limit $q \to \infty$, where the Fano resonance becomes a Lorentzian.}
\end{center}
\end{figure}

Fig.~\ref{fig:fanopics} shows $\rhofano$ for some representative values of $q$. Fig.~\ref{fig:fanopics} (a) shows $\rhofano$ for generic $q>0$, with a characteristic asymmetric Fano resonance. In these cases, $\rhofano$ has a minimum and maximum:
\begin{center}
\begin{tabular}{cccc}
minimum: & $\rhofano=0$ & at & $\omega=\omega_0 - q \frac{\Gamma}{2}$ \\ \\
maximum: & $\rhofano=1+q^2$ & at & $\omega=\omega_0 + \frac{1}{q} \frac{\Gamma}{2}$.
\end{tabular}
\end{center}
At $\omega=\omega_0$, which is between the minimum and maximum, $\rhofano=q^2$. Taking $q<0$ simply reflects the Fano resonance described above about the $\omega=0$ axis, so we will restrict to $q>0$ henceforth.

For the special values $q=0$, $1$, and $\infty$, the Fano resonance becomes symmetric. Fig.~\ref{fig:fanopics} (b) shows $\rhofano$ for $q=0$, meaning purely non-resonant scattering. In this case, the maximum moves to $\omega=+\infty$ while the minimum moves to $\omega=\omega_0$, leaving only a symmetric dip called an \textit{anti-resonance}. Fig.~\ref{fig:fanopics} (c) shows $\rhofano$ for $q=1$, meaning equal probabilities of resonant and non-resonant scattering. In this case, the minimum and maximum are symmetric about $\omega=\omega_0$. Fig.~\ref{fig:fanopics} (d) shows $\rhofano/q^2$ for $q \to \infty$, meaning purely resonant scattering. In this case, the minimum moves to $\omega=-\infty$ and the maximum moves to $\omega=\omega_0$, leaving the Lorentzian peak of the resonant state itself.

\begin{figure}[h!]
\begin{center}
\includegraphics[width=0.6\textwidth]{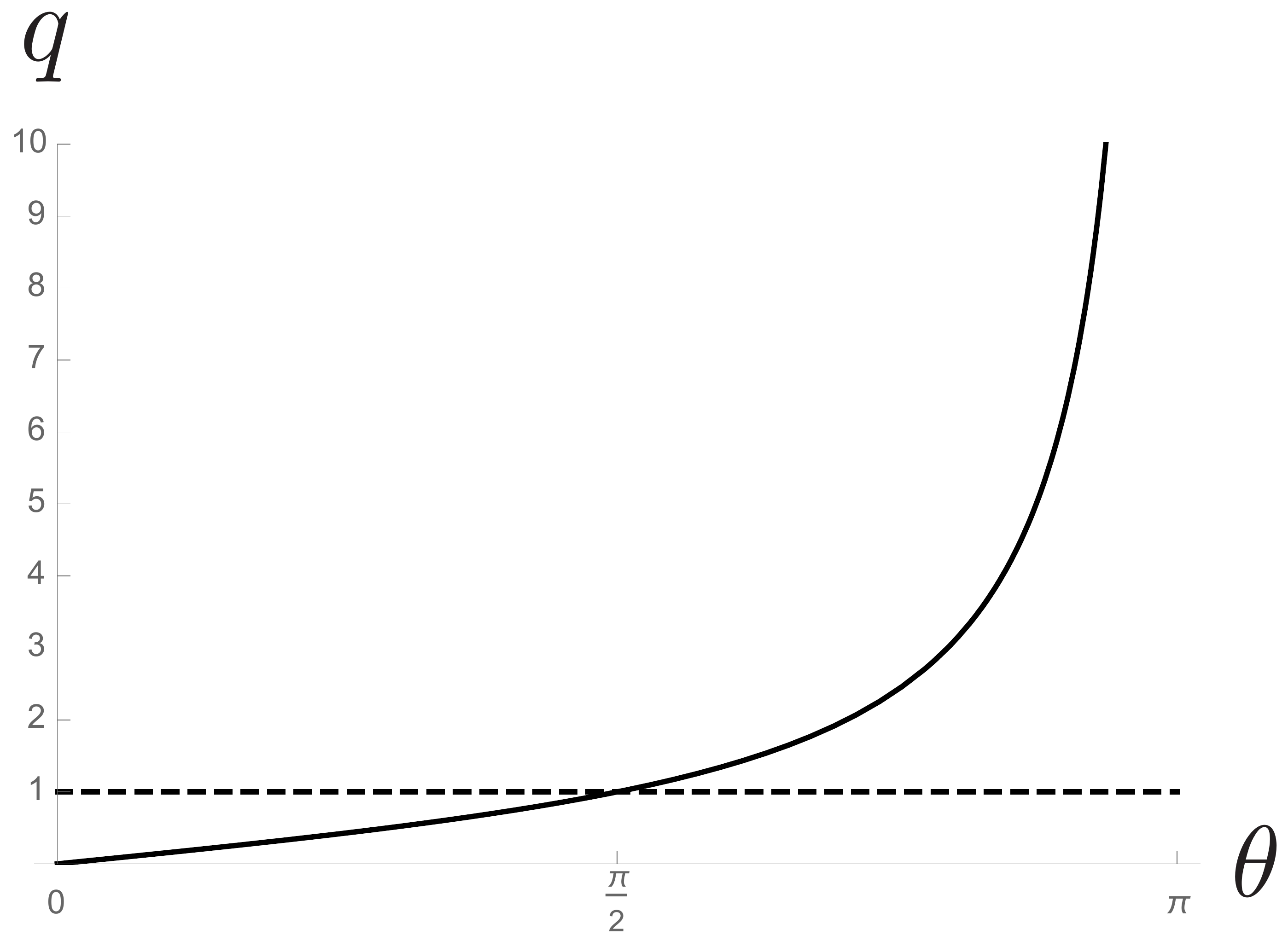}
\caption{\label{fig:qplot}The Fano/asymmetry parameter $q$ as a function of $\theta$ (solid black line), from eq.~\eqref{eq:qtheta}, for a simple pole in a retarded Green's function with complex residue $Z = |Z| e^{i\theta}$. The value $q=1$ (dashed black line) produces a symmetric Fano resonance, as in fig.~\ref{fig:fanopics} (c).}
\end{center}
\end{figure}

Near a simple pole at $\omega^*=\omega_R + i \omega_I$ in the complex $\omega$ plane, the retarded Green's function is $G = \frac{Z}{\omega-\omega^*}$, with residue $Z$. As is well-known, a real-valued $Z$ leads to a Lorentzian resonance in $\rho$ (where the latter is restricted to real $\omega$). However, a complex-valued residue, $Z=Z_R + i Z_I$ with $Z_I\neq 0$, leads to a Fano resonance:
\beq
\label{eq:fanofromG}
\rho = - 2 \,\textrm{Im}\, G = \frac{-2Z_R\,\omega_I}{\left(\omega-\omega_R\right)^2+\left(\omega_I\right)^2}+\frac{-2 Z_I\left(\omega-\omega_R\right)}{\left(\omega-\omega_R\right)^2+\left(\omega_I\right)^2} = -1 + \rhofano,
\eeq
where in the final equality we added and subtracted $1$, and used the form of $\rhofano$ in eq.~\eqref{eq:rhofano2}, with the identifications $\omega_0 = \omega_R$ and $\Gamma/2 = |\omega_I|$ and
\beq
q^2 - 1 = -\frac{2Z_R}{\omega_I}, \qquad 2q = - \frac{2Z_I}{|\omega_I|}.
\eeq
The ratio of these two equations leads to $q^2 -  \textrm{sign}\left(\omega_I\right)2\frac{Z_R}{Z_I} q - 1 =0$. Unitarity requires $\textrm{sign}\left(\omega_I\right)=-1$, in which case the solutions for $q$ are
\beq
\label{eq:qZ}
q = -\frac{Z_R}{Z_I} \pm \sqrt{1+\frac{Z_R^2}{Z_I^2}},
\eeq
or equivalently, using $Z = |Z| e^{i \theta}$,
\beq
\label{eq:qtheta}
q = -\cot \theta \pm \csc \theta.
\eeq
We can obtain the solution with the minus (lower) sign from the solution with the plus (upper) sign by shifting $\theta \to \theta + \pi$, so we will restrict to the upper (plus) sign and to the interval $\theta \in [0,\pi]$, where $q>0$. Fig.~\ref{fig:qplot} shows $q$ as a function of $\theta$, and the table below shows how various limits of $\theta$ lead to the symmetric Fano resonances in fig.~\ref{fig:fanopics}.

\begin{table}[h!]
\begin{center}
\begin{tabular}{|c|c|c|c|c|}
\hline
$\theta$ & $Z_R$ & $Z_I$ & $q$ & Fig.~\ref{fig:fanopics} \\ \hline \hline
$0$ & $|Z|$ & $0$ & $0$ & (b) \\ \hline
$\pi/2$ &  $0$ & $|Z|$ & $1$ & (c)  \\ \hline
$\pi$ & $-|Z|$ & $0$ & $\infty$ & (d) \\ \hline
\end{tabular}
\end{center}
\end{table}

In sections~\ref{sec:highT} and~\ref{sec:lowT} we will see that generically the spectral functions of $\mathcal{O}$ and $\mathcal{O}^{\dagger}$ exhibit Fano resonances, in both the unscreened and screened phases, with various $q$. In our case, the continuum arises from the $(0+1)$-dimensional scale invariance associated with the $AdS_2$ subspace, inherited from the $(1+1)$-dimensional scale invariance associated with $AdS_3$, and which forces any spectral function to be a power law in $\omega$, \textit{i.e.}\ a continuum. Resonances can then only occur if scale invariance is broken, which we achieve via our marginally relevant Kondo coupling. In our model, the asymmetry is related to particle-hole symmetry breaking, that is, $q$ will depend on $Q$.

\section{Unscreened Phase}
\label{sec:highT}
\setcounter{equation}{0}

In this section we use the results of sections~\ref{sec:review} and~\ref{sec:holorg} to determine the excitation spectrum of our system in the unscreened phase, by locating the poles of $\godo$ and $\good$ in the plane of complex frequency $\omega$ (subsection~\ref{sec:normpoles}), and the corresponding peaks in $\rodo$ and $\rood$ for real $\omega$ (subsection~\ref{sec:normspec}).

Some results for the poles appear already in refs.~\cite{Erdmenger:2013dpa}, in the unscreened phase and at small $\omega$. Indeed, a key result of ref.~\cite{Erdmenger:2013dpa} was that in the unscreened phase, and for any $Q$ (including $Q=0$), as $T\to T_c^+$ a pole moves towards the origin of the complex $\omega$ plane, reaching the origin at precisely $T=T_c$. If we then take $T<T_c$ but remain in the unscreened phase, then the pole moves into the upper half of the complex $\omega$ plane, $\textrm{Im} \,\omega >0$, signaling the instability towards the screened phase.

Further results appeared in ref.~\cite{Erdmenger:2016vud}, including in particular our central result, the analytic (\textit{i.e.}~non-numerical) result for $\godo$. In ref.~\cite{Erdmenger:2016vud}, we discussed the movement of poles in $\godo$ as $T \to T_c^+$, presented an analytic formula for $T_c$ in terms of $T_K$ and $Q$, showed that $\rodo$ generically has Fano resonances, derived an analytic form for the pole producing the Fano resonance for $T$ just above $T_c$, and showed that $q \to \infty$ as $Q \to \infty$, producing symmetric Fano resonances (Lorentzians).

In this section we will not only reproduce these results, but also extend them, in particular by exploring in far greater detail the $T$ and $Q$ dependence of the poles in $\godo$ and peaks in $\rodo$. Moreover, we will present analytic results for poles in the $T \gg T_c$ limit, which demonstrate conclusively the appearance of Fano resonances in $\rodo$ when $T \gg T_c$.

As derived in section~\ref{sec:holorg}, we have $\< \co(\o)^\dag\co^\dag(-\o)\>=0$ and $\< \co(\o)\co(-\o)\>=0$, and from eq.~\eqref{eq:unscreened2pt}
\beq
\label{eq:odo1}
\godo= N \frac{\hat{\mathcal{R}}_{\Phi^{\dagger}\Phi}}{1-\kappa\,\hat{\mathcal{R}}_{\Phi^{\dagger}\Phi}},
\eeq
where from eq.~\eqref{ren-responses-symm} we have
\beq
\label{eq:odo2}
\hat{\mathcal{R}}_{\Phi^{\dagger}\Phi} = H\left(-\frac{1}{2} + i Q - i \omega z_H\right) + H\left(-\frac{1}{2} - i Q\right) - \ln (z_H\Lambda/2),
\eeq
where $H(n)$ denotes the $n^{\textrm{th}}$ harmonic number.

We can write eqs.~\eqref{eq:odo1} and~\eqref{eq:odo2} in terms of field theory quantities using $z_H = 1/(2 \pi T)$ and by replacing $\Lambda$ with the Kondo temperature $T_K$, following refs.~\cite{Erdmenger:2013dpa,O'Bannon:2015gwa}, as follows. In the metric of eq.~\eqref{ads3metric}, we re-scale to produce dimensionless coordinates,
\beq
\label{eq:rescaledcoords}
(z/z_H,t/z_H,x/z_H) \to (z,t,x),
\eeq
which leaves the metric in eq.~\eqref{ads3metric} invariant, except for $h(z)  = 1 - z^2/z_H^2 \to 1-z^2$, so the boundary remains at $z=0$ but the horizon is now at $z=1$. We also re-scale $a_t(z)z_H \to a_t(z)$, which is then dimensionless. After the re-scaling, $\Phi(z)$'s asymptotic expansion is
\beq
\label{eq:rescaledphiexp}
\Phi(z) = \alpha_T \, z^{1/2} \ln z + \beta_T \, z^{1/2} + \ldots,
\eeq
where $\ldots$ represents terms that vanish faster than those shown when $z \to 0$, and are completely determined by the terms shown, via the equations of motion. The boundary condition $\alpha = \kappa \beta$ discussed below eq.~\eqref{sources} is now $\alpha_T = \kappa_T \beta_T$, with $\kappa_T \beta_T = z_H^{1/2} \kappa \beta$, and where
\beq
\kappa_T \equiv \frac{\kappa}{1 + \kappa \ln \left(z_H\Lambda\right)},
\eeq
is our running holographic Kondo coupling, with UV cutoff $\Lambda$. If $\kappa<0$, then if $T$ increases, meaning $z_H = 1/(2 \pi T) \to 0$, then $\kappa_T$ exhibits asymptotic freedom, $\kappa_T \to 0^-$. We thus identify $\kappa<0$ as an anti-ferromagnetic holographic Kondo coupling. If $\kappa<0$ and $T$ decreases, so $z_H = 1/(2 \pi T)$ increases, then $\kappa_T$ diverges by definition at the Kondo temperature,
\beq
\label{eq:TKdef}
T_K \equiv \frac{\Lambda}{2\pi} \, e^{1/\kappa}.
\eeq
Using eq.~\eqref{eq:TKdef} in eq.~\eqref{eq:odo2} to replace $\Lambda$ with $T_K$, we thus find
\beq
\label{eq:odo3}
\godo = -\frac{N}{\kappa} - \frac{N}{\kappa^2}\frac{1}{H\left(-\frac{1}{2} + i Q - i \frac{\omega}{2\pi T}\right) + H\left(-\frac{1}{2} - i Q\right) + \ln\left(\frac{2T}{T_K}\right)}.
\eeq
The form of $\good$ is the same as $\godo$, but with $Q \to -Q$.

\subsection{Unscreened Phase: Poles in the Green's Function}
\label{sec:normpoles}

Clearly $\godo$ in eq.~\eqref{eq:odo3} has a pole whenever
\beq
\label{eq:normpoles}
H\left(-\frac{1}{2} + i Q - i \frac{\omega}{2\pi T}\right) + H\left(-\frac{1}{2} - i Q\right) + \ln\left(\frac{2T}{T_K}\right)=0.
\eeq
Given values for $Q$ and $T/T_K$, we can thus find the poles of $\godo$ by solving eq.~\eqref{eq:normpoles} for $\omega/(2\pi T)$.

Because the form of $\good$ is the same as $\godo$, but with $Q \to -Q$, if $\omega = \textrm{Re}\left(\omega\right) + i \,\textrm{Im}\left(\omega\right)$ is a pole of $\godo$, then $-\overline{\omega}= -\textrm{Re}\left(\omega\right) + i \, \textrm{Im}\left(\omega\right)$ will be a pole of $\good$. In other words, the poles of $\godo$ and $\good$ come in pairs mirrored about the imaginary axis in the $\omega$-plane.

\begin{figure}[h!]
\begin{center}
\includegraphics[width=0.8\textwidth]{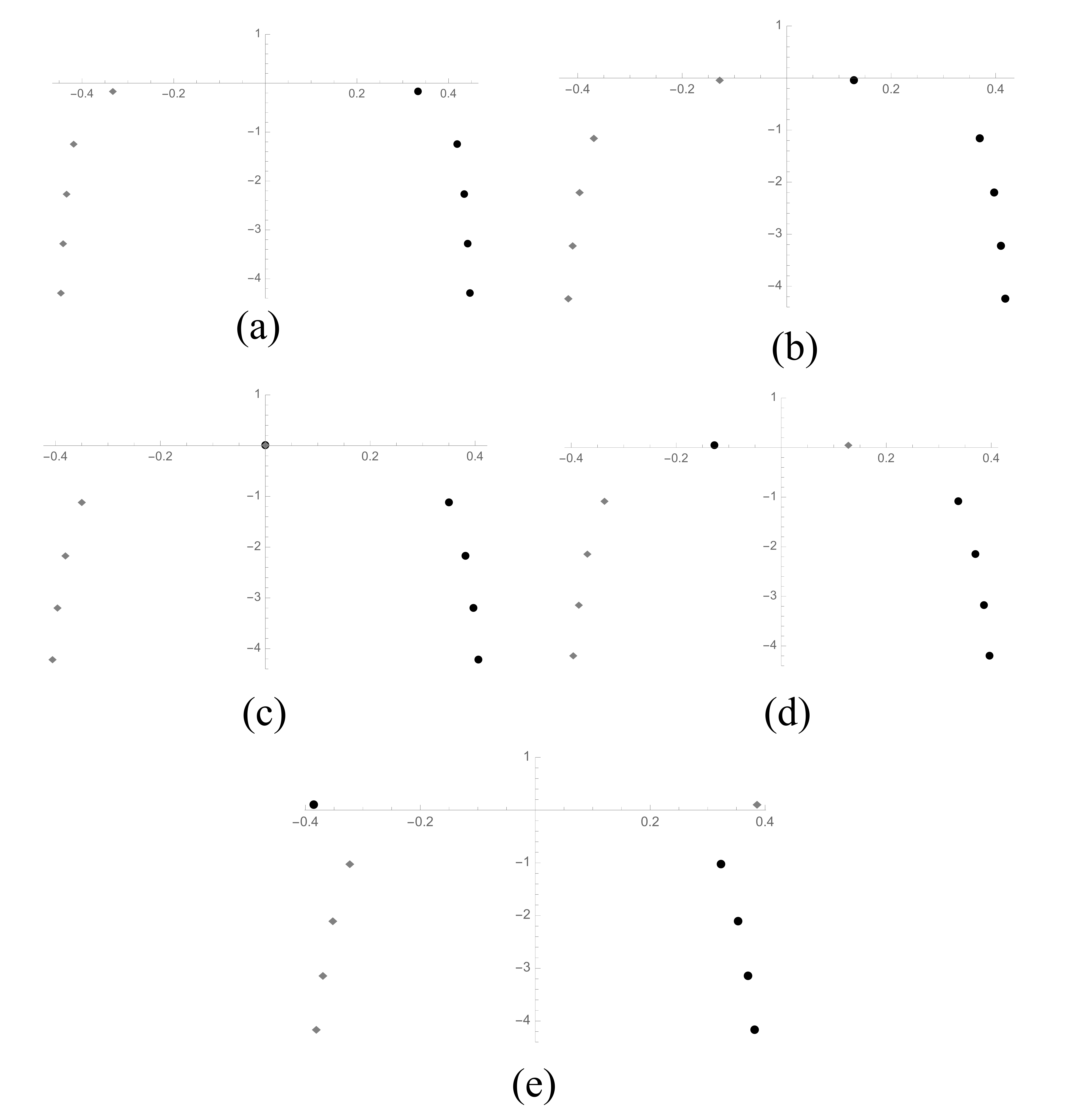}
\caption{\label{fig:Q=halfnormalQNMs}The positions of poles in $\godo$ (black dots) and $\good$ (gray diamonds) in the complex $\omega/(2\pi T)$ plane, determined by solving eq.~\eqref{eq:normpoles} numerically, for $Q=0.5$ and $T/T_K$ equal to (a) $4.92$, (b) $1.34$, (c) $0.895$, (d) $0.671$, and (e) $0.447$. As $T/T_K$ decreases, the ``lowest'' poles, meaning the poles closest to the origin at $T/T_K=4.92$ (a), move towards the origin (b), reach the origin at $T/T_K=0.895$ (c), and then pass into the upper half of the complex $\omega/(2\pi T)$ plane (d and e), producing an instability.}
\end{center}
\end{figure}

Fig.~\ref{fig:Q=halfnormalQNMs} shows our numerical results for the positions of poles of $\godo$ (black dots) and $\good$ (gray diamonds) in the complex $\omega/(2 \pi T)$ plane, for the representative value $Q=0.5$, for five temperatures: $T/T_K = 4.92,1.34,0.895,0.671,0.447$. Fig.~\ref{fig:Q=halfnormalQNMs} shows that each of $\godo$ and $\good$ has a sequence of poles descending down into the complex plane, \textit{i.e.}\ with decreasing imaginary part, spaced apart from one another by an amount $\omega/(2 \pi T) \approx 1$, and with $\textrm{Re}\left(\omega/2\pi T\right)\to Q$ as $\textrm{Im}\left(\omega/2\pi T\right)\to-\infty$.

As $T/T_K$ decreases, the most significant change in fig.~\ref{fig:Q=halfnormalQNMs} occurs in the position of the ``lowest'' poles, meaning the poles nearest the origin at $T/T_K=4.92$ (fig.~\ref{fig:Q=halfnormalQNMs} (a). As $T/T_K$ decreases, the lowest poles move towards the origin (fig.~\ref{fig:Q=halfnormalQNMs} (b)), reach the origin at the critical temperature $T/T_K=0.895$ (fig.~\ref{fig:Q=halfnormalQNMs} (c)), and then move into the upper half of the complex $\omega/(2 \pi T)$ plane (fig.~\ref{fig:Q=halfnormalQNMs} (d) and (e)), signaling an instability. For any other non-zero $Q$, the plots of the pole positions are qualitatively similar to those in fig.~\ref{fig:Q=halfnormalQNMs}. In particular, as $T/T_K$ decreases the lowest poles always pass through the origin and into the upper half of the complex plane, signaling an instability.

\begin{figure}[h!]
\begin{center}
\includegraphics[width=0.9\textwidth]{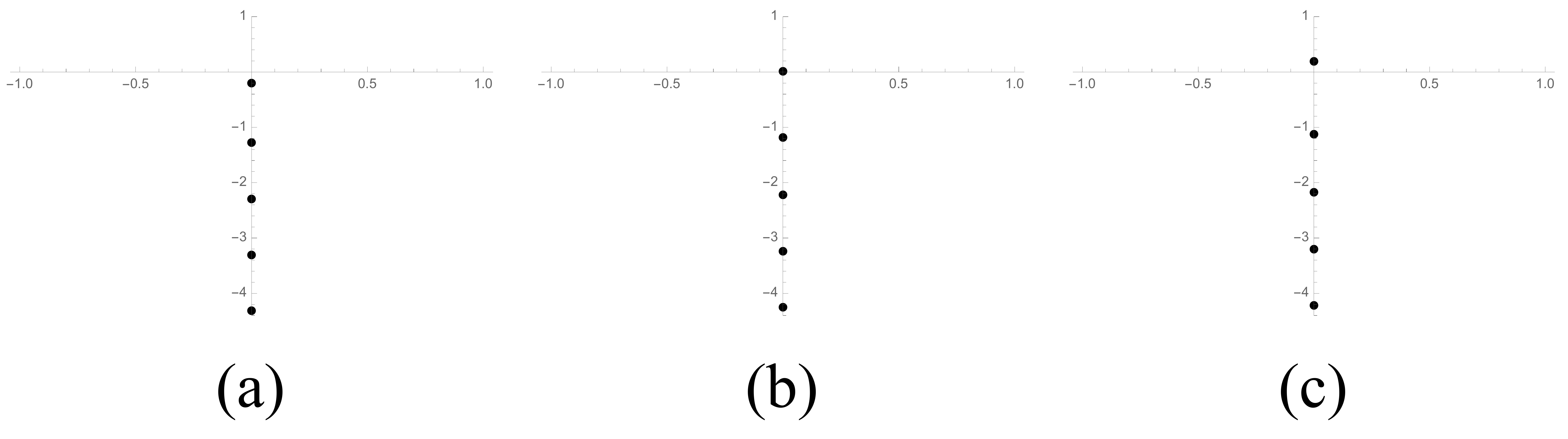}
\caption{\label{fig:Q=0normalQNMs}The positions of poles in $\godo$ (black dots) and $\good$ (gray diamonds) in the complex $\omega/(2\pi T)$ plane, determined by solving eq.~\eqref{eq:normpoles} numerically for $Q=0$ and $T/T_K$ equal to (a) $44$, (b) $8$, and (c) $4$. Compared to the $Q>0$ case in fig.~\ref{fig:Q=halfnormalQNMs}, the poles now all lie on the imaginary axis (the black dots and gray diamonds overlap), but otherwise exhibit similar behavior: as $T/T_K$ decreases, the lowest poles from (a) move up, reach the origin at the critical temperature $T/T_K=8$ in (b) and then cross into the upper half of the complex $\omega/(2 \pi T)$ plane in (c), signaling an instability.}
\end{center}
\end{figure}

However, $Q=0$ is slightly different. When $Q=0$, so that $H\left(-1/2 - i Q\right) = H\left(-1/2\right) = -1.368\ldots$ is real-valued, the only imaginary term in eq.~\eqref{eq:normpoles} is in the argument of the harmonic number, which is $\propto \,\textrm{Re}\left(\frac{\omega}{2\pi T}\right)$. As a result, solutions of eq.~\eqref{eq:normpoles} must have $\textrm{Re}\left(\frac{\omega}{2\pi T}\right)=0$. Clearly, when $Q=0$ the particle-hole symmetry $\textrm{Re}\left(\omega\right)\to-\textrm{Re}\left(\omega\right)$ is restored. Fig.~\ref{fig:Q=0normalQNMs} shows our numerical results for the positions of poles of $\godo$ (black dots) and $\good$ (gray diamonds) in the complex $\omega/(2 \pi T)$ plane for $Q=0$, for the temperatures $T/T_K = 44$, $8$, and $4$. All the poles are now on the imaginary axis, but otherwise we observe similar behavior to the $|Q|>0$ cases: as $T/T_K$ decreases, the lowest poles in fig.~\ref{fig:Q=0normalQNMs} (a) pass through the origin, now at a critical temperature $T/T_K=8$ in fig.~\ref{fig:Q=0normalQNMs} (b), and then cross into the upper half of the complex plane in fig.~\ref{fig:Q=0normalQNMs} (c).

\begin{figure}[h!]
\begin{center}
\includegraphics[width=0.6\textwidth]{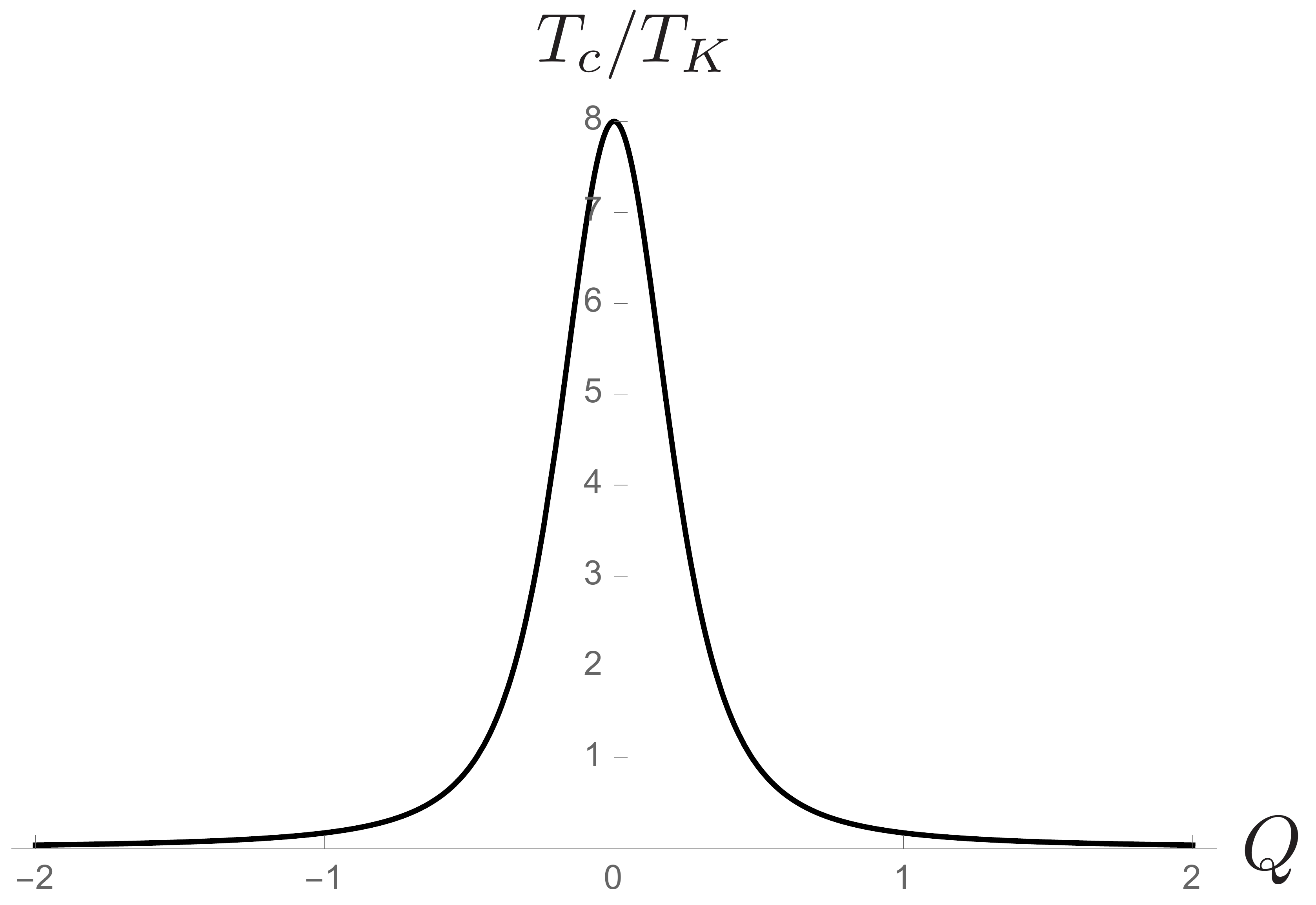}
\caption{\label{fig:criticalT}The critical temperature $T_c$ in units of $T_K$, as a function of $Q$, from eq.~\eqref{eq:criticalT}.}
\end{center}
\end{figure}

Since the instability always appears as poles passing through the origin and into the upper half of the complex plane, we can determine the critical temperature $T_c$ easily, as the temperature where the poles reach the origin: in eq.~\eqref{eq:normpoles} we set $\omega=0$ and then solve for $T/T_K=T_c/T_K$, with the result
\beq
\label{eq:criticalT}
\ln \left(\frac{T_c}{T_K}\right) = - H\left(-\frac{1}{2} + i Q\right)- H\left(-\frac{1}{2} - i Q\right) - \ln 2 = - 2 \,\textrm{Re}\left[H\left(-\frac{1}{2} + i Q\right)\right] - \ln 2.
\eeq
Fig.~\ref{fig:criticalT} shows $T_c/T_K$ as a function of $Q$, which has a maximum $T_c/T_K=8$ at $Q=0$, decreases monotonically as $|Q|$ increases, and goes to zero as $|Q|\to\infty$.

As mentioned in section~\ref{sec:intro}, our results for the movement of $\omega^*$ differ dramatically from those of the standard (non-holographic) Kondo model, at large $N$ and at leading order in perturbation theory in $\lambda$~\cite{Coleman2015}. In that model, the poles are determined by a condition identical to eq.~\eqref{eq:normpoles}, but without the $\ln\left(2T/T_K\right)$ term. As a result, the lowest pole sits exactly at $\omega = 0$ for all $T$. The $\ln\left(2T/T_K\right)$ term is thus repsonsible for the non-trivial movement of $\omega^*$, relative to the standard Kondo model. Indeed, the $\ln\left(2T/T_K\right)$ term in eq.~\eqref{eq:normpoles} can be viewed as arising from the renormalization of $\lambda$, \textit{i.e.}\ as a strong coupling effect arising from working non-perturbatively in both $\lambda$ and the 't Hooft coupling.

We have been able to compute the position and residue of the poles analytically (without numerics) in two limits: $T\gg T_c$ and $T$ just above $T_c$ ($T \gtrsim T_c$). In each case, we find a residue $Z$ with non-zero imaginary part, indicating that $\rodo$ will exhibit Fano resonances, as we will confirm in subsection~\ref{sec:normspec}.

In terms of $T/T_c$ (instead of $T/T_K$), $\godo$ takes a particularly simple form: using eq.~\eqref{eq:criticalT} to re-write eq.~\eqref{eq:odo3}, we find
\beq
\label{eq:odo4}
\godo = -\frac{N}{\kappa} - \frac{N}{\kappa^2}\frac{1}{H\left(-\frac{1}{2} + i Q - i \frac{\omega}{2\pi T}\right) - H\left(-\frac{1}{2} + i Q\right) + \ln\left(\frac{T}{T_c}\right)}.
\eeq

If $T \gg T_c$, or equivalently $\ln\left(T/T_c\right)\gg 1$, then $H\left(-\frac{1}{2} + i Q - i \frac{\omega}{2\pi T}\right)$ must also be large for $\godo$ to have a pole. The Harmonic numbers $H(n)$ grow large either when $n \to \infty$ with $|\textrm{Arg}\left(n\right)|<\pi$, where $H(n) \to \ln(n)$, or when $n$ approaches a negative integer, as apparent from the series representation
\beq
\label{eq:seriesrep}
H(n) = \sum_{k=1}^{\infty} \left ( \frac{1}{k} - \frac{1}{n+k}\right).
\eeq
We are interested in poles near the origin of the complex $\omega$-plane, rather than poles at large $|\omega|$, since the former have a larger effect on the spectral function, so we will only consider the poles where $H\left(-\frac{1}{2} + i Q - i \frac{\omega}{2\pi T}\right)$ has argument equal to a negative integer. Explicitly, in the $\godo$ in eq.~\eqref{eq:odo4}, near such a pole we use eq.~\eqref{eq:seriesrep} to take
\beq
H\left(-\frac{1}{2} + i Q - i \frac{\omega}{2\pi T}\right) - H\left(-\frac{1}{2} + i Q \right) \approx \frac{-1}{-\frac{1}{2} + i Q - i\frac{\omega}{2\pi T} + k},
\eeq
with $k=1,2,3,\ldots$. In that approximation, and with $\ln\left(T/T_c\right)\gg 1$, the $\godo$ in eq.~\eqref{eq:odo4} becomes
\beq
\label{eq:highTgodo}
\godo \approx -\frac{N}{\kappa} - \frac{N}{\kappa^2} \frac{1}{\frac{-1}{-\frac{1}{2} + i Q - i\frac{\omega}{2\pi T} + k} + \ln\left(T/T_c\right)}.
\eeq
The pole's position $\omega^* = \omega_R^*+i \omega_I^*$ and residue $Z=Z_R + i Z_I$ are then given by
\beq
\label{eq:highTpole}
\frac{\omega^*}{2 \pi T} = Q + i \left(-k + \frac{1}{2} + \frac{1}{\ln\left(T/T_c\right)}\right), \qquad Z = -\frac{N}{\kappa^2}\frac{i \left(2\pi T\right)}{\left(\ln\left(\frac{T}{T_c}\right)\right)^2},
\eeq
where the lowest pole has $k=1$, and the higher poles have $k=2,3,\ldots$. The residue $Z$ in eq.~\eqref{eq:highTpole} is purely imaginary, $Z_R=0$, so (recalling the table in section~\ref{sec:fano}) we expect $\rodo$ will have a $q=1$ symmetric Fano resonance.

\begin{figure}[h!]
\begin{center}
\includegraphics[width=0.9\textwidth]{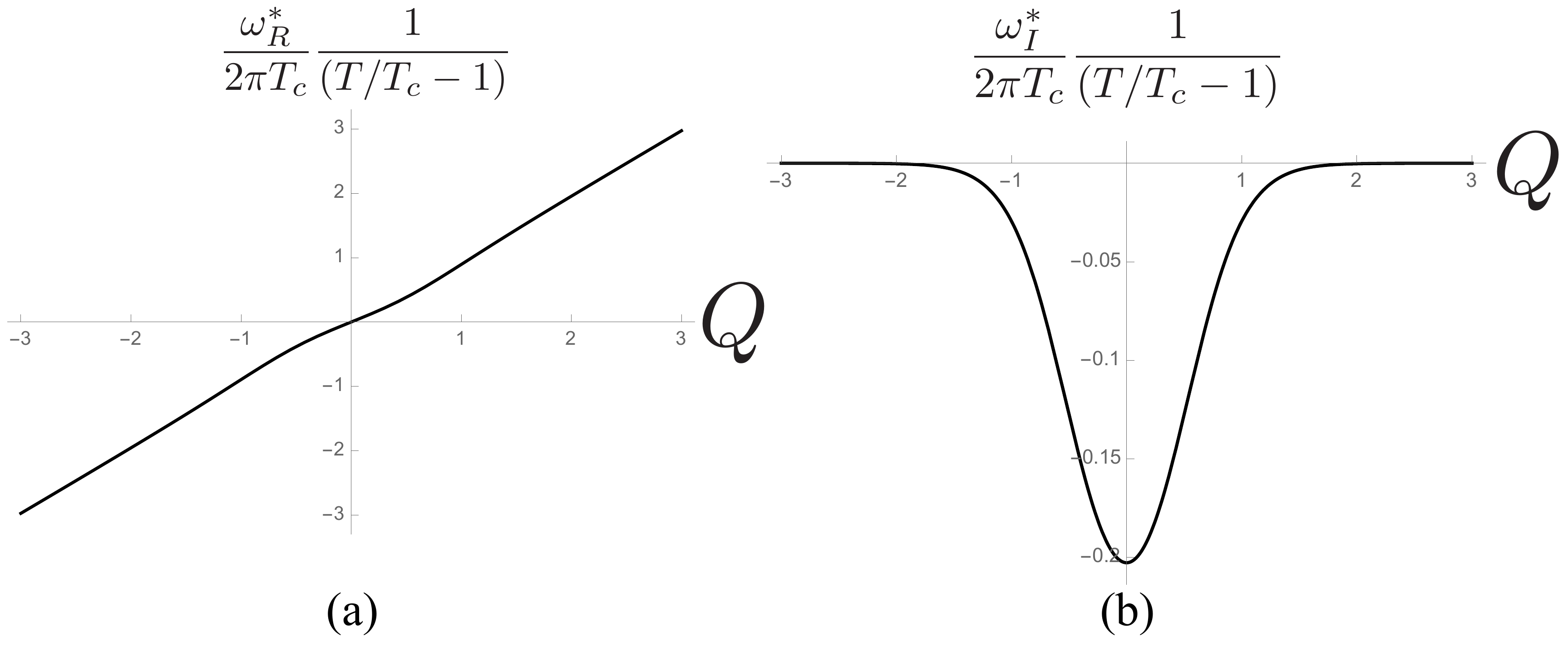}
\caption{\label{fig:normpoleslope}The slope of $(T/T_c-1)$ of the lowest pole in $\godo$ for $T$ just above $T_c$, as functions of $Q$,  from eq.~\eqref{eq:normscaling}. (a) The slope of the real part of the pole, $\omega_R^*$, in units of $2 \pi T_c$. (b) The slope of the imaginary part of the pole, $\omega_I^*$, in units of $2\pi T_c$.}
\end{center}
\end{figure}

Eq.~\eqref{eq:odo4} makes obvious the pole at $\omega=0$ when $T=T_c$. For $T$ just above $T_c$, $T\gtrsim T_c$, we can obtain this pole's position and residue by expanding eq.~\eqref{eq:odo4} in $T$ around $T_c$ and simultaneously in $\omega$ around $\omega=0$. For the expansion in $\omega$ we use
\beq
H\left(-\frac{1}{2} + i Q - i \frac{\omega}{2\pi T}\right) - H\left(-\frac{1}{2} + i Q\right) = -\psi'\left(\frac{1}{2} + i Q\right) \frac{i\omega}{2\pi T} + \mathcal{O}\left(\left(\frac{\omega}{2\pi T}\right)^2\right),
\eeq
where $\psi'(n)=\partial_n \psi(n)$ denotes the first derivative of the digamma function $\psi(n)$. The pole's position $\omega^* = \omega_R^*+i \omega_I^*$ and residue $Z=Z_R + i Z_I$ are then given by
\beq
\label{eq:normscaling}
\frac{\omega^*}{2\pi T_c}=\frac{-i}{\psi'\left(\frac{1}{2} + i Q\right)}\left(T/T_c-1\right), \qquad Z = \frac{-i}{\psi'\left(\frac{1}{2} + i Q\right)} \left(2\pi T_c\right) \frac{N}{\kappa^2},
\eeq
as derived in ref.~\cite{Erdmenger:2016vud}. As $T\to T_c^+$, both $\omega_R^*$ and $\omega_I^*$ vanish linearly, \textit{i.e.} as $T/T_c-1$, with slopes determined by $Q$ alone. Fig.~\ref{fig:normpoleslope} shows these slopes as functions of $Q$. In particular, fig.~\ref{fig:normpoleslope} (b) shows that the magnitude of $\omega_I$'s slope is largest when $Q=0$ and decreases monotonically as $|Q|$ increases.

\begin{figure}[h!]
\begin{center}
\includegraphics[width=0.6\textwidth]{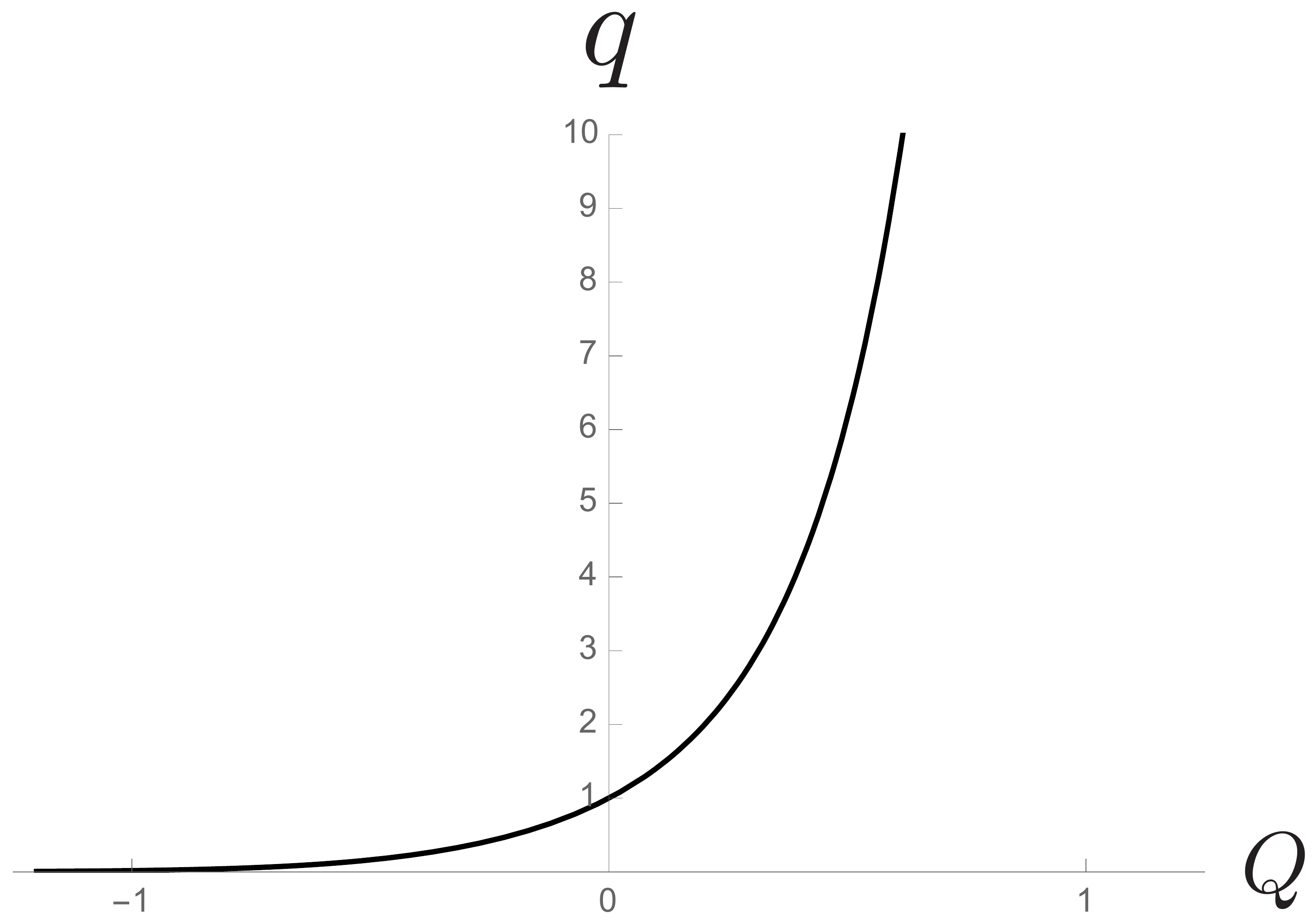}
\caption{\label{fig:qvsQnearTc}The Fano/asymmetry parameter $q$ as a function of $Q$ for $T \gtrsim T_c$, obtained by plugging the residue $Z$ in eq.~\eqref{eq:normscaling} into eq.~\eqref{eq:qZ} for $q$. The limits $Q \to - \infty$, $0$, $+\infty$ produce symmetric Fano resonances with $q\to 0$, $1$, $+\infty$, respectively.}
\end{center}
\end{figure}

The residue $Z$ in eq.~\eqref{eq:normscaling} is in general complex-valued, so when $T \gtrsim T_c$, the lowest pole in $\godo$ will produce a Fano resonance in $\rodo$. Plugging the $Z$ in eq.~\eqref{eq:normscaling} into eq.~\eqref{eq:qZ} gives us the Fano/asymmetry parameter $q$ as a function of $Q$, shown in fig.~\ref{fig:qvsQnearTc}. Symmetric Fano resonances will occur when $Q\to-\infty$, $0$, $+\infty$, where $q\to 0$, $1$, and $\infty$, respectively, corresponding to a Fano anti-resonance, symmetric Fano resonance, and Lorentzian resonance (figs.~\ref{fig:fanopics} (b), (c), and (d)), respectively.

\subsection{Unscreened Phase: Spectral Function}
\label{sec:normspec}

The spectral function $\rodo$ in the unscreened phase is trivial to compute from $\godo$ in eq.~\eqref{eq:odo4}:
\beq
\rodo = -2 \, \textrm{Im} \, \godo = 2 \, \frac{N}{\kappa^2} \, \textrm{Im} \, \left[ \frac{1}{H\left(-\frac{1}{2} + i Q - i \frac{\omega}{2\pi T}\right) - H\left(-\frac{1}{2} + i Q\right) + \ln\left(\frac{T}{T_c}\right)} \right],
\eeq
where we now restrict to real-valued $\omega$. In our case, $\rodo$ vanishes when $\omega \to 0$ or $|\omega| \to \infty$, in the latter case vanishing as $\left(\ln|\omega|\right)^{-2}$, ultimately because the Harmonic numbers are asymptotically logarithmic, as mentioned above. Such $\left(\ln|\omega|\right)^{-2}$ behavior means our $\rodo$ cannot be exactly $\rhofano$ in eq.~\eqref{eq:fanospec}, since $\rhofano$ involves only powers of $\omega$, with no logarithms. Nevertheless, we have shown in subsection~\ref{sec:normpoles} that the lowest pole in $\godo$ generically has residue with non-zero imaginary part, so we expect Fano resonances in $\rodo$ at $\omega$ near the real part of the lowest pole's position, $\omega^*_R$.

\begin{figure}[h!]
\begin{center}
\includegraphics[width=0.9\textwidth]{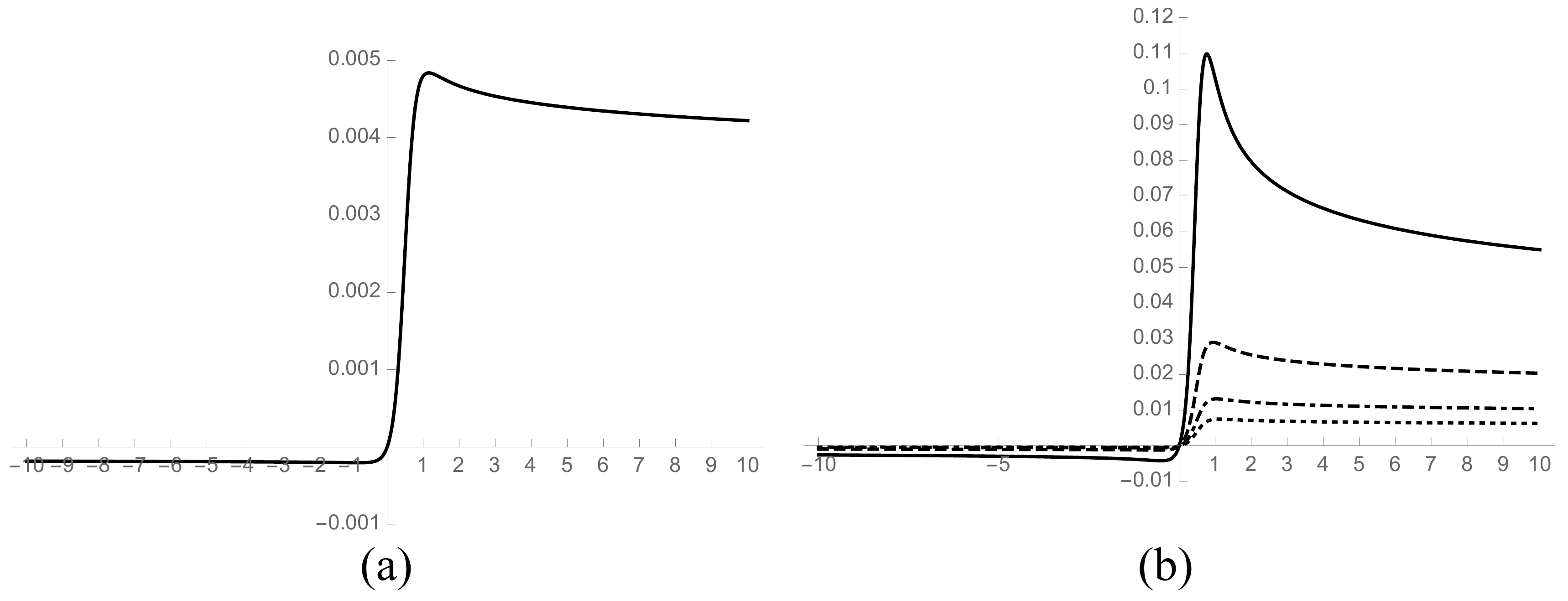}
\caption{\label{fig:symmhighTrodo}The spectral function, $\rodo/\left(N/\kappa^2\right)$, as a function of $\omega/(2 \pi T)$ for the representative value $Q=0.5$ and in the $T \gg T_c$ regime, namely for (a) $T/T_c = 10^{15}$ and (b) $T/T_c=10^{12}$ (dotted), $10^9$ (dot-dashed), $10^6$ (dashed), and $10^3$ (solid).}
\end{center}
\end{figure}

Fig.~\ref{fig:symmhighTrodo} shows $\rodo/\left(N/\kappa^2\right)$ as a function of $\omega/(2\pi T)$ for the representative value $Q=0.5$ and in the $T \gg T_c$ regime, namely from $T/T_c=10^{15}$ (fig.~\ref{fig:symmhighTrodo} (a)) down to $T/T_c=10^3$ (fig.~\ref{fig:symmhighTrodo} (b)). From the $T \gg T_c$ results in eqs.~\eqref{eq:highTgodo} and~\eqref{eq:highTpole}, we expect $\rodo$ to have a $q=1$ symmetric Fano resonance when $\omega$ equals the real part of the lowest pole's position, $\omega_R^*$, which is $\omega^*_R = Q$ when $T \gg T_c$. Sure enough, for sufficiently high $T/T_c$, as in fig.~\ref{fig:symmhighTrodo} (a), $\rodo$ has an approximately $q=1$ symmetric Fano resonance at $\omega \approx \omega_R^* \approx Q$. As $T/T_c$ decreases through twelve orders of magnitude, the asymmetry of the resonance appears to increase, although the position changes by only $\approx5\%$: $\omega^*_R \approx 0.499$ when $T/T_c = 10^{15}$, while $\omega_R^* \approx 0.475$ when $T/T_c = 10^3$. We have confirmed numerically that as $T/T_c$ decreases through the values in fig.~\ref{fig:symmhighTrodo}, the peak value of the resonance grows as $1/\left(\ln(T/T_c)\right)^2$, consistent with the $T\gg T_c$ results for $\omega^*_I$ and $Z_I$ in eq.~\eqref{eq:highTpole}. Crucially, the resonance in fig.~\ref{fig:symmhighTrodo} is not at the particle-hole symmetric value $\omega=0$, and so is not the Kondo resonance---as expected, since the Kondo resonance is generically absent at large-$N$ in the unscreened phase.

\begin{figure}[h!]
\begin{center}
\includegraphics[width=0.9\textwidth]{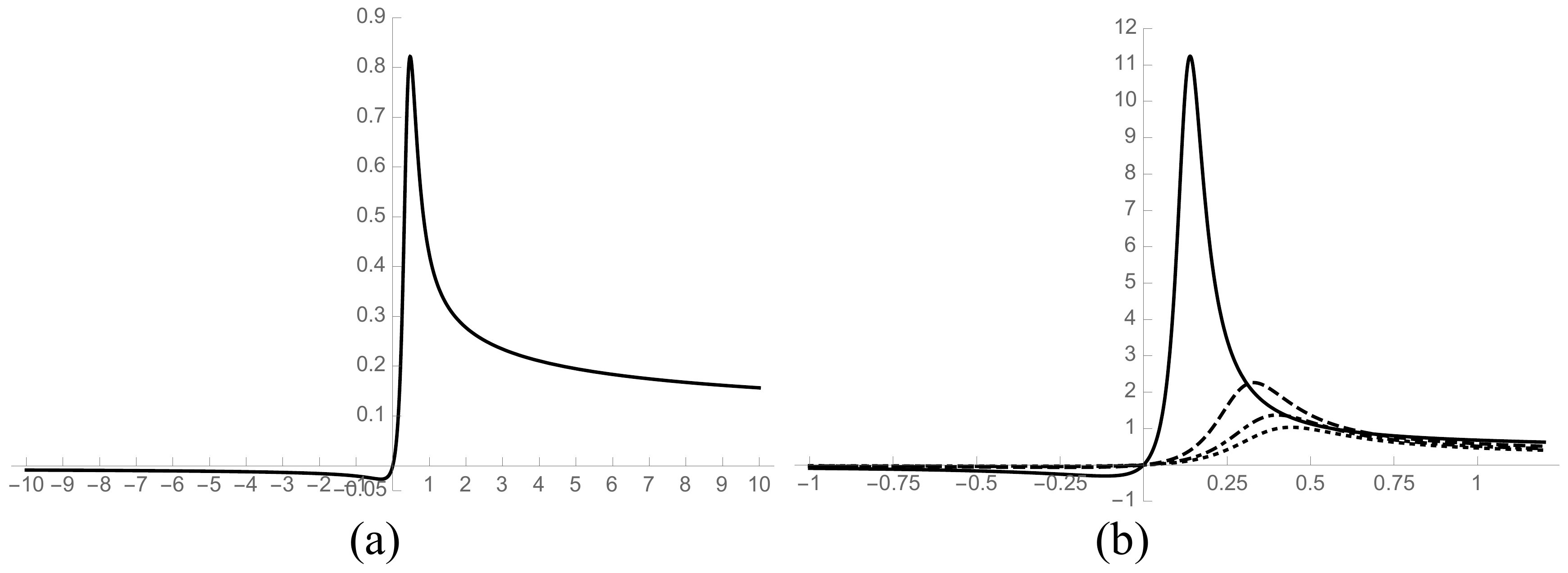}
\caption{\label{fig:symmmedTrodo}The spectral function, $\rodo/\left(N/\kappa^2\right)$, as a function of $\omega/(2 \pi T)$ for $Q=0.5$ and (a) $T/T_c = 10$, (b) $7.5$ (dotted), $5.5$ (dot-dashed), $3.5$ (dashed), and $1.5$ (solid).}
\end{center}
\end{figure}

Fig.~\ref{fig:symmmedTrodo} shows $\rodo/\left(N/\kappa^2\right)$ as a function of $\omega/(2\pi T)$ for $Q=0.5$ from $T/T_c=10$ (fig.~\ref{fig:symmmedTrodo} (a)) down to $T/T_c=1.5$ (fig.~\ref{fig:symmmedTrodo} (b)), including $T/T_c = 5.5$, corresponding to $T/T_K = 4.92$ (fig.~\ref{fig:Q=halfnormalQNMs} (a)), and $T/T_c = 1.5$, corresponding to $T/T_K=1.34$ (fig.~\ref{fig:Q=halfnormalQNMs} (b)). In fig.~\ref{fig:symmmedTrodo}, as $T/T_c$ decreases, we see four changes in the resonance. First, the peak of the resonance moves towards $\omega=0$, following the position of the lowest pole in $\godo$. For example, compare the position of the peak in $\rodo$ at $T/T_c=5.5$ or $1.5$ in fig.~\ref{fig:symmmedTrodo} (b) (dot-dashed or solid curve, respectively) to the position of the lowest pole in $\godo$ in fig.~\ref{fig:Q=halfnormalQNMs} (a) or (b), respectively. Second, the resonance grows taller, by about an order of magnitude for the values of $T/T_c$ in fig.~\ref{fig:symmmedTrodo}. Third, the peak grows narrower, also by about an order of magnitude for the values of $T/T_c$ in fig.~\ref{fig:symmmedTrodo}. Fourth, the Fano/asymmetry parameter $q$ increases. For example, $q\approx 2$ for $T/T_c=10$ (fig.~\ref{fig:symmmedTrodo} (a)) and $q\approx 5$ for $T/T_c=1.5$ (fig.~\ref{fig:symmmedTrodo} (b)).

\begin{figure}[h!]
\begin{center}
\includegraphics[width=0.9\textwidth]{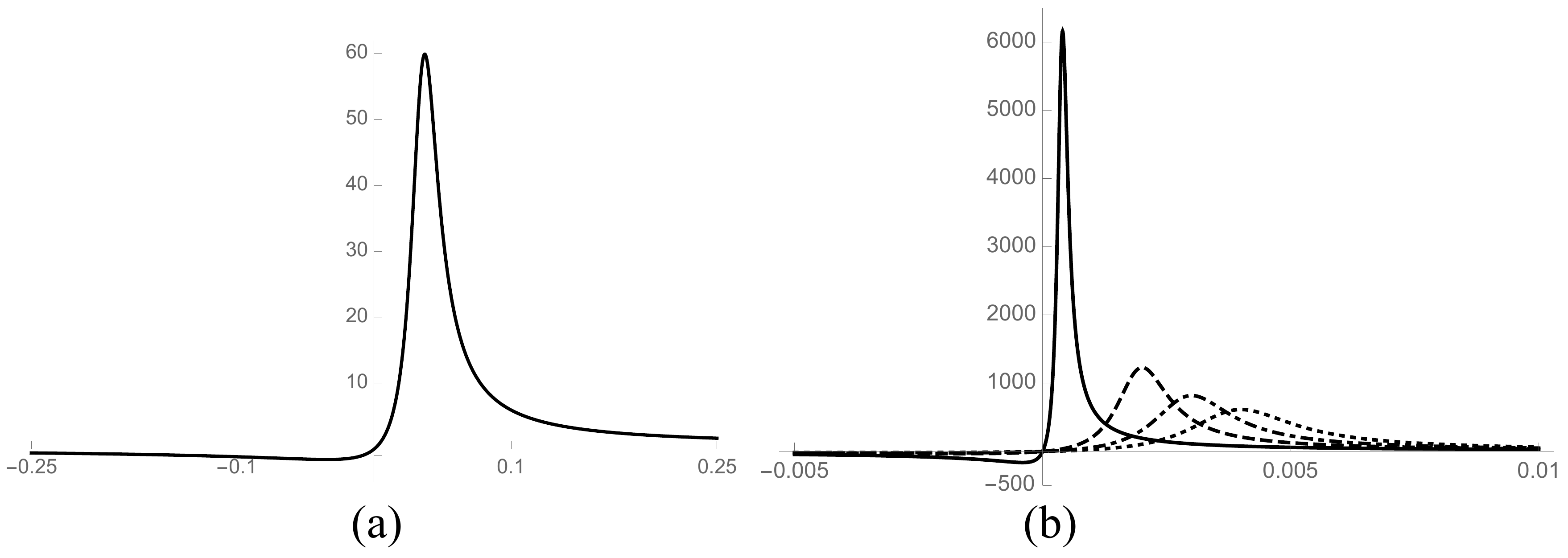}
\caption{\label{fig:symmlowTrodo}The spectral function, $\rodo/\left(N/\kappa^2\right)$, as a function of $\omega/(2 \pi T)$ for $Q=0.5$ and (a) $T/T_c = 1.1$, (b) $1.01$ (dotted), $1.0075$ (dot-dashed), $1.005$ (dashed), and $1.001$ (solid).}
\end{center}
\end{figure}

Fig.~\ref{fig:symmlowTrodo} shows $\rodo/\left(N/\kappa^2\right)$ as a function of $\omega/(2\pi T)$ for $Q=0.5$ and in the $T \gtrsim T_c$ regime, namely for $T/T_c=1.1$ (fig.~\ref{fig:symmlowTrodo} (a)) down to $T/T_c=1.001$ (fig.~\ref{fig:symmlowTrodo} (b)). The four trends observed in fig.~\ref{fig:symmmedTrodo} appear again in fig.~\ref{fig:symmlowTrodo}. First, the resonance moves towards $\omega=0$, following the real part of the position of the lowest pole in $\godo$ in the $T \gtrsim T_c$ regime, given by $\omega^*$ in eq.~\eqref{eq:normscaling}, which in particular has $\omega_R^* \propto (T/T_c-1)$. Second, the resonance grows taller. Indeed, plugging the $T \gtrsim T_c$ results for $\omega^*$ and $Z$ of eq.~\eqref{eq:normscaling} into eq.~\eqref{eq:fanofromG} reveals that the peak of the resonance increases as $(T/T_c -1)^{-1}$. Such power-law growth, rather than logarithmic growth, again indicates that the resonance is not a Kondo resonance. Third, the resonance grows narrower, with a width proportional to the imaginary part of the lowest pole in $\godo$ in the $T \gtrsim T_c$ regime, which from eq.~\eqref{eq:normscaling} has $\omega_I^* \propto (T/T_c-1)$. Fourth, the Fano/asymmetry parameter $q$ increases. For example, $q\approx 5.8$ for $T/T_c=1.1$ (fig.~\ref{fig:symmmedTrodo} (a)) and $q\approx 6.2$ for $T/T_c=1.001$ (fig.~\ref{fig:symmmedTrodo} (b)).

\begin{figure}[h!]
\begin{center}
\includegraphics[width=0.9\textwidth]{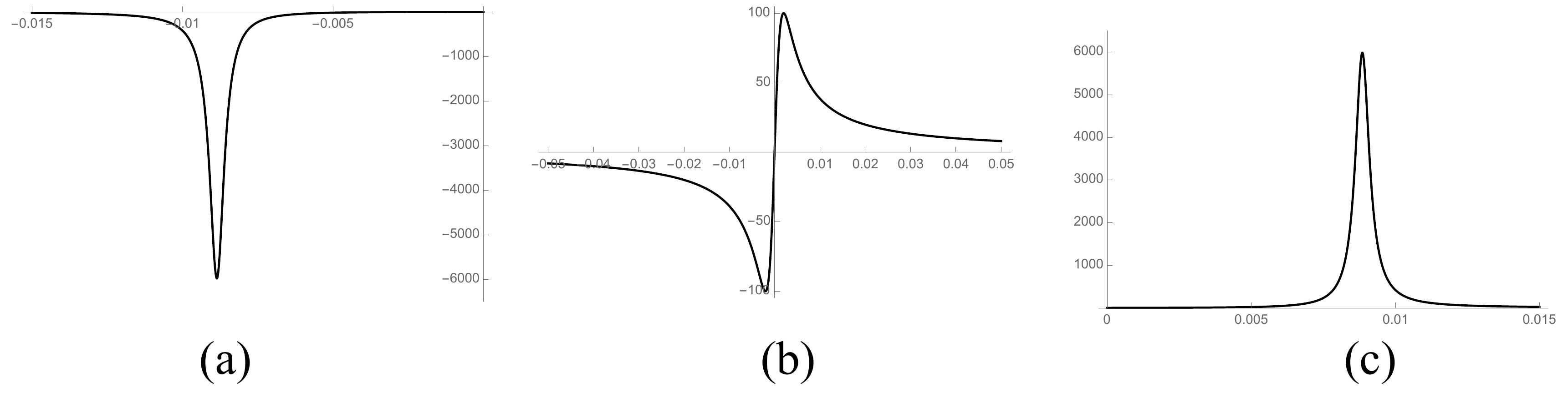}
\caption{\label{fig:symmsymmfano}The spectral function, $\rodo/\left(N/\kappa^2\right)$, as a function of $\omega/(2 \pi T)$ for $T/T_c=1.01$ and (a) $Q=-1$, (b) $Q=0$, and (c) $Q=+1$.}
\end{center}
\end{figure}

In the $T \gtrsim T_c$ regime, we expect symmetric Fano resonances when $Q \to -\infty$, $0$, $\infty$, as discussed below eq.~\eqref{eq:normscaling} and in fig.~\ref{fig:qvsQnearTc}. We indeed find such behavior, already at relatively small values of $|Q|$. Fig.~\ref{fig:symmsymmfano} shows $\rodo/\left(N/\kappa^2\right)$ as a function of $\omega/(2\pi T)$ for $T/T_c = 1.01$ and (a) $Q=-1$, (b) $Q=0$, and (c) $Q=+1$. We clearly see symmetric Fano (anti-)resonances with (a) $q \approx 0.0164$, (b) $q=1$, and (c) $q\approx 60.9$, respectively, all consistent with eq.~\eqref{eq:normscaling} and fig.~\ref{fig:qvsQnearTc}.

\begin{figure}[h!]
\begin{center}
\includegraphics[width=0.9\textwidth]{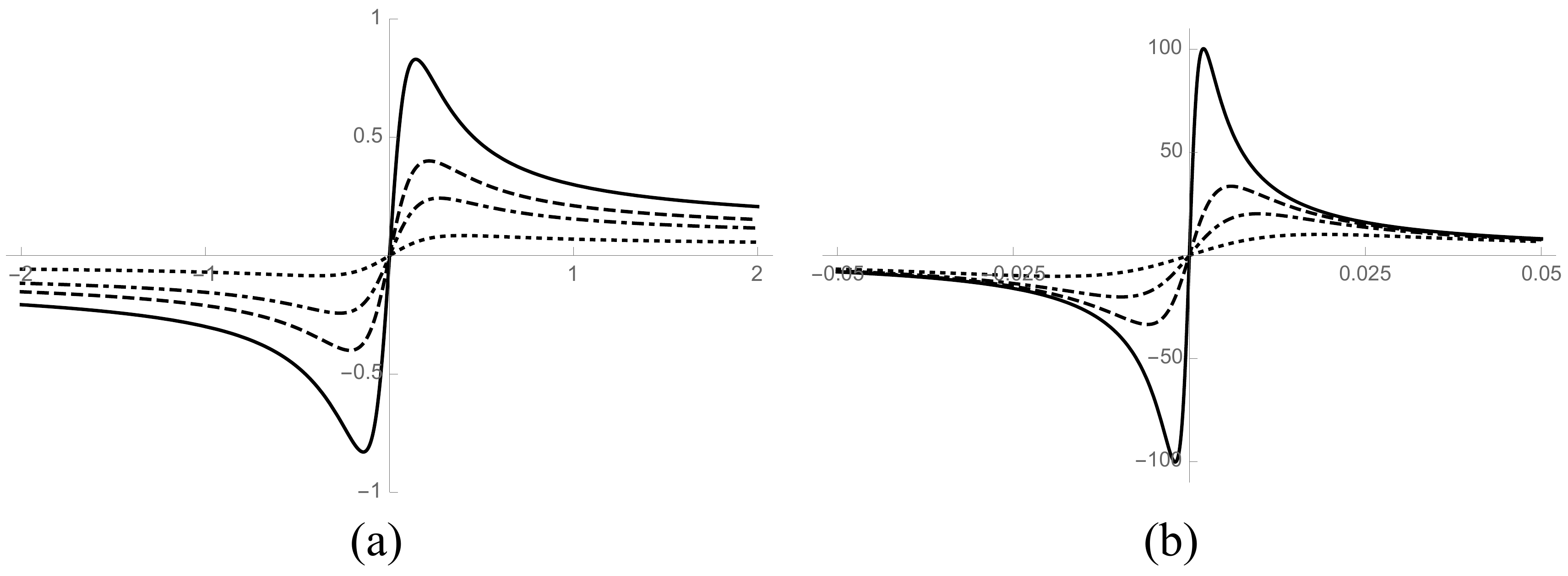}
\caption{\label{fig:symmQzero}The spectral function, $\rodo/\left(N/\kappa^2\right)$, as a function of $\omega/(2 \pi T)$ for $Q=0$ and (a) $T/T_c=100$ (dotted), $10$ (dot-dashed), $5$ (dashed), and $2.5$ (solid), and (b) $T/T_c=1.1$ (dotted), $1.05$ (dot-dashed), $1.03$ (dashed), and $1.01$ (solid).}
\end{center}
\end{figure}

For the special value $Q=0$ nothing breaks the particle-hole symmetry $\textrm{Re}\,\omega \to -\textrm{Re}\,\omega$, and all poles of $\godo$ have vanishing real part, as shown for example in fig.~\ref{fig:Q=0normalQNMs}. When $Q=0$ we thus expect a $q=1$ symmetric Fano resonance at $\omega=0$ for all $T/T_c$. Fig.~\ref{fig:symmQzero} shows $\rodo/\left(N/\kappa^2\right)$ as a function of $\omega/(2\pi T)$ for $Q=0$ and $T/T_c$ from $T/T_c = 100$ down to $2.5$ (fig.~\ref{fig:symmQzero} (a)) and from $T/T_c = 1.1$ down to $1.01$ (fig.~\ref{fig:symmQzero} (b)). We indeed find $q=1$ symmetric Fano resonances at $\omega=0$ for all $T/T_c$.

\begin{figure}[h!]
\begin{center}
\includegraphics[width=0.9\textwidth]{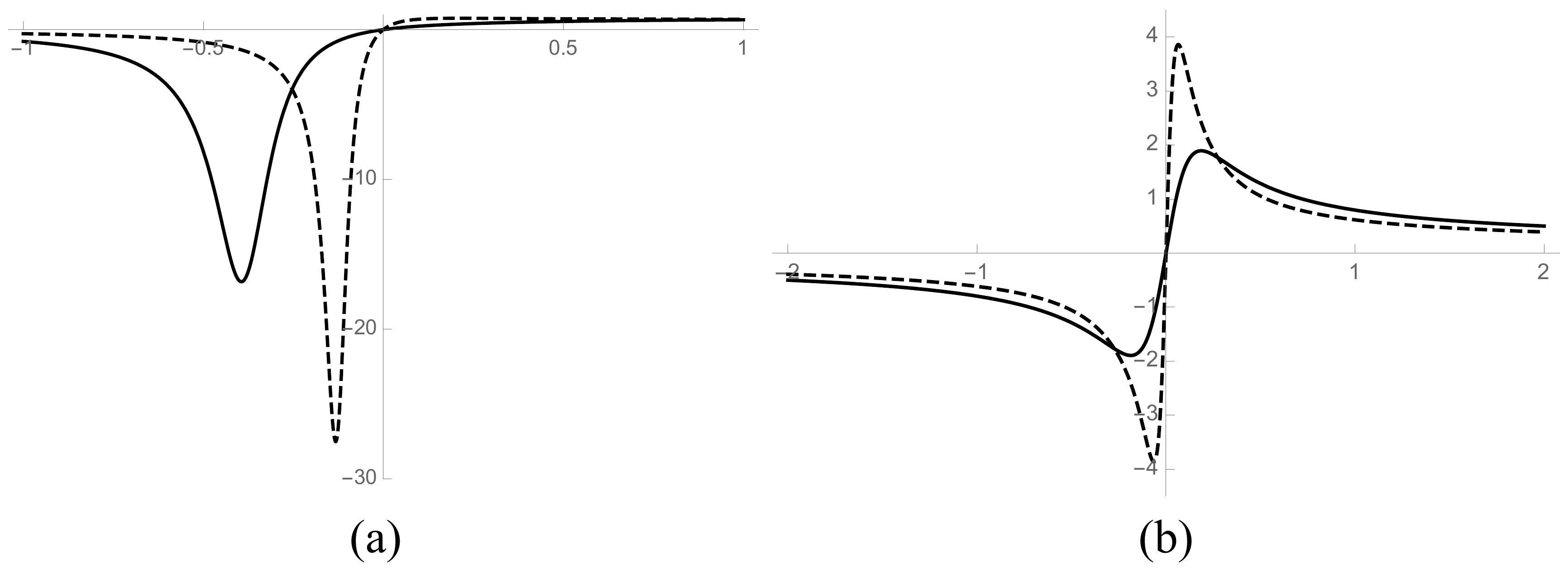}
\caption{\label{fig:symmbelowTc}The spectral function, $\rodo/\left(N/\kappa^2\right)$, as a function of $\omega/(2 \pi T)$ for $T/T_c=0.75$ (dashed) and $0.5$ (solid), for (a) $Q=0.5$ and (b) $Q=0$.}
\end{center}
\end{figure}

We can also consider $\rodo$ in the unscreened phase when $T<T_c$, bearing in mind that the unscreened phase is unstable when $T<T_c$ because $\godo$ has a pole with $\textrm{Im}\, \omega^* >0$, as discussed above. Fig.~\ref{fig:symmbelowTc} (a) shows $\rodo/\left(N/\kappa^2\right)$, as a function of $\omega/(2 \pi T)$ for $Q=0.5$ and for $T<T_c$, namely for $T/T_c = 0.75$ and $0.5$, corresponding to $T/T_K=0.671$ and $0.447$, respectively (figs.~\ref{fig:Q=halfnormalQNMs} (d) and (e), respectively). We find a mirror version of the four trends observed for $T >T_c$ in figs.~\ref{fig:symmmedTrodo} and~\ref{fig:symmlowTrodo}. First, the resonance moves away from $\omega=0$, with peak position at $\omega \approx \omega^*_R$. Second, the resonance grows shorter. Third, the resonance grows wider. Fourth, the value of $q$ decreases. In particular, $q \approx 0.11$ for $T/T_c = 0.75$ and $q \approx 0.06$ for $T/T_c = 0.5$. Fig.~\ref{fig:symmbelowTc} (b) shows $\rodo/\left(N/\kappa^2\right)$, as a function of $\omega/(2 \pi T)$ for $Q=0$ and $T/T_c = 0.75$ and $0.5$. In that case, as expected we find a $q=1$ symmetric Fano resonance at $\omega=0$ whose height decreases as $T$ decreases. All of these behaviors are consistent with the motion of the lowest pole in $\godo$ in the complex $\omega$ plane described in subsection~\ref{sec:normpoles}.

In summary, we have learned two key lessons from the poles in $\godo$ and corresponding resonances in $\rodo$ in the unscreened phase. First, we \textit{do not} see a Kondo resonance, consistent with the expectations of large-$N$ Kondo models, where the Kondo effect (screening, phase shift, etc.) occurs only in the screened phase. Second, the resonances we find are all Fano resonances, consistent with our interpretation that $(0+1)$-dimensional scale invariance implies a continuum, and our Kondo coupling then breaks scale invariance and produces a resonance that is necessarily immersed the continuum.

\section{Screened Phase}
\label{sec:lowT}
\setcounter{equation}{0}

In this section we use the results of sections~\ref{sec:review} and~\ref{sec:holorg} to determine the excitation spectrum of our system in the screened phase $(T<T_c)$ by locating the poles in $\godo$ in the plane of complex $\omega$ (subsection~\ref{sec:condensedpoles}), and the corresponding peaks in $\rodo$ for real $\omega$ (subsection~\ref{sec:condensedspec}).

The main results of this section appeared in ref.~\cite{Erdmenger:2016vud}, namely that for $T$ just below $T_c$ ($T \lesssim T_c$), a pole of the form $\omega^* \propto - i \langle \mathcal{O}\rangle^2$ appears in $\godo$, giving rise to a $q=1$ symmetric Fano resonance in $\rodo$, which is a signature of a Kondo resonance at large $N$. In this section we will present some additional details about these results. Moreover, in appendix~\ref{appendix:condensedlowestqnm} we show, without using numerics, that $\omega^* \propto - i \langle \mathcal{O}\rangle^2$, but only for $Q=-1/2$, although our methods should easily generalize to any $Q$.

As derived in eq.~\eqref{eq:screened2ptresult}, in the screened phase
\beq
\label{eq:odo12}
\godo= N \frac{\hat{\mathcal{R}}_{22}/4}{1-\kappa\hat{\mathcal{R}}_{22}/2},
\eeq
and $\godo = \good = \goo = \godod$, so we will henceforth discuss only $\godo$. In the unscreened phase we had the analytic (\textit{i.e.}~non-numerical) result for $\hat{\mathcal{R}}_{\Phi^{\dagger}\Phi}$ in eq.~\eqref{eq:odo2}, however, in this section our solutions for $\hat{\mathcal{R}}_{22}$ will be numerical.

\subsection{Screened Phase: Poles in the Green's Function}
\label{sec:condensedpoles}

\begin{figure}[h!]
\begin{center}
\includegraphics[width=0.8\textwidth]{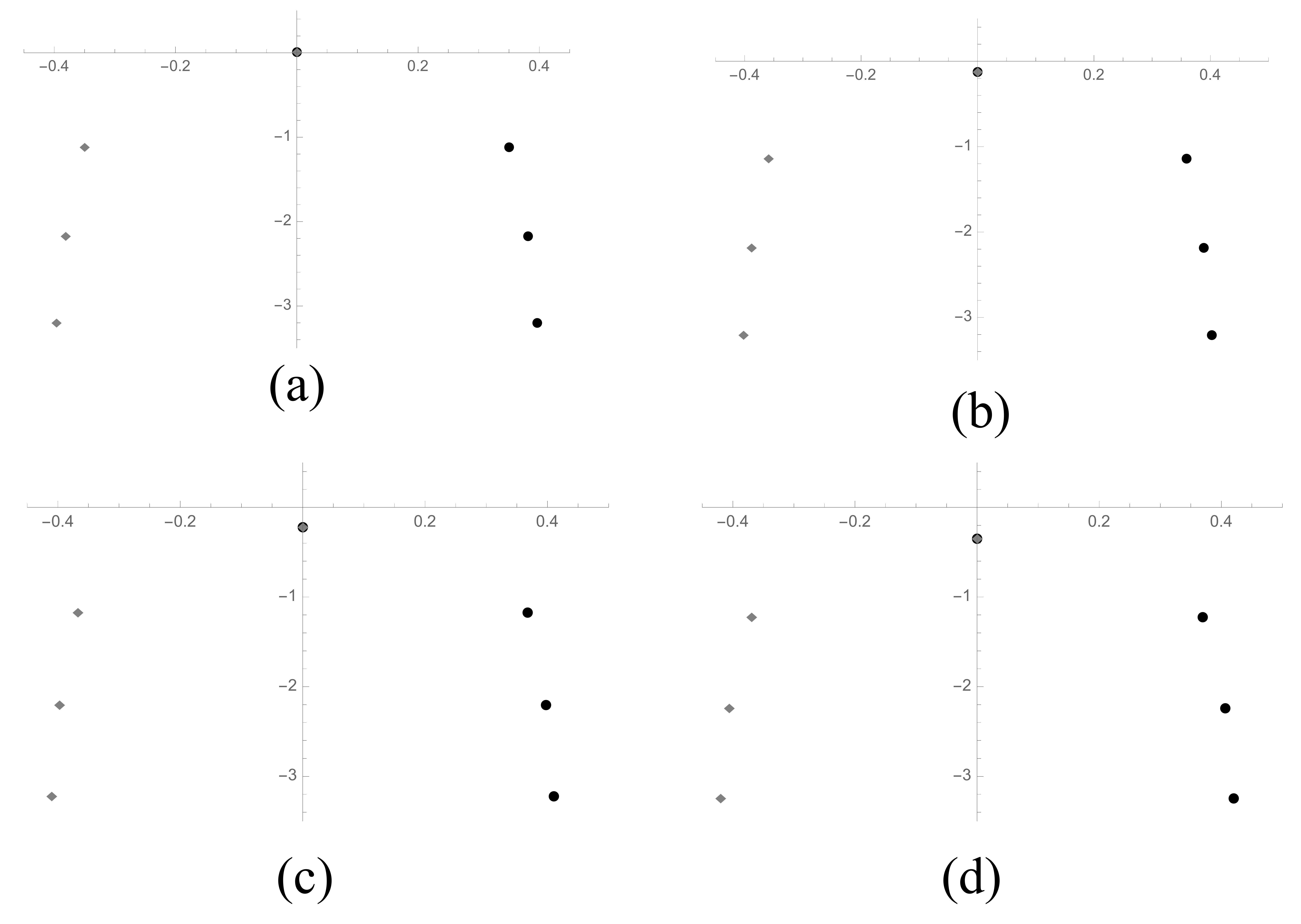}
\caption{\label{fig:Q=halfcondensedQNMs}Our numerical results for the positions of poles in $\godo$ in the complex $\omega/(2\pi T)$ plane, for $Q=0.5$ and $T/T_c$ equal to (a) $1$, (b) $0.588$, (c) $0.389$, and (d) $0.200$. As $T/T_c$ decreases, the most significant change occurs in the position of the lowest pole, which moves straight down the imaginary axis.}
\end{center}
\end{figure}

Clearly $\godo$ in eq.~\eqref{eq:odo12} has a pole whenever $1-\kappa\hat{\mathcal{R}}_{22}/2=0$. (Via eq.~\eqref{eq:screened2ptresult}, several other two-point functions have the same poles as well, namely $\good$, $\goo$, $\godod$, $ \<\co(\o)\ca_t(-\o)\>_\k$ and $\<\co^\dag(\o)\ca_t(-\o)\>_\k$.) Given values of $Q$ and $T/T_K$, we can thus find the poles in $\godo$ by solving the equation $1-\kappa\hat{\mathcal{R}}_{22}/2=0$ for $\omega/(2 \pi T)$, which we have done numerically. Our numerical results for the positions of the poles appear in fig.~\ref{fig:Q=halfcondensedQNMs}, for $Q=0.5$ and with $T/T_c=1$ in fig.~\ref{fig:Q=halfcondensedQNMs} (a), $0.588$ (b), $0.389$ (c), and $0.200$ (d). When $T/T_c=1$, the poles' positions agree with those we found in the unscreened phase in subsection~\ref{sec:normpoles}, including in particular the lowest pole, $\omega^*$, sitting at the origin of the complex $\omega/(2 \pi T)$ plane. As $T/T_c$ decreases the most significant change occurs in $\omega^*$, which moves straight down the imaginary axis. For any other non-zero $Q$, the plots of the pole positions are qualitatively similar to those in fig.~\ref{fig:Q=halfcondensedQNMs}, except for $Q=0$, where all the higher poles are on the imaginary axis. In particular, for all $Q$, including $Q=0$, as $T$ decreases the most significant change occurs in $\omega^*$, which moves straight down the imaginary axis.

\begin{figure}[h!]
\begin{center}
\includegraphics[width=0.6\textwidth]{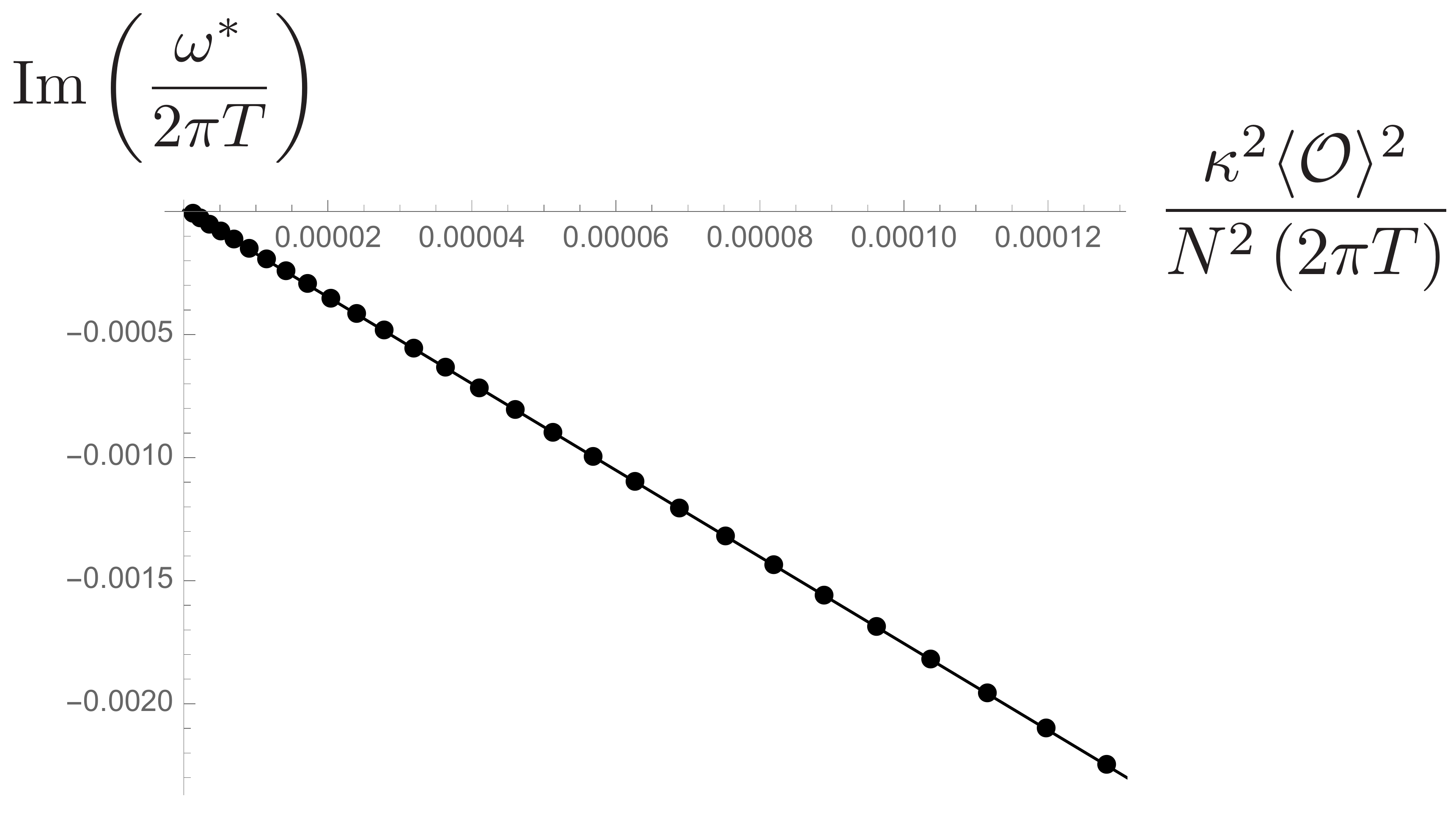}
\caption{\label{fig:condensedlowestQNM}In the screened phase, the lowest pole in $\godo$, $\omega^*$, is purely imaginary (see fig.~\ref{fig:Q=halfcondensedQNMs}). The black dots denote $\textrm{Im}\,\omega^*/(2 \pi T)$ as a function of  $\kappa^2 \langle \mathcal{O}\rangle^2/(N^2 (2 \pi T))$ for $Q=0.5$. The solid black line is a numerical linear fit with slope $\approx -17.6$ and intercept $\approx 5 \times 10^{-6}$. The agreement between the data and the fit shows that $\omega^* \propto - i \langle \mathcal{O}\rangle^2$.}
\end{center}
\end{figure}

For $T$ just below $T_c$, $T \lesssim T_c$, we find that $\omega^*$ is determined by $\langle \mathcal{O}\rangle$. More specifically, fig.~\ref{fig:condensedlowestQNM} shows that $\omega^* \,\propto - \, i \langle \mathcal{O} \rangle^2$ when $T\lesssim T_c$. In appendix~\ref{appendix:condensedlowestqnm}, for the case $Q=-1/2$ we show analytically (\textit{i.e.}\ without numerics) that $\omega^* \,\propto - \, i \langle \mathcal{O} \rangle^2$ for $T \lesssim T_c$. Given the mean-field scaling discussed in sec.~\ref{sec:review}, $\langle \mathcal{O}\rangle \,\propto \,\left(T_c-T\right)^{1/2}$ when $T \lesssim T_c$, we thus have $\omega^* \,\propto - i |T-T_c|$ when $T \lesssim T_c$.

As mentioned in section~\ref{sec:intro}, a pole in $\godo$ of the form $\omega^* \propto - i \langle \mathcal{O}\rangle^2$ is precisely the manifestation of the Kondo resonance that we expect at large $N$~\cite{Coleman2015}. In other words, in addition to the dynamically generated scale $T_K$, impurity screening, a phase shift, and so forth, our holographic Kondo model also correctly captures an essential spectral feature of the Kondo effect, namely the Kondo resonance.

\subsection{Screened Phase: Spectral Function}
\label{sec:condensedspec}

Knowing the result of subsection~\ref{sec:normspec}, that our spectral function $\rodo$ generically exhibits a Fano resonance associated with the lowest pole $\omega^*$ in $\godo$, and knowing the result of subsection~\ref{sec:condensedpoles}, that in the screened phase $\omega^*$ is purely imaginary and simply moves down the imaginary axis as $T$ decreases, we can anticipate how $\rodo$ will behave in the screened phase. Given that $\omega^*$ is purely imaginary, and hence does not break particle-hole symmetry $\textrm{Re}\,\omega \to - \textrm{Re}\,\omega$, we expect $\rodo$ to exhibit a $q=1$ symmetric Fano resonance at $\omega=0$. Moreover, given that $\omega^*$ moves straight down the imaginary $\omega$ axis as $T$ decreases, we expect the Fano resonance's width to increase as $T$ decreases.

\begin{figure}[h!]
\begin{center}
\includegraphics[width=0.7\textwidth]{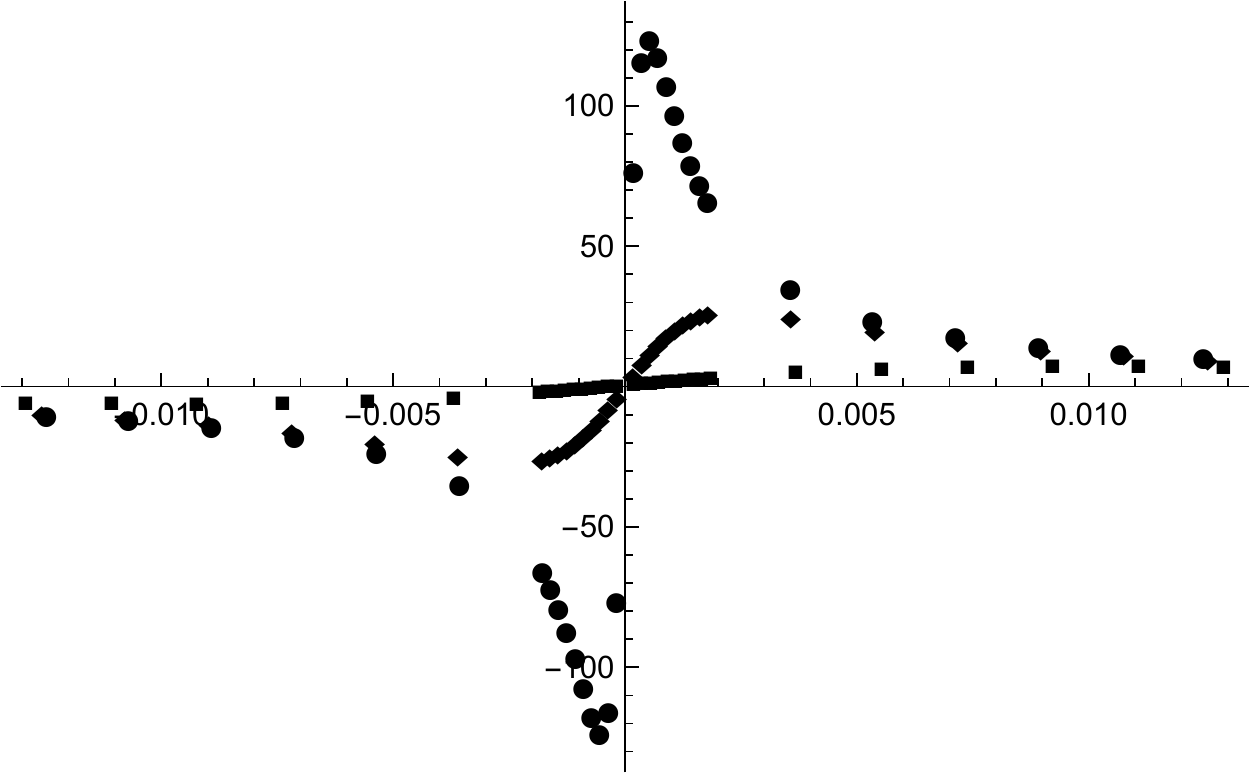}
\caption{\label{fig:condensedspec} The spectral function, $\rodo/\left(N/\kappa^2\right)$, as a function of $\omega/(2 \pi T)$ for the representative value $Q=0.5$ and in the screened phase, for $T/T_c=0.998$ (dots), $0.991$ (diamonds), and $0.964$ (squares). For all $T/T_c$ we find a $q=1$ symmetric Fano resonance at $\omega=0$ whose height decreases and width increases as $T/T_c$ decreases.}
\end{center}
\end{figure}

Our numerical results for $\rodo$ in the screened phase confirm these expectations. Fig.~\ref{fig:condensedspec} shows our numerical results for $\rodo/(N/\kappa^2)$ in the screened phase as a function of real-valued $\omega/(2 \pi T)$ for $Q=0.5$ and $T/T_c \approx 0.998$, $0.991$, and $0.964$. We indeed find only $q=1$ symmetric Fano resonances whose width increases as $T$ decreases. We also find that the resonance's height decreases rapidly as $T$ decreases: in fig.~\ref{fig:condensedspec}, $T/T_c$ decreases by only about $4\%$, from $T/T_c\approx 0.998$ down to $T/T_c \approx 0.964$, but the height of the peak drops by roughly two orders of magnitude. As $T$ decreases further (not shown in fig.~\ref{fig:condensedspec}), $\rodo$ continues to flatten, and indeed, as $T$ approaches zero, $\rodo$ appears to approach zero for all $\omega$. All of these features of $\rodo$ appear for other values of $Q$ as well, including $Q=0$.

In the standard (non-holographic) large-$N$ Kondo model with Abrikosov pseudo-fermions, the Kondo resonance has width $\propto \langle \mathcal{O}\rangle^2$~\cite{Coleman2015}. For $T \lesssim T_c$, the mean-field behavior $\langle \mathcal{O} \rangle \propto (T_c-T)^{1/2}$ then implies the width is $\propto T_c-T$. When $T \to 0$, $\langle \mathcal{O}\rangle$ reaches a finite value $\propto T_K^{1/2}$ at the minimum of its wine-bottle effective potential. The Kondo resonance then has width $\propto T_K$, similarly to finite $N$.

Our model also exhibits mean-field behavior, and hence a width $\propto T_c-T$ when $T \lesssim T_c$. However, in our screened phase, as $T$ decreases our manifestation of the Kondo resonance, \textit{i.e.}\ the $q=1$ symmetric Fano resonance in $\rodo$, flattens out, and ultimately disappears, so that at $T=0$ apparently $\rodo$ is featureless. What accounts for the difference? In our model, $\langle \mathcal{O}\rangle$'s effective potential is apparently unbounded: we found numerically that $\langle \mathcal{O} \rangle$ grows without bound as $T$ decreases, because $\Phi$ grows without bound. Indeed, as $T$ decreases, eventually the solutions for $a_t(z)$ and $\Phi(z)$ violate the probe limit: the stress-energy tensor grows without bound, and eventually cannot be neglected in Einstein's equation. That is unsurprising, given that in our bulk action eq.~\eqref{action}, $\Phi$'s potential is unbounded, being only a mass term, $M^2 \Phi^{\dagger} \Phi$. Presumably, stabilizing $\Phi$'s potential, for example with a $(\Phi^{\dagger}\Phi)^2$ term, would stabilize $\langle \mathcal{O}\rangle$, and hence stabilize the width of our resonance.

\section{Discussion and Outlook}
\label{sec:discussion}
\setcounter{equation}{0}

We studied the poles in retarded Green's functions and the associated peaks in spectral functions in the holographic Kondo model of refs.~\cite{Erdmenger:2013dpa,O'Bannon:2015gwa,Erdmenger:2015spo,Erdmenger:2015xpq}. We had three main results. First was the holo-ren of our model, which provided the covariant counterterms required to compute the renormalized free energy and one- and two-point functions in our model. Second, at all $T$, we found that generically the poles in our Green's functions have residue with non-zero imaginary part, giving rise to Fano resonances in spectral functions. Fano resonances occur when a resonance appears in a continuum (in energy) of states. Our continuum comes from $(0+1)$-dimensional scale invariance, inherited from $(1+1)$-dimensional scale invariance of our holographic CFT. Our resonances are possible because we break scale invariance via our marginally-relevant Kondo coupling. Third, in the screened phase, where $\langle \mathcal{O}\rangle\neq0$, and with $T$ just below $T_c$, we found a pole in $\godo$ of the form $\omega^* \propto \, - i \langle \mathcal{O}\rangle^2$, precisely as expected for the Kondo resonance at large $N$~\cite{Coleman2015}. In contrast, in the unscreened phase $\omega^*$ passed through the origin as $T$ decreased through $T_c$, which was clearly a strong coupling effect: in the standard (non-holographic) Kondo model at large $N$ and at leading order in perturbation theory in $\lambda$, in the unscreened phase $\omega^*$ sits at the origin of the complex $\omega$ plane for all $T$~\cite{Coleman2015}.

For the future, some obvious, immediate tasks involve improvements to our model. For example, giving our bulk scalar $\Phi$ a quartic self-interaction could not only prevent $\Phi$ from diverging at low $T$, and hence maintain the validity of the probe limit at low $T$, but could also prevent our Kondo resonance from disappearing as $T$ decreases, as we discussed in subsection~\ref{sec:condensedspec}. Indeed, adding a quartic term would introduce an additional dimensionful parameter in our model, which could presumably be fixed by demanding that our Kondo resonance has width $\propto \, T_K$ when $T=0$.

However, as discussed in refs.~\cite{Erdmenger:2013dpa,O'Bannon:2015gwa}, all holographic quantum impurity models to date, including ours, have a fundamentally worrying issue: the spin symmetry group is the \textit{gauge} group, $SU(N)$. Holography provides direct access only to gauge-invariant quantities. As a result, many important quantities that are not spin singlets, such as the magnetization and spin susceptibility, are prohibitively difficult if not impossible to calculate using holography. The obvious route to address this issue is to develop holographic quantum impurity models in which spin is a \textit{global} symmetry.

We have seen that even a minimal holographic quantum impurity model can produce Fano resonances. Indeed, Fano resonances require simple, common ingredients, and thus are very generic. We therefore expect Fano resonances in practically any holographic quantum impurity model, under the key condition that conformal symmetry is broken at the impurity. (Otherwise, all two-point functions at the impurity are determined by the conformal symmetry, as we mentioned in the section~\ref{sec:intro}.) In fact, more generally we expect asymmetric Fano resonances in practically any holographic system with a UV fixed point, breaking of scale invariance, and breaking of particle-hole symmetry.

Most importantly, we expect our holographic Kondo model, other similar holographic quantum impurity models, and variations of SYK models, to be useful in addressing many of the open questions mentioned in section~\ref{sec:intro}, about EE, quantum quenches, etc. We expect Fano resonances in particular to play a crucial role in developing a precise ``dictionary'' between theoretical models and experiments.

\section*{Acknowledgments}

We thank Ian Affleck, Natan Andrei, Piers Coleman, Mario Flory, Antal Jevicki, Henrik Johannesson, Andrew Mitchell, Max Newrzella, and Philip Phillips for helpful conversations and correspondence. A.~O'B. is a Royal Society University Research Fellow. J.~P. is supported by the Clarendon Fund and St John's College, Oxford, and by the European Research Council under the European Union's Seventh Framework Programme (ERC Grant agreement 307955). C.H. is supported by the Ramon y Cajal fellowship RYC-2012-10370, the
Asturian grant FC-15-GRUPIN14-108 and the Spanish national grant MINECO-16-FPA2015-63667-P.

\appendix

\renewcommand{\thesection}{\Alph{section}}
\renewcommand{\theequation}{\Alph{section}.\arabic{equation}}

\section*{Appendices}
\setcounter{section}{0}

\section{Near-Horizon Expansions in the Screened Phase}
\label{appendix:IRexp}
\setcounter{equation}{0}

In this appendix we determine the near-horizon expansions of the response functions $\car_{11}$, $\car_{12}$, $\car_{22}$, and $\car$ defined in eq.~\eqref{building-blocks} in the screened phase. We use these expansions to fix in-going boundary conditions at the horizon when we solve eq.~\eqref{riccati-screened-z} numerically for the response functions.

In this appendix we will switch from the holographic radial coordinate $z$ in eq.~\eqref{ads3metric} to the coordinate $\zeta \equiv z_H -z$, so that near the horizon, and using eq.~\eqref{r-coord} to translate from $z$ to $r$ of eq.~\eqref{gf-metric},
\be
\g=-\frac{2}{z_H^3}\zeta+\co(\zeta^2),\qquad
\pa_r=\sqrt{2z_H\zeta}\left(1+\co(\zeta)\right)\pa_{\zeta},\qquad
\frac12\g^{-1}\dot\g=\sqrt{\frac{z_H}{2\zeta}}\left(1+\co(\zeta)\right).
\ee
Near the horizon, eqs.~\eqref{eoms-bg} for the background fields $a_t^0$ and $\phi_0$ thus become 
\be
\pa_{\zeta}^2a_t^0-\frac{2}{z_H}\pa_{\zeta} a_t^0 -\frac{\f_0^2}{z_H\zeta}a_t^0= 0,\qquad 
\pa_{\zeta}^2\f_0+\frac{1}{\zeta}\pa_{\zeta}\f_0+\frac{\left(a_t^0\right)^2z_H^2}{4\zeta^2}\f_0-\frac{M^2}{2z_H\zeta}\f_0= 0,
\ee
with regular solutions
\be
a_t^0=a_{(1)}\Big(\zeta+\frac{(2+\f_0^2)}{2z_H}\zeta^2+\co(\zeta^3)\Big),\qquad
\f_0=\f_{(0)}\Big(1+\frac{M^2}{2z_H}\zeta+\co(\zeta^2)\Big),
\ee
with integration constants $a\sub{1}$ and $\f\sub{0}$, which we determine in our numerical solutions by matching with the integration constants in the near-boundary expansions. 

Using the above, we can determine the near-horizon expansions of the fluctuations $\delta a_t$, $\delta \f$, and $\delta \psi$ in eq.~\eqref{eoms-fl-FT}. Near the horizon, the constraint eq.~\eqref{eq:ar-fl-FT} becomes
\be
-\frac{z_H^3}{2\zeta}i\om\pa_{\zeta}\d a_t+2\f_{(0)}^2\pa_{\zeta}\d\j=0.
\ee
with solutions
\be
\label{eq:flucnearhorizon}
\d a_t= c_1 \, \zeta^{2+c_2}\left(1+\co(\zeta)\right),\qquad \d\j=c_1\frac{i\om z_H^3(2+c_2)}{4\f_{(0)}^2(1+c_2)}\zeta^{1+c_2}\left(1+\co(\zeta)\right), 
\ee
with integration constants $c_1$ and $c_2$. Inserting eq.~\eqref{eq:flucnearhorizon} into eqs.~\eqref{eq:Phi-fl-FT} and~\eqref{eq:Phi-dag-fl-FT} gives us
\begin{subequations}
	\bal
	&\left(\zeta^2\pa_{\zeta}^2+\frac{(2+c_2)z_H^2}{4(1+c_2)}\om^2\right)\d a_t= \frac{2}{z_H}\f_{(0)}a\sub{1}\zeta^2\d\f,\\
	&\left(\zeta^2\pa_{\zeta}^2+\zeta\pa_{\zeta}+\frac{z_H^2\om^2}{4}\right)\d\f= a\sub{1}\frac{\om^2z_H^5(2+c_2)}{8\f_{(0)}(1+c_2)}\d a_t,
	\eal
\end{subequations}
to leading order, with the linearly-independent solutions
\begin{subequations}
\label{eq:flucnearhorizonsolutions}
\bal
\d a_t &= c_1 \zeta^{2+c_2}\(1+\co(\zeta)\),\\
\d\f &= c_1\frac{z_H}{2\f_{(0)}a\sub{1}}\left(\frac{c_2+2}{c_2+1}\right)\left((c_2+1)^2+\left(\frac{z_H\o}{2}\right)^2\right)\zeta^{c_2}\left(1+\co(\zeta)\right),\quad c_2=\pm i\frac{z_H\om}{2},\\\NO\\
\d a_t &= c_1 \zeta^{2+c_2}\(1+\co(\zeta)\),\\
\d\f &= c_1\frac{z_H^5a_{(1)}\om^2}{8\f\sub{0}}\left(\frac{c_2+2}{c_2+1}\right)\left((c_2+2)^2+\left(\frac{z_H\o}{2}\right)^2\right)^{-1}\zeta^{2+c_2}\left(1+\co(\zeta)\right),\quad c_2=-1\pm i\frac{z_H\om}{2}.
\eal
\end{subequations}
The most general in-going solution is a linear combination of the solutions with
\beq
c_2=-i\frac{z_H\om}{2}\equiv c_{\textrm{in}}^{(1)}, \qquad c_2=-1-i\frac{z_H\om}{2}\equiv c_{\textrm{in}}^{(2)}.
\eeq

Near the horizon, the definitions of the response functions in eq.~\eqref{min-response-screened} become
\begin{subequations}
\label{eq:flucresponsenearhorizon}
	\bal
	&\d\dot a_t=\car_{11}(\d a_t+i\o\d\j)-\frac{2\zeta}{z_H^3}\car_{12}\d\f,\\
	&\d\dot\f=\frac12\car_{22}\d\f+\frac12\Big(\car_{12}-\frac{z_H^3}{2\zeta}\car\Big)(\d a_t+i\o\d\j),\\
	&\d\dot\j=\frac{i\o z_H^3}{4\zeta \f_{(0)}^{2}}\d\dot a_t.
	\eal
\end{subequations}
Inserting the two linearly-independent in-going solutions of eq.~\eqref{eq:flucnearhorizonsolutions} into eq.~\eqref{eq:flucresponsenearhorizon} leads to four algebraic equations for the leading near-horizon behavior of the response functions,
\begin{subequations}
	\bal
	&\frac{\om^2z_H^3}{4\f_{(0)}^2}\car_{11}+\frac{(c_{\textrm{in}}^{(1)}+1)^2+\left(\frac{z_H\o}{2}\right)^2}{z_H^2\f_{(0)}a_{(1)}}\car_{12}= -\sqrt{2z_H}(c_{\textrm{in}}^{(1)}+1)\zeta^{\frac12},\\
	&\frac{\om^2z_H^3}{4\f_{(0)}^2}\car_{11}+\frac{z_H^2\om^2a_{(1)}}{4\f_{(0)}}\left((c_{\textrm{in}}^{(2)}+2)^2+\left(\frac{z_H\o}{2}\right)^2\right)^{-1}\zeta^2\car_{12}=-\sqrt{2z_H}(c_{\textrm{in}}^{(2)}+1)\zeta^{\frac12},\\
	&\frac12\car_{22}-\frac{a_{(1)}z_H^2\om^2}{4\f_{(0)}}\left((c_{\textrm{in}}^{(1)}+1)^2+\left(\frac{z_H\o}{2}\right)^2\right)^{-1}\zeta\Big(\car_{12}-\frac{z_H^3}{2\zeta}\car\Big)= \sqrt{2z_H}\zeta^{-\frac12}c_{\textrm{in}}^{(1)},\\
	&\frac12\car_{22}-\frac{(c_{\textrm{in}}^{(2)}+2)^2+\left(\frac{z_H\o}{2}\right)^2}{z_H^2\f_{(0)}a_{(1)}}\zeta^{-1}\Big(\car_{12}-\frac{z_H^3}{2\zeta}\car\Big)=\sqrt{2z_H}(c_{\textrm{in}}^{(2)}+2)\zeta^{-\frac12},
	\eal
\end{subequations}
with solutions
\bsub
\label{Rs_IR_NLO}
\begin{align}
	\car_{11} &= i\frac{2\sqrt{2}\phi_{(0)}^2}{\o z_H^{3/2}}\sqrt{\zeta}
	\left\{1 + \zeta\(\frac{3}{4z_H} - \frac{2 i\phi_{(0)}^2}{\o z_H^2(1 - i \o z_H)} - \frac{i M^2 \o}{1 - i \o z_H}\)+\co(\zeta^2)\right\}\,, \\
	\mathcal{R}_{22} &= -i\frac{\sqrt{2}\o z_H^{3/2}}{\sqrt{\zeta}}
	\left\{1 - \zeta\(\frac{3}{4z_H} - \frac{i M^2}{\o z_H^2(1 - i \o z_H)}\)+\co(\zeta^2)\right\}\,, \\
	\mathcal{R}_{12} &= -\frac{\sqrt{2}z_H^{5/2}a_{(1)}\phi_{(0)}}{1 - i \o z_H}\sqrt{\zeta}\Bigg\{1 - \zeta\Bigg(\frac{i \o}{4(2 - i \o z_H)} 
	+ \frac{M^2 \o(3i + \o z_H)}{2(1 - i \o z_H)(2 - i \o z_H)} \notag \\
	&\qquad\qquad\qquad\qquad\qquad\qquad\qquad\qquad
	+ \frac{\big((2i + \o z_H)(1 + \o^2 z_H^2) + 2i\big)\phi_{(0)}^2}{2\o z_H^2(1 - i \o z_H)(2 - i \o z_H)}\Bigg)+\co(\zeta^2)\Bigg\},
\end{align}
\esub
which are the main results of this appendix. Inserting eq.~\eqref{Rs_IR_NLO} into the general solution for $\car$ in eq.~\eqref{R} then gives us
\be
\car= C(\o)\frac{\sqrt{2}}{z_H^{3/2}}\(\zeta^{\frac12+iz_H\o}+\co(\zeta^{3/2})\),
\ee
and hence in-going boundary conditions require that $C(\o)=0$ and thus $\car=0$, as advertised in subsection~\ref{subsec:screenedresponse}. As a result, the Riccati equations in eq.~\eqref{riccati-screened-r} simplify to those in eq.~\eqref{riccati-screened-z}.

\section{Near-Boundary Expansions}
\label{appendix:UVexp}
\setcounter{equation}{0}

In this appendix we determine the general Fefferman-Graham (FG) asymptotic expansions of the $AdS_2$ fields in our model. As mentioned at the beginning of section~\ref{sec:holorg} these FG expansions involve a number of subtleties, related to the special form of the FG expansion of gauge fields in $AdS_2$. In particular, the leading asymptotic mode of the gauge field is the charge $Q$ instead of the chemical potential $\m$, unlike gauge fields in higher-dimensional AdS spacetimes, and moreover the value of $Q$ affects the FG expansion of the scalar field $\F$. As a result, a well-defined space of asymptotic solutions requires keeping $Q$ fixed, which corresponds to an asymptotic second class constraint on the space of solutions. Such a constraint is unusual, compared to many holographic systems, although the constraint required for Lifshitz asymptotics in Einstein-Proca theory~\cite{Chemissany:2014xpa,Chemissany:2014xsa} is analogous.

A direct result of the constraint is that, if we allow fluctuations about a background solution to have non-zero variation of $Q$, then the background and fluctuations need not have the same FG expansions. Indeed, in that case, higher order fluctuations are increasingly dominant asymptotically, relative to both the background solutions and to the lower order fluctuations. As a result, the small fluctuation approximation breaks down asymptotically, and we are forced to work with a cut-off near the boundary, until fluctuations proportional to $\d Q$ are set to zero. In addition, generically no well-defined asymptotic solutions to the full non-linear equations of motion exist, so we must consider the FG expansions of the background and of the fluctuations separately. Below we determine the FG expansions both for the background and the fluctuations, discussing separately fluctuations with $\d Q\neq 0$ and $\d Q=0$. 

\paragraph{Note about Notation:} In this appendix and in appendix \ref{appendix:hol-ren}, $\co\sbtx{log}(x)$ denotes a quantity that asymptotes to zero like $x \log^k(x)$ as $x\to 0^+$, with $k$ a non-negative integer.  

\subsection{Expansions of the Background and the Second Class Constraint}

Upon choosing a gauge with $A_t=0$, the equations of motion for $a_t$, $\phi$, and $\psi$, eqs.~\eqref{eoms-fg}, become 
\begin{subequations}
	\label{gen-bg}
	\bal
	&\ddot a_t-\frac12\g^{-1}\dot\g\dot a_t-2\f^2(a_t-\pa_t\j)=0,\\
	&\ddot\f+\frac12\g^{-1}\dot\g\dot\f-\dot\j^2\f+\g^{-1}\pa_t^2\f-\g^{-1}(a_t-\pa_t\j)^2\f-M^2\f=0,\\
	&\pa_r(\f^2\dot\j)+\frac12\g^{-1}\dot\g\f^2\dot\j-\g^{-1}\pa_t\left(\f^2(a_t-\pa_t\j)\right)=0,\\
	&\g^{-1}\pa_t\dot a_t=2\f^2\dot\j.\label{eq:gen-bg-constraint}
	\eal
\end{subequations}
Given the asymptotic form of the metric, $\g\sim -e^{2r}$ as $r\to+\infty$, as long as $\f\to 0$ asymptotically (\textit{i.e.}\ the dual operator is relevant), then the gauge field's leading asymptotic behavior is $a_t\sim e^r Q(t)$, with $Q(t)$ an arbitrary function of time $t$. Moreover, $Q^2$ enters $\phi$'s equation as a mass term, so that $\phi$ has an ``effective mass'' $M^2 - Q^2$, hence $Q^2$ affects the FG expansion of $\phi$. A well-defined space of asymptotic solutions thus requires the (second class) constraint that $Q$ is fixed. The charge $Q$ is not automatically conserved by the equations of motion, due to the coupling to the charged scalar field. Charge conservation, therefore, can only be imposed as a boundary condition.   

As in ref.~\cite{Erdmenger:2013dpa}, we fix $Q$ such that $\mathcal{O}$ has dimension $1/2$, so that our Kondo coupling $\mathcal{O}^{\dagger}\mathcal{O}$ is classically marginal. The scalar's effective mass must thus \textit{saturate} the $AdS_2$ Breitenlohner-Freedman bound:   
\be\label{Q-condition}
M^2-Q^2=-\frac14.
\ee
We want to determine the FG expansions with $Q$ satisfying the constraint eq.~\eqref{Q-condition}. Crucially, in the first three equations in \eqref{gen-bg}, terms containing time derivatives affect only sub-leading orders in the FG expansion: for the leading non-normalizable and normalizable orders, we can thus ignore all time derivatives in eqs.~\eqref{gen-bg}. For similar reasons, we can take $\g = - e^{2r}$ for the purpose of determining the FG expansions. With these simplifications, eqs.~\eqref{gen-bg} become
\bsub
\bal
&\ddot a_t-\dot a_t-2\f^2a_t=0,\\
&\ddot\f+\dot\f-\dot\j^2\f+e^{-2r}a_t^2\f-M^2\f=0,\\
&\pa_r(\f^2\dot\j)+\f^2\dot\j=0,
\eal
\esub
and hence the FG expansions of the $AdS_2$ fields are 
\bsub
\label{UV-asymptotics-nl}
\bal
a_t &=e^rQ-2Q\Big(\frac13\a^2r^3+(\a^2-\a\b)r^2+(2\a^2-2\a\b+\b^2)r\Big)+\m(t)+\cdots,\\
\f &=e^{-r/2}\left(-\a(t) r+\b(t)\right)+\cdots,\\
\j &=\j_-(t)+\j_+(t) r^{-1}+\cdots,
\eal
\esub
where $\m(t)$, $\a(t)$, $\b(t)$ and $\j_\pm(t)$ are arbitrary functions of time, and $\ldots$ represent terms that vanish as $r \to \infty$ faster than those shown, and are completely determined by those shown, via the equations of motion. Inserting eqs.~\eqref{UV-asymptotics-nl} into eq.~\eqref{eq:gen-bg-constraint}, which is the constraint imposed by the $AdS_2$ $U(1)$ gauge invariance, and using eq.~\eqref{Q-condition}, we find $\j_+=0$ and $\frac12\a^{-2}\pa_tQ=0$. The FG expansions are thus parameterized by the arbitrary functions $\m(t)$, $\a(t)$, $\b(t)$ and $\j_-(t)$. Moreover, $\m(t)$ is defined only up to a $U(1)$ gauge transformation, $\m(t)\to\m(t)+\pa_t\l(t)$. We will refer to eqs.~\eqref{UV-asymptotics-nl} as ``background FG expansions,'' because $Q$ is required to satisfy eq.~\eqref{Q-condition}. Fluctuations are allowed to violate eq.~\eqref{Q-condition}, which leads to different FG expansions, as we will see.

\subsection{Expansions of the Response Functions}

In the unscreened phase, we want to find the FG expansions of the response functions $\car_{\F^\dag\F}$ and $\car_{\F\F^\dag}$, using the Riccati equations in eq.~\eqref{riccati-unscreened-r}. As above, to do so we may ignore terms involving time derivatives, \textit{i.e.}\ frequency $\omega$, and we may set $\gamma = - e^{2r}$, in eq.~\eqref{riccati-unscreened-r}, leading to
\be
\dot\car_{\F^\dag\F}+\car_{\F^\dag\F}+\car_{\F^\dag\F}^2+\frac14=0,\qquad \dot\car_{\F\F^\dag}+\car_{\F\F^\dag}+\car_{\F\F^\dag}^2+\frac14=0,
\ee
and hence the FG expansions of $\car_{\F^\dag\F}$ and $\car_{\F\F^\dag}$ are
\begin{subequations}
	\label{Rff-UV-sym}
	\bal
	&\car_{\F^\dag\F}=-\frac12+\frac{1}{r-\Hat\car_{\F^\dag\F}}=-\frac12+\frac1r+\frac{\Hat\car_{\F^\dag\F}}{r^2}+\cdots,\\
	&\car_{\F\F^\dag}=-\frac12+\frac{1}{r-\Hat\car_{\F\F^\dag}}=-\frac12+\frac1r+\frac{\Hat\car_{\F\F^\dag}}{r^2}+\cdots,
	\eal
\end{subequations}
where $\Hat\car_{\F^\dag\F}$ and $\Hat\car_{\F\F^\dag}$ are functions of $\o$, and $\ldots$ represent terms that vanish as $r \to \infty$ faster than those shown, and are completely determined by those shown, via eq.~\eqref{riccati-unscreened-r}.

In the screened phase, we instead need to solve instead the Riccati equations eqs.~\eqref{riccati-screened-r}, with $\car=0$, as required by in-going boundary conditions at the horizon, as shown in appendix~\ref{appendix:IRexp}. Again ignoring terms involving time derivatives, and setting $\g=-e^{2r}$, eqs.~\eqref{riccati-screened-r} become
\bsub
\label{RGeqs-simple-UV}
\bal
&\dot\car_{11}-\car_{11}+\car_{11}^2-\frac12 e^{2r}\car_{12}^2-2\f_0^2= 0,\label{RGeqs-simple-UV-R11}\\
&\dot\car_{12}+\car_{12}+\car_{11}\car_{12}+\frac12\car_{12}\car_{22}+4e^{-2r}\f_0a_t^0=0,\label{RGeqs-simple-UV-R12}\\
&\dot\car_{22}+\car_{22}-e^{2r}\car_{12}^2+\frac12\car_{22}^2+\frac12=0.\label{RGeqs-simple-UV-R22}
\eal
\esub
These equations admit two distinct classes of asymptotic solutions, depending on whether $\d Q(t)\neq 0$ or $\delta Q(t)=0$. We present both of these solutions in turn.

For fluctuations with $\d Q\neq 0$, the defining relations in eqs.~\eqref{min-response-screened} and the asymptotic solution for $a_t$ in eq.~\eqref{UV-asymptotics-nl} imply that asymptotically $\car_{11}\sim 1$. Eqs.~\eqref{RGeqs-simple-UV} then determine the leading asymptotic behavior of the response functions: $\car_{11}=1+\co\sbtx{log}(e^{-r})$, $\car_{12}=\co\sbtx{log}(e^{-3r/2})$, $\car_{22}=-1+\co(1/r)$. In eq.~\eqref{RGeqs-simple-UV-R22}, the term $\propto \car_{12}^2$ is exponentially subleading relative to the other terms, and hence can be ignored. The resulting equation for $\car_{22}$ then admits an exact solution, with asymptotic expansion
\be\label{R22-UV}
\car_{22}=-1+\frac{2}{r-\Hat\car_{22}/2}+\co\sbtx{log}(e^{-r}),
\ee 
where $\Hat\car_{22}$ is an undetermined function of $\omega$. Eqs.~\eqref{RGeqs-simple-UV-R11} and~\eqref{RGeqs-simple-UV-R12} then determine
\bsub
\bal
\car_{12} & =-\frac{4e^{-3r/2}}{r-\Hat\car_{22}/2}\int \tx dr \Big(r-\Hat\car_{22}/2\Big)e^{-r/2}\f_0(r)a_t^0(r)+\co\sbtx{log}(e^{-5r/2}),\label{R12-UV}\\
\car_{11} & =1+e^{-r}\int \tx dr e^r\Big(\frac12 e^{2r}\car_{12}^2+2\f_0^2\Big)+\co\sbtx{log}(e^{-2r}).\label{R11-UV}
\eal	
\esub
Expanding these then leads to the FG expansions
\bsub
\label{UV-exp-R}
\bal 
\car_{11}=\,&1+e^{-r}\left(\frac{8Q^2\a^2_0}{45}r^5-\frac{Q^2\a_0}{9}(\Hat\car_{22}\a_0+6\b_0)r^4+\frac{1}{18}\left((12-Q^2\Hat\car_{22}^2)\a^2_0+12Q^2(\Hat\car_{22}\a_0+\b_0)\b_0\right)r^3\right.\NO\\
&\left.\hskip.5in+\Big(\frac{Q\a_0}{36}(24\Hat\car_{12}+Q\a_0\Hat\car_{22}^3)-2\a_0\b_0-Q^2\b_0^2\Hat\car_{22}\Big)r^2\right.\NO\\
&\left.\hskip0.5in+\frac{1}{72}\left(Q\a_0\Hat\car_{22}(24\Hat\car_{12}+Q\a_0\Hat\car_{22}^3)-12Q(12\Hat\car_{12}+Q\a_0\Hat\car_{22}^3)\b_0+36(4+Q^2\Hat\car_{22}^2)\b_0^2\right)r\right.\NO\\
&\left.\hskip0.5in+\Hat\car_{11}+\co(1/r)\rule{0.0cm}{0.5cm}\right)+\co\sbtx{log}(e^{-2r}),\\
\car_{12}=\,&e^{-\frac{3r}{2}}\Big(\frac{4Q\a_0}{3}r^2-\frac{Q}{3}(\a_0\Hat\car_{22}+6\b_0)r-\frac{Q\Hat\car_{22}}{6}(\a_0\Hat\car_{22}-6\b_0)+\frac{\Hat\car_{12}}{r}+\frac{\Hat\car_{12}\Hat\car_{22}}{2r^2}\NO\\
&\hskip0.5in+\frac{\Hat\car_{12}\Hat\car_{22}^2}{4r^3}+\co(1/r^4)\Big)+\co\sbtx{log}(e^{-5r/2}),\\
\car_{22}=\,&-1+\frac{2}{r}+\frac{1}{r^2}\Hat\car_{22}+\frac{1}{2r^3}\Hat\car_{22}^2+\frac{1}{4r^4}\Hat\car_{22}^3+\frac{1}{8r^5}\Hat\car_{22}^4+\co(1/r^6),
\eal
\esub
where $\Hat\car_{11}$, $\Hat\car_{12}$ and $\Hat\car_{22}$ are undetermined functions of the frequency $\om$. If we plug eqs.~\eqref{UV-exp-R} into the defining relations eqs.~\eqref{min-response-screened}, then these asymptotic expansions lead to linear fluctuations that are asymptotically \textit{more} divergent that the background solutions in eqs.~\eqref{UV-asymptotics-nl}---an effect of the asymptotic second class constraint eq.~\eqref{Q-condition}, which is violated infinitesimally by the linear fluctuations with $\d Q\neq 0$. The second class constraint also causes the integration constant $\Hat\car_{22}$ to enter the asymptotic expansions of $\car_{11}$ and $\car_{12}$ before their corresponding integration constants $\Hat\car_{11}$ and $\Hat\car_{12}$. We must therefore determine the asymptotic expansions of $\car_{22}$ and $\car_{12}$ beyond the order where $\Hat\car_{22}$ and $\Hat\car_{12}$ appear linearly, since these terms enter in the expansions of $\car_{11}$ and $\car_{12}$. 

While fluctuations with $\delta Q \neq 0$ have three integration constants, $\Hat\car_{11}$, $\Hat\car_{12}$ and $\Hat\car_{22}$, fluctuations with $\delta Q =0$ have only one, as we will now show. For fluctuations with $\d Q=0$, the three response functions have the leading order behavior $\car_{11}=\co\sbtx{log}(e^{-r})$, $\car_{12}=\co\sbtx{log}(e^{-3r/2})$, and $\car_{22}= -1+\co(1/r)$. Eq.~\eqref{RGeqs-simple-UV} then implies that $\car_{22}$ is again given by eq.~\eqref{R22-UV}, while
\bsub
\label{eq:responsedQzero}
\bal
\car_{11}&=-e^r\int_r^\infty \tx dr'e^{-r'}\(\frac12 e^{2r'}\car_{12}^2+2\f_0^2\)+\co\sbtx{log}(e^{-2r}),\\
\car_{12}&=\frac{4e^{-r/2}}{r-\Hat\car_{22}/2}\int_r^\infty \tx dr'e^{-3r'/2}\(r'-\Hat\car_{22}/2\)\f_0(r')a_t^0(r')+\co\sbtx{log}(e^{-5r/2}).
\eal
\esub
Expanding eq.~\eqref{eq:responsedQzero} using eq.~\eqref{UV-asymptotics-nl} then gives the FG expansions
\bsub
\label{UV-exp-R-dQ=0}
\bal
\car_{11}&=\,e^{-r}\Big(-(1+4Q^2)\a^2_0r^2+\a_0\(2(1+4Q^2)\b_0-(1+20Q^2)\a_0\)r\\
&\hskip0.2in+\Big((1+28Q^2)\a_0\b_0-(1+4Q^2)\b^2_0-\frac12(1+4(2\Hat\car_{22}+21)Q^2)\a^2_0\Big)+\co(1/r)\Big)+\co\sbtx{log}(e^{-2r}),\NO\\
\car_{12}&=\,e^{-\frac{3r}{2}}\Big(-4Q\a_0 r+4Q(\b_0-2\a_0)+\frac{4Q\b_0-2(\Hat\car_{22}+4)Q\a_0}{r}+\co\Big(\frac{1}{r^2}\Big)\Big)+\co\sbtx{log}(e^{-\frac{5r}{2}}),\\
\car_{22}&=\,-1+\frac{2}{r}+\frac{1}{r^2}\Hat\car_{22}+\frac{1}{2r^3}\Hat\car_{22}^2+\frac{1}{4r^4}\Hat\car_{22}^3+\frac{1}{8r^5}\Hat\car_{22}^4+\co(1/r^6).
\eal
\esub
Inserting the expansions for $\car_{11}$, $\car_{12}$ and $\car_{22}$ for either $\d Q\neq 0$ or $\d Q=0$ into eqs.~\eqref{building-blocks}, then gives
\be\label{Rff-UV}
\car_{\F^\dag\F}=-\frac12+\frac1r+\frac{\Hat\car_{\F^\dag\F}}{r^2}+\cdots,\quad
\car_{\F\F^\dag}=-\frac12+\frac1r+\frac{\Hat\car_{\F\F^\dag}}{r^2}+\cdots,\quad
\car_{\F\F}=\frac{\Hat\car_{\F\F}}{r^2}+\cdots,
\ee
which is of the same form as the unscreened case, eq.~\eqref{Rff-UV-sym}, but now with the constraints
\be
\label{hat-responses-scalar}
	\Hat\car_{\F^\dag\F}=\Hat\car_{\F\F^\dag}=\Hat\car_{\F\F}+1/\kappa=\frac14(\Hat\car_{22}+2/\kappa),
\ee
where $\kappa=\beta_0/\alpha_0$ comes from the background solution for the scalar, as discussed below eq.~\eqref{sources}.

\section{Further Details of Holographic Renormalization}
\label{appendix:hol-ren}
\setcounter{equation}{0}

In this appendix we summarize some technical results related to the holo-ren in subsection~\ref{sec:holorgsub}. In particular, we determine asymptotically the functions $g_0(v)$, $g_1(v)$, and $g_2(v)$, defined in eq.~\eqref{counterterms-exp}, up to the relevant order for renormalizing the two-point functions, and we obtain explicit expressions for the renormalized response functions that enter in the two-point functions. 

\subsection{Determining the Boundary Counterterms}

We write $g_0$ and $g_1$ as in eq.~\eqref{eq:g0g1param}: $g_0=-u_o+h_0$ and $g_1=-1+h_1$. Plugging these into~eq.~\eqref{g-eqs} and expanding in $v$, and using the fact that $h_0$, $h_1$, $g_2$ and $g_3$ are all $\co\sbtx{log}(v)$ as $v \to 0$, we find
\be\label{hj-eqs}
\hskip-0.2cmh_0-v(h_0'^2+1/4)=\co\sbtx{log}(v^2),\quad h_1-v(2h_0'h_1'+2)= \co\sbtx{log}(v^2),\quad g_2-v(h_1'^2+2h_0'g_2')=\co\sbtx{log}(v^2),
\ee
where primes denote $\partial_v$ (see appendix \ref{appendix:UVexp} for the definition of $\co\sbtx{log}$). A simple power-counting argument using the near-boundary asymptotic expansion of the scalar field in eq.~\eqref{UV-asymptotics-nl} suffices to show that in general only terms up to order $\co\sbtx{log}(v)$ can potentially contribute to near-boundary divergences, so we can neglect all the right-hand-sides in eqs.~\eqref{hj-eqs}. The resulting equations can then be solved exactly.

The most general solution for $h_0(v)$ can be expressed implicitly in the form  
\be
\frac{1}{1-\l(v)}+\log(1-\l(v))=q_0+\log2-\frac12\log v,\qquad \l(v)\equiv\sqrt{\frac{4h_0(v)}{v}-1},
\ee 
where $q_0$ is an integration constant. Expanding this solution for small $v$, we obtain
\bal\label{h0-exp}
h_0(v)&=v\Big(\frac12+\frac{1}{\log v}+\frac{2q+1}{(\log v)^2}+\frac{4q^2}{(\log v)^3}+\frac{8q^2(q-1)}{(\log v)^4}+\frac{16q^2\left(q^2-\frac73q+1\right)}{(\log v)^5}\NO\\
&\hskip3.in+\frac{32q^2\left(q^3-\frac{47}{12}q^2+4q-1\right)}{(\log v)^6}+\cdots\Big),
\eal
where $q\equiv \log(-\log v)+c_0$. The equations for $h_1(v)$ and $g_2(v)$ are linear, with general solutions
\be
h_1=\vth(v)\Big(q_1-\int^v_0 \frac{\tx d\bar v}{\vth(\bar v)h_0'(\bar v)}\Big),\quad
g_2=\vth(v)\Big(q_2-\frac12\int^v_0 \frac{\tx d\bar v\; h_1'{}^2(\bar v)}{\vth(\bar v)h_0'(\bar v)}\Big),\quad \vth\equiv \exp\Big(\int^v_0 \frac{\tx d\bar v}{2\bar v h_0'(\bar v)}\Big),
\ee
where $\bar{v}$ is a dummy integration variable, and $q_1$ and $q_2$ are integration constants. Expanding these solutions at small $v$ gives us
\bsub
\label{h12-exp}
\bal
\vth(v)=&\frac{v}{(\log v)^2}\Big(1+\frac{4q}{\log v}+\frac{4q(3q-2)}{(\log v)^2}+\frac{8q\left(4q^2-7q+2\right)}{(\log v)^3}+\frac{16q\left(5q^3-\frac{47}{3}q^2+12q-2\right)}{(\log v)^4}\NO\\
&\hskip2.2in+\frac{80q\left(36q^4-171q^3+238q^2-108q+12\right)}{15(\log v)^5}+\cdots\Big),\\
h_1(v)=&-\frac{2}{3}v\log v\Big(1-\frac{2q}{\log v}+\frac{4q}{(\log v)^2}+\frac{4q(q-2)}{(\log v)^3}+\frac{8q\left(\frac23q^2-3q+2\right)}{(\log v)^4}\NO\\
&\hskip2.5in+\frac{8q\left(3q^3-22q^2+36q-12\right)}{3(\log v)^5}+\cdots\Big)+q_1\vth(v),\\
g_2(v)=&-\frac{4}{45}v(\log v)^3\Big(1-\frac{6q-\frac52}{\log v}+\frac{12q^2+2q-5}{(\log v)^2}-\frac{8q^3+26q^2-6q-10+\frac{15q_1}{2}}{(\log v)^3}\\
&\hskip0.0in+\frac{16q^3+52q^2-12q-5(4-3q_1)}{(\log v)^4}+\frac{2q\left(12q^3+4q^2-174q+45q_1-24\right)}{3(\log v)^5}+\cdots\Big)+q_2\vth(v).\NO
\eal
\esub

The integration constants $q_0$, $q_1$, $q_2$ correspond respectively to the constants $\Hat\car_{22}$, $\Hat\car_{12}$ and $\Hat\car_{11}$ in the near-boundary expansions of the response functions in eq.~\eqref{UV-exp-R}. This can be deduced as follows. Combining \eqref{building-blocks} and \eqref{responses-G}, and using the expansion in eq.~\eqref{counterterms-exp} and eqs.~\eqref{hj-eqs}, we obtain 
\bsub
\label{eq:appcresponseeq}
\bal
\car_{11}^\cg &=1+h_1+2Q^2g_2+\co\sbtx{log}(e^{-2r}),\\
\car_{12}^\cg &=-2h_1'e^{-2r}\f a_t+\co\sbtx{log}(e^{-5r/2}),\\
\car_{22}^\cg &=-4\f^2h_0''=-2+\frac{1}{2h'_0}+\co\sbtx{log}(e^{-r}),\label{lastequal}
\eal
\esub
where the last equality in eq.~\eqref{lastequal} follows from the first in eq.~\eqref{hj-eqs}. As in eq.~\eqref{responses-G}, the superscript $\cg$ indicates that these response functions are obtained from eq.~\eqref{HJ-ansatz}, not the full on-shell action. Moreover, taking $\p^\cg_\f=\p^\cg_\F+\p^\cg_{\F^\dag}$ (see eq.~\eqref{canonical-relations}) with the $\pi_{\phi}$ in eq.~\eqref{momenta-phase} gives
\be\label{flow-eq}
\dot v=-2vh'_0+\co\sbtx{log}(e^{-2r}).
\ee
Eqs.~\eqref{eq:appcresponseeq} and~\eqref{flow-eq}, together with eqs.~\eqref{hj-eqs}, suffice to show that $\car_{11}^\cg$, $\car_{12}^\cg$ and $\car_{22}^\cg$ satisfy the corresponding eqs.~\eqref{RGeqs-simple-UV}, with the important caveat that $\f_0$ in eqs.~\eqref{RGeqs-simple-UV} is replaced by $\f$, \textit{i.e.}~the solution that satisfies the first order eq.~\eqref{flow-eq}. Since $\f_0$ and $\f$ have the same asymptotic behavior, apart from the values of the coefficients $\a$ and $\b$, $\car_{11}^\cg$, $\car_{12}^\cg$ and $\car_{22}^\cg$ have near-boundary expansions of the same form as those of $\car_{11}$, $\car_{12}$ and $\car_{22}$, and hence they should have the same integration constants. This implies that $q_0$, $q_1$ and $q_2$ are related to $\Hat\car_{22}$, $\Hat\car_{12}$ and $\Hat\car_{11}$, respectively, although the explicit map between these integration constants is rather complicated.

However, the fact that $\car_{11}^\cg$, $\car_{12}^\cg$ and $\car_{22}^\cg$ satisfy eqs.~\eqref{RGeqs-simple-UV} with $\f_0$ replaced by $\f$, does have implications for the boundary counterterms. We have just argued that the near boundary expansion of $\car_{22}^\cg$ is of the same form as that of $\car_{22}$ in eq.~\eqref{UV-exp-R}, but with some integration constant $\Hat\car_{22}^\cg$ that is related to $q_0$. Since $v=\f^2$, eq.~\eqref{flow-eq} implies that $\f$ has a near-boundary expansion of the form in~\eqref{UV-asymptotics-nl} with
\be\label{beta-condition}
\b=\(\Hat\car_{22}^\cg/2-2\)\a.
\ee
If we want to use $\car_{11}^\cg$, $\car_{12}^\cg$ and $\car_{22}^\cg$ as counterterms to renormalize $\car_{11}$, $\car_{12}$ and $\car_{22}$, respectively, we must set $\b=\b_0$ and $\a=\a_0$, since these are the values appearing in the near boundary expansions in eq.~\eqref{UV-exp-R}. However, eq.~\eqref{beta-condition} then forces us to set $\Hat\car_{22}^\cg=4+2/\k$. This poses no problem for renormalizing $\car_{22}$, but as we pointed out earlier, an unusual feature of the asymptotic expansions in eq.~\eqref{UV-exp-R} is that $\car_{11}$ and $\car_{12}$ contain divergences that involve $\Hat\car_{22}$, which is a dynamical quantity determined by the near-horizon conditions. Setting $\Hat\car_{22}^\cg=4+2/\k$ will thus not renormalize $\car_{11}$ and $\car_{12}$. This is similar to cases where a source for an irrelevant operator is turned on perturbatively, much like our $\d Q$, and additional multi-trace counterterms are required \cite{vanRees:2011fr}. In our case this means $\car_{11}^\cg(v)$ and $\car_{12}^\cg(v)$ should be considered functions of $\car_{22}^{\rm ren}=\car_{22}+2-\frac{1}{2(h_0^{\rm ct})'}$ as well, \textit{i.e.}~$\car_{11}^\cg(v_0;\car_{22}^{\rm ren})$ and $\car_{12}^\cg(v_0;\car_{22}^{\rm ren})$, where $v_0=\f_0^2$ should evaluated on the background. These functions can be determined by demanding they satisfy exactly the same equations as $\car_{11}$ and $\car_{12}$, eq.~\eqref{RGeqs-simple-UV}.

As discussed in section~\ref{sec:holorg}, an additional complication arises due to the logarithmic dependence of the functions $g_0(v)$, $g_1(v)$, and $g_2(v)$ on $v$, which forces us to introduce explicit cutoff dependence in the counterterms, to ensure they are local functions of the scalar source. For example, keeping only terms that contribute to the near-boundary divergences we set
\be\label{final-counterterms}
g_0^{ct}(v)=v\left(1/2-1/r\right)-u_o,
\ee
which suffices to renormalize the on-shell action (evaluated with $\d Q=0$), as well as $\car_{22}$.
 
We will not give the explicit expressions for the counterterms $\car_{12}^{\rm ct}$ and $\car_{11}^{\rm ct}$ here, but they can be constructed as outlined above, and they allow us to obtain the renormalized quantities
\bsub
\label{R-infinity}
\bal
\car_{11}^{\infty} &\equiv\lim_{r\to\infty}\(e^{r}(\car_{11}+\car_{11}^{\rm ct})\)=\Hat\car_{11}+\cc_{11}(\Hat\car_{22},\a_0,\b_0),\\
\car_{12}^{\infty} &\equiv\lim_{r\to\infty}\(re^{3r/2}(\car_{12}+\car_{12}^{\rm ct})\)=\Hat\car_{12}+\cc_{12}(\Hat\car_{22},\a_0,\b_0),\\
\car_{22}^{\infty} &\equiv\lim_{r\to\infty}\(r^2(\car_{22}+\car_{22}^{\rm ct})\)=\Hat\car_{22},
\eal
\esub
where $\cc_{11}(\Hat\car_{22},\a_0,\b_0)$ and $\cc_{12}(\Hat\car_{22},\a_0,\b_0)$ are determined by the specific choice for the counterterm functions.  

\subsection{Renormalized Response Functions }

To determine the renormalized response functions, and hence the corresponding two-point functions, we need to consider the variation of the one-point functions. Moreover, if we want to allow $\d Q\neq 0$, then the variations of the one-point functions must be considered at a radial cutoff, and the cutoff should be removed only in the end.    

A general variation of the $AdS_2$ gauge field momentum at a radial cutoff yields
\bal
\d\p_a^t=-N\sqrt{-\g}\g^{-1}\d\dot a_t&=-N\sqrt{-\g}\g^{-1}\(\car_{aa}\d a_t+\g \car_{a\F}\d\F+\g \d\F^\dag\car_{a\F^\dag}\),\NO\\
&=-N\sqrt{-\g}\g^{-1}\(\car_{aa}^{\rm ren}\d a_t^{\rm ren}+\g \car_{a\F}^{\rm ren}\d\F+\g \d\F^\dag\car^{\rm ren}_{a\F^\dag}\),
\eal
where we have used the definitions in eqs.~\eqref{ren-variables-cutoff} and introduced the renormalized response functions
\bal
\car^{\rm ren}_{aa}&=\frac{\car_{aa}}{1+\car_{aa}\(\cg_{u}^{\rm ct}+2u\cg_{uu}^{\rm ct}\)},\NO\\
\car^{\rm ren}_{a\F}&=\frac{\car_{a\F}-\car_{aa}\frac1N\sqrt{-\g}\;\g^{-1}\cg_{u v}^{\rm ct}\p_a^t\F^\dag}{1+\car_{aa}\(\cg_{u}^{\rm ct}+2u\cg_{uu}^{\rm ct}\)},\qquad
\car^{\rm ren}_{a\F^\dag}=\frac{\car_{a\F^\dag}-\car_{aa}\frac1N\sqrt{-\g}\;\g^{-1}\cg_{u v}^{\rm ct}\p_a^t\F}{1+\car_{aa}\(\cg_{u}^{\rm ct}+2u\cg_{uu}^{\rm ct}\)}.
\eal
Using the fact that $\car=0$ for solutions that satisfy ingoing boundary conditions at the horizon, we easily find that the response functions $\car_{aa}$, $\car_{a\F}$ and $\car_{a\F^\dag}$ are related to those introduced in eqs.~\eqref{response-screened} and~\eqref{min-response-screened} as $\car_{aa}=\car_{11}$, $\car_{a\F}=\g^{-1}\car_{\F^\dag a}$, $\car_{a\F^\dag}=\g^{-1}\car_{\F a}$.

However, since the one-point function associated with the $AdS_2$ gauge field is given by $a_t^{\rm ren}$, we need to express $\d a_t^{\rm ren}$ in terms of the variations of the other variables. Namely,
\be
\d a_t^{\rm ren}=\frac{-\g}{\car^{\rm ren}_{aa}}\(\car^{\rm ren}_{a\F}\d\F+\car^{\rm ren}_{a\F^\dag}\d\F^\dag+\frac{1}{N\sqrt{-\g}}\d\p_a^t\)=-\(\car^{\rm ren}_{\p_a^t\F}\d\F+\car^{\rm ren}_{\p_a^t\F^\dag}\d\F^\dag+\car_{\p_a^t\p_a^t}^{\rm ren}\d\p_a^t\),
\ee
where
\bsub
\bal
\car^{\rm ren}_{\p_a^t\F}&=\frac{\g \car^{\rm ren}_{a\F}}{\car^{\rm ren}_{aa}}=\Big(\g\car_{a\F}-\frac1N\sqrt{-\g}\;\cg_{u v}^{\rm ct}\p_a^t\F^\dag\Big)(1+\co\sbtx{log}(e^{-r})),\\
\car^{\rm ren}_{\p_a^t\F^\dag}&=\frac{\g \car^{\rm ren}_{a\F^\dag}}{\car^{\rm ren}_{aa}}=\Big(\g\car_{a\F^\dag}-\frac1N\sqrt{-\g}\;\cg_{u v}^{\rm ct}\p_a^t\F\Big)(1+\co\sbtx{log}(e^{-r})),\\
\car_{\p_a^t\p_a^t}^{\rm ren}&=-\frac{\sqrt{-\g}}{N\car^{\rm ren}_{aa}}=-\frac{\sqrt{-\g}}{N}\Big(1+\car_{aa}\(\cg_{u}^{\rm ct}+2u\cg_{uu}^{\rm ct}\)\Big)(1+\co\sbtx{log}(e^{-r})),
\eal
\esub
and we have used that $\car_{aa}=1+\co\sbtx{log}(e^{-r})$. These renormalized response functions at the radial cutoff are directly related with the physical two-point functions in section \ref{sec:holorg}.

Similarly, the generic variation of the renormalized scalar canonical momenta at the radial cutoff gives
\bsub
\label{mom-var-new}
\bal
\d\p_{\F^\dag}^{\rm ren}&=-N\sqrt{-\g}\(\d\dot\F+\d(\cg_{v}^{\rm ct}\F)\)\NO\\
&=-N\sqrt{-\g}\(\car_{\F^\dag\F}\d\F+\car_{\F^\dag\F^\dag}\d\F^\dag+\g^{-1}\car_{\F^\dag a}\d a_t+\d(\cg_{v}^{\rm ct}\F)\)\NO\\
&=-N\sqrt{-\g}\(\car^{\rm ren}_{\F^\dag\F}\d\F+\car^{\rm ren}_{\F^\dag\F^\dag}\d\F^\dag\)+\car^{\rm ren}_{\F^\dag\p_a^t}\d\p_a^t,\\\NO\\
\d\p_{\F}^{\rm ren}&=-N\sqrt{-\g}\(\d\dot\F^\dag+\d(\cg_{v}^{\rm ct}\F^\dag)\)\NO\\
&=-N\sqrt{-\g}\(\car_{\F\F}\d\F+\car_{\F\F^\dag}\d\F^\dag+\g^{-1}\car_{\F a}\d a_t+\d(\cg_{v}^{\rm ct}\F^\dag)\)\NO\\
&=-N\sqrt{-\g}\(\car^{\rm ren}_{\F\F}\d\F+\car^{\rm ren}_{\F\F^\dag}\d\F^\dag\)+\car^{\rm ren}_{\F\p_a^t}\d\p_a^t,
\eal
\esub
where 
\bal
\car^{\rm ren}_{\F\F}&=\car_{\F\F}+\cg^{\rm ct}_{vv}(\F^\dag)^2+\co\sbtx{log}(e^{-r}), &&
\car^{\rm ren}_{\F^\dag\F^\dag}=\car_{\F^\dag\F^\dag}+\cg^{\rm ct}_{vv}\F^2+\co\sbtx{log}(e^{-r}), \\
\car^{\rm ren}_{\F\F^\dag}&=\car_{\F\F^\dag}+\(\cg^{\rm ct}_v+v\cg^{\rm ct}_{vv}\)+\co\sbtx{log}(e^{-r}),&&
\car^{\rm ren}_{\F^\dag\F}=\car_{\F^\dag\F}+\(\cg^{\rm ct}_v+v\cg^{\rm ct}_{vv}\)+\co\sbtx{log}(e^{-r}),\NO\\
\car^{\rm ren}_{\F\p_a^t}&=\Big(\car_{\F a}-\frac{\sqrt{-\g}}{N}\cg^{\rm ct}_{uv}\p_a^t\F^\dag\Big)\(1+\co\sbtx{log}(e^{-r})\), &&
\car^{\rm ren}_{\F^\dag\p_a^t}=\Big(\car_{\F^\dag a}-\frac{\sqrt{-\g}}{N}\cg^{\rm ct}_{uv}\p_a^t\F\Big)\(1+\co\sbtx{log}(e^{-r})\).\NO
\eal
These renormalized response functions at the radial cutoff are also directly related with the physical two-point functions in section \ref{sec:holorg}.

Finally using eq.~\eqref{building-blocks} and the limits in eq.~\eqref{R-infinity}, we can remove the radial cutoff to obtain the renormalized response functions
\begin{subequations}
	\label{ren-responses}
	\bal
	\Hat\car_{\F\F^\dag}^{\rm ren}&=\lim_{r\to\infty} \left(r^2 \car_{\F\F^\dag}^{\rm ren}\right)=\Hat\car_{\F\F^\dag}=\frac14(\Hat\car_{22}+2\b_0/\a_0),\\
	\Hat\car_{\F^\dag\F}^{\rm ren}&=\lim_{r\to\infty} \left(r^2 \car_{\F^\dag\F}^{\rm ren}\right)=\Hat\car_{\F^\dag\F}=\frac14(\Hat\car_{22}+2\b_0/\a_0),\\
	\Hat\car_{\F\F}^{\rm ren}&=\lim_{r\to\infty} \left(r^2 \car_{\F\F}^{\rm ren}\right)=\Hat\car_{\F\F}=\frac14(\Hat\car_{22}-2\b_0/\a_0),\\
	\Hat\car_{\F^\dag\F^\dag}^{\rm ren}&=\lim_{r\to\infty} \left(r^2 \car_{\F^\dag\F^\dag}^{\rm ren}\right)=\Hat\car_{\F^\dag\F^\dag}=\frac14(\Hat\car_{22}-2\b_0/\a_0),\\
	\Hat\car^{\rm ren}_{\F\p_a^t}&=\lim_{r\to\infty} \Big(re^{-r/2} \car_{\F\p_a^t}^{\rm ren}\Big)=\Hat\car_{\F\p_a^t}=-\frac12\(\Hat\car_{12}^\infty-\o/\a_0\),\\
	\Hat\car^{\rm ren}_{\F^\dag\p_a^t}&=\lim_{r\to\infty} \Big(re^{-r/2} \car_{\F^\dag\p_a^t}^{\rm ren}\Big)=\Hat\car_{\F^\dag\p_a^t}=-\frac12\(\Hat\car_{12}^\infty+\o/\a_0\),\\
	\Hat\car^{\rm ren}_{\p_a^t\F^\dag}&=\lim_{r\to\infty} \Big(re^{-3r/2} \car_{\F\p_a^t}^{\rm ren}\Big)=\Hat\car_{\p_a^t\F^\dag}=-\frac12\(\Hat\car_{12}^\infty-\o/\a_0\),\\
	\Hat\car^{\rm ren}_{\p_a^t\F}&=\lim_{r\to\infty} \Big(re^{-3r/2} \car_{\F^\dag\p_a^t}^{\rm ren}\Big)=\Hat\car_{\p_a^t\F}=-\frac12\(\Hat\car_{12}^\infty+\o/\a_0\),\\
	\Hat\car^{\rm ren}_{\p_a^t\p_a^t}&=\lim_{r\to\infty} \Big( \car_{\p_a^t\p_a^t}^{\rm ren}\Big)=\Hat\car_{\p_a^t\p_a^t}=\frac1N \Hat\car_{11}^\infty,
	\eal
\end{subequations}
where $\Hat\car_{11}^\infty$ and $\Hat\car_{12}^\infty$ are defined in eq.~\eqref{R-infinity}. Eqs.~\eqref{ren-responses} are valid for the screened phase only. In the unscreened phase, the scalar's response functions $\Hat\car_{\F^\dag\F}$ and $\Hat\car_{\F\F^\dag}$ are integration constants determined by imposing boundary conditions on the horizon, while all other response functions vanish.

\section{Analytic Derivation of the Lowest Pole in the Screened Phase}
\label{appendix:condensedlowestqnm}
\setcounter{equation}{0}

In this appendix we present an analytic (\textit{i.e.}\ non-numerical) derivation of the behavior $\omega^* \propto - i \langle \mathcal{O}\rangle^2$ of the lowest pole in the screened phase, for $T \lesssim T_c$.

In this appendix we use the metric in eq.~\eqref{ads3metric}, but with the re-scaling in eq.~\eqref{eq:rescaledcoords} to produce dimensionless coordinates,
\beq
\label{eq:rescaledcoords}
(z/z_H,t/z_H,x/z_H) \to (z,t,x),
\eeq
which leaves the metric in eq.~\eqref{ads3metric} invariant, except for $h(z)  = 1 - z^2/z_H^2 \to 1-z^2$, so the boundary remains at $z=0$ but the horizon is now at $z=1$. We also re-scale $a_t(z)z_H \to a_t(z)$, which is then dimensionless. After the re-scaling, $\Phi(z)$'s asymptotic expansion is that of eq.~\eqref{eq:rescaledphiexp},
\beq
\Phi(z) = \alpha_T \, z^{1/2} \ln z + \beta_T \, z^{1/2} + \ldots,
\eeq
where here and below $\ldots$ represents terms that vanish faster than those shown when $z \to 0$, and the boundary condition $\alpha = \kappa \beta$ becomes $\alpha_T = \kappa_T \beta_T$. We additionally re-scale to produce a dimensionless frequency: $\omega z_H = \omega/(2 \pi T) \to \omega$. Moreover, in this appendix we exclusively use $Q=-1/2$.

We introduce fluctuations of the defect fields, for example $a_t(z,t) = a_t^0(z) + \delta a_t(z,t)$, where $a_t^0(z)$ is the background solution and $\delta a_t(z,t)$ is the fluctuation, and similarly $\Phi(z,t)=\Phi_0(z)+\delta \Phi(z,t)$, and $\Phi^{\dagger}(z,t)=\Phi_0^{\dagger}(z)+\delta \Phi^{\dagger}(z,t)$. In the screened phase, $\Phi_0(z)\neq0$ and $\Phi_0^{\dagger}(z)\neq0$. In this appendix we will assume the background solution is real, $\Phi_0(z) = \Phi_0^{\dagger}(z)$. Next we Fourier transform using $\partial_t \to - i \omega$, and use the same notation for the Fourier transforms of the fluctuations, for example $\delta a_t(z,\omega)$. Linearizing the equations of motion about in the fluctuations then gives the fluctuation equations (the equivalent of eq.~\eqref{eoms-fl-FT}, but in the coordinates of eq.~\eqref{eq:rescaledcoords}),
\begin{subequations}
\label{eq:appfluceoms}
\bea
\delta\Phi''+\frac{h'}{h}\delta\Phi'+\frac{(\omega+ a_t^0)^2}{h^2}\delta\Phi+\frac{\omega+2 a_t^0}{h^2}\Phi_0 \delta a_t & = & 0,\\
\delta\Phi^{\dagger\,''}+\frac{h'}{h}\delta\Phi^{\dagger\,'}+\frac{(\omega- a_t^0)^2}{h^2}\delta\Phi^{\dagger}-\frac{\omega-2 a_t^0}{h^2}\Phi^{\dagger}_0 \delta a_t & = & 0, \\
\delta a_t''+\frac{2}{z}\delta a_t'-\frac{2\Phi^{\dagger}\Phi}{z^2 h}\delta a_t+\frac{\Phi_0\delta\Phi^{\dagger}}{z^2 h}\left(\omega-2 a_t^0\right) -\frac{\Phi^{\dagger}_0\delta\Phi}{z^2 h}(\omega+2 a_t^0) & = & 0,\\
\omega z^2 \delta a_t'+h \left[\Phi_0(\delta\Phi'- \delta\Phi^{\dagger\,'})-\Phi'_0(\delta\Phi- \delta\Phi^{\dagger})\right] & = & 0,
\eea
\end{subequations}
where prime denotes $\partial_z$, for example $\Phi' \equiv \partial_z \Phi$.

We want the QNMs, that is, solutions for the fluctuations that are normalizable at the boundary $z=0$ and in-going at the horizon $z=1$, which exist only for particular $\omega$~\cite{Birmingham:2001pj,Kovtun:2005ev}. The asymptotic expansions of the fluctuations are
\beq
\delta a_t = \frac{\delta Q}{z} + \delta \mu + \ldots, \qquad \delta \Phi = \delta \alpha_T \, z^{1/2} \log z + \delta \beta_T \, z^{1/2} + \ldots.
\eeq
To guarantee normalizability, and specifically to guarantee that the asymptotic expansions of the fluctuations do not have terms more divergent than the asymptotic expansions of the background solutions, we must impose $\delta Q =0$, which requires $\delta \alpha_T = \kappa \, \delta\beta_T$, with the same value of $\kappa$ as the background solution $\Phi_0(z)$.

We parameterize the solutions of eq.~\eqref{eq:appfluceoms} as
\beq
\delta \Phi(z,\omega) = h^{-i \omega/2} \,p(z) \,y(z,\omega), \qquad \delta \Phi^{\dagger}(z,\omega) = h^{-i \omega/2} \,p(z) \,y^{\dagger}(z,\omega), \qquad \delta a_t(z,\omega) = h^{1-i \omega/2} \, a(z,\omega),
\eeq
where the powers of $h$ are determined by the in-going boundary condition at the horizon, $p(z)$ is the background solution $\Phi_0(z)$ with $\alpha = 1$, so that asymptotically
\beq
p(z) = z^{1/2} \log z + \frac{1}{\kappa_T} \, z^{1/2} + \ldots,
\eeq
and now we must solve for $y(z,\omega)$, $y^{\dagger}(z,\omega)$, and $a(z,\omega)$, which must be regular at both the boundary $z=0$ and the horizon $z=1$.

We want the QNM solutions for $T$ near $T_c$, where the condensate $\langle \mathcal{O} \rangle \, \propto \, \alpha/\kappa$ is small, or equivalently $\Phi_0(z)$ is negligible. We thus treat $p(z)$ as a small correction to the solution in the unscreened phase, that is, we use the background solution with $\Phi_0(z)=0$ and $Q=-1/2$, where
\beq
\label{eq:appbgasol}
a_t^0(z) = - \frac{1}{2} \left(-1 + \frac{1}{z}\right), 
\eeq
and then determine $p(z)$ by solving the equation of motion for the scalar, linearized about the solution with $\Phi_0(z)=0$ and eq.~\eqref{eq:appbgasol}, which gives
\beq
p(z) = -\left|\Gamma\left(\frac{1+i}{2}\right)\right|^2 \sqrt{\frac{z}{z+1}} \, P_{\frac{i-1}{2}}\left(\frac{3z-1}{z+1}\right),
\eeq
where $P_{\nu}$ is a Legendre function of the first kind.

When $T\lesssim T_c$, we know from subsection~\ref{sec:condensedpoles} that the lowest QNM frequency $\omega^*$ is near the origin of the complex $\omega$ plane, and hence is also small. We thus expand $y(z,\omega)$, $y^{\dagger}(z,\omega)$, and $a(z,\omega)$ in both $\omega$ and also $\alpha \,\propto \,\kappa \langle \,\mathcal{O}\rangle$,
\beq
y(z,\omega) = \sum_{n,m=0}^{\infty} \omega^n \alpha^m y_{nm}(z), \qquad y^{\dagger}(z,\omega) = \sum_{n,m=0}^{\infty} \omega^n \alpha^m y^{\dagger}_{nm}(z), \qquad a(z,\omega) = \sum_{n,m=0}^{\infty} \omega^n \alpha^m a_{nm}(z),
\eeq
so that now we must solve for the coefficients $y_{nm}(z)$, $y_{nm}^{\dagger}(z)$, and $a_{nm}(z)$. For $n=0$ and $m=0$,
\beq
y_{00}'' + \left[\frac{2p'}{p}+\frac{h'}{h}\right] y_{00}' = 0, \qquad \omega z^2 \left[a_{00}'+\frac{h'}{h} a_{00}\right]=0,
\eeq
and $y^{\dagger}_{00}(z)$ obeys the same equation as $y_{00}(z)$. The only solutions regular at both the boundary $z=0$ and the horizon $z=1$ are $y_{00}'(z)=0$, ${y^{\dagger}_{00}}'(z)=0$, and $a_{00}(z)=0$. For higher values of $n$ and $m$, the equations for the coefficients are inhomogeneous,
\beq
y_{nm}'' + \left[\frac{2p'}{p}+\frac{h'}{h}\right] y_{nm}' = I_{nm}, \qquad z^2 \left[a_{nm}'+\frac{h'}{h} a_{nm}\right]= A_{nm},
\eeq
where $y^{\dagger}_{nm}$ obeys the same equation as $y_{nm}$, but with source $I_{nm}^{\dagger}$. The sources $I_{nm}$ and $A_{nm}$ depend only on solutions at lower order in $n$ and $m$. For example, $I_{n0}=I^{\dagger}_{n0}=-\frac{2a_t^0}{h} \,a_{(n-1)0}$, which implies $y'_{n0}=y^{\dagger \,\prime}_{n0}$, which in turn implies $A_{n0}=0$. Furthermore, $A_{0m}=0$ so that $a_{0m}=0$. Determining the sources $I_{nm}$, $I^{\dagger}_{nm}$, and $A_{nm}$ is straightforward but unilluminating, so we will not present explicit results for them. However, the most singular behavior possible at the horizon $z=1$ is $I_{nm} \propto \,(z-1)^{-1}$, and similarly for $I^{\dagger}_{nm}$ and $A_{nm}$. As a result, solutions regular at the horizon $z=1$ have the form
\beq
y_{nm}'(z) =- \frac{1}{h(z)\,p^2(z)}\int_z^1 d\bar{z} \, h(\bar{z}) \,p(\bar{z})^2 \, I_{nm}(\bar{z}), \qquad a_{nm}(z) = -\frac{1}{h(z)} \int_z^1 d\bar{z} \, \frac{h(\bar{z})}{\bar{z}^2} A_{nm}(\bar{z}),
\eeq
where $\bar{z}$ is a dummy variable, and ${y^{\dagger}_{nm}}'(z)$ obeys the same equation as $y_{nm}'(z)$, but with $I_{nm} \to I_{nm}^{\dagger}$. Regularity of $a(z,\omega)$ at the boundary requires
\beq
\label{eq:apparegcond}
\int_0^1 d\bar{z} \, h(\bar{z}) \,p(\bar{z})^2 \left(I_{01}(\bar{z})-I_{01}^{\dagger}(\bar{z})\right) = 0,
\eeq
and a second condition, identical to eq.~\eqref{eq:apparegcond}, but with $I_{01} \to I_{02}$ and $I_{01}^{\dagger} \to I_{02}^{\dagger}$. Regularity of $y(z,\omega)$ at the boundary requires
\beq
\label{eq:appphiregcond}
\int_0^1 d \bar{z} \,h(\bar{z}) \,p(\bar{z})^2 \left[ \omega \, I_{01}(\bar{z}) + \omega^2 \, I_{02}(\bar{z}) + \alpha^2 \, I_{20}(\bar{z}) \right] = 0,
\eeq
while regularity of $y^{\dagger}(z,\omega)$ at the boundary requires a condition identical to eq.~\eqref{eq:appphiregcond}, but with $I_{01} \to I_{01}^{\dagger}$, $I_{02} \to I_{02}^{\dagger}$, and $I_{20} \to I_{20}^{\dagger}$. However, using $I_{20}=I^{\dagger}_{20}$, as mentioned above, and the second regularity condition for $a(z,\omega)$, we can show that the regularity condition for $y^{\dagger}(z,\omega)$ is equivalent to that for $y(z,\omega)$ in eq.~\eqref{eq:appphiregcond}. We are thus left with only eq.~\eqref{eq:appphiregcond}, which will be satisfied only for certain values of $\omega$. In particular, in our regime of interest, with small $\omega$ and $\alpha$, the solution of eq.~\eqref{eq:appphiregcond} gives the lowest QNM frequency,
\beq
\label{eq:applowestqnm}
\omega^* \approx - \alpha^2 \,\frac{\int_0^1 d \bar{z} \,h(\bar{z}) \,p(\bar{z})^2 I_{20}(\bar{z})}{\int_0^1 d \bar{z} \,h(\bar{z}) \,p(\bar{z})^2 I_{01}(\bar{z})}.
\eeq
Performing the integrals in eq.~\eqref{eq:applowestqnm} numerically, we find $\omega^* \approx - i \, 17 \, \alpha^2$. Given $\alpha \propto \,\kappa \langle \mathcal{O}\rangle$, we have thus shown that for $Q=-1/2$, and in the $T \lesssim T_c$ regime of the screened phase, $\omega^* \propto \, - i \langle \mathcal{O}\rangle^2$, as advertised in subsection~\ref{sec:condensedpoles}.

\addcontentsline{toc}{section}{References}


\bibliography{kondo2pointv2}
\bibliographystyle{utphys}

\end{document}